\pdfoutput=1
\documentclass[12pt, a4paper]{article}

\usepackage[nosort]{cite}
\usepackage{epsf, amssymb} 
\usepackage{mathrsfs, epstopdf} 
\usepackage{amsmath}
\usepackage{hyperref, graphicx}
\usepackage{mathrsfs}
\usepackage[bf,footnotesize]{caption2}
\usepackage{epsfig}
\usepackage{multirow}
\usepackage{color}
\usepackage{braket}
\usepackage{appendix}
\usepackage{lscape}
\usepackage{slashed}

\hypersetup{
colorlinks=true,
linkcolor=red,
anchorcolor=black,
citecolor=blue,
filecolor=cyan,
menucolor=red,
runcolor=filecolor,
urlcolor=magenta,
bookmarks=true,
bookmarksnumbered=true
}

\def\dst{\displaystyle}

\def\dst{\displaystyle}

\def\lsim{\mathrel{\rlap{\lower3pt\hbox{\hskip0pt$\sim$}}
   \raise1pt\hbox{$<$}}}         
\def\gsim{\mathrel{\rlap{\lower4pt\hbox{\hskip1pt$\sim$}}
   \raise1pt\hbox{$>$}}}         

\newcommand{\eps}{\epsilon}
\newcommand{\hhref}[1]{\href{http://arxiv.org/abs/#1}{arXiv:#1}}

\newcommand{\bea}{\begin{eqnarray}}
\newcommand{\eea}{\end{eqnarray}}
\newcommand{\be}{\begin{equation}}
\newcommand{\ee}{\end{equation}}
\newcommand{\bit}{\begin{itemize}}
\newcommand{\eit}{\end{itemize}}

\begin{document}

\setcounter{page}{0}
\thispagestyle{empty}

\begin{flushright}
CERN-PH-TH/2013-161 
\end{flushright}

\vskip 8pt

\begin{center}
{\bf \Large
Strong Higgs Interactions at a Linear Collider
}
\end{center}

\vskip 10pt

\begin{center}
{\bf Roberto~Contino$^{\,a}$, Christophe~Grojean$^{\, b, c}$, Duccio~Pappadopulo$^{\, d, e}$, Riccardo~Rattazzi$^{\, f}$  and Andrea~Thamm$^{\, c, f}$}  
\footnote{Email: \url{roberto.contino@roma1.infn.it},~ \url{christophe.grojean@cern.ch},\\ ~\url{ pappadopulo@berkeley.edu}, ~ \url{riccardo.rattazzi@epfl.ch},~ \url{andrea.thamm@epfl.ch} }

\end{center}

\vskip 20pt

\begin{center}

$^{a}$ {\small \it Dipartimento di Fisica, Universit\`a di Roma ``La Sapienza" and INFN, Sezione di Roma, Italy}
\vskip 5pt
$^{b}$ {\small \it ICREA at IFAE, Universitat Aut\`onoma de Barcelona, E-08193 Bellaterra, Spain}
\vskip 5pt
$^{c}$ {\small \it CERN, Physics Department, Theory Unit, Geneva, Switzerland}
\vskip 5pt
$^{d}$ {\small \it Department of Physics, University of California, Berkeley, CA 94720, USA}
\vskip 5pt
$^{e}$ {\small \it Theoretical Physics Group, Lawrence Berkeley National Laboratory, Berkeley, CA 94720, USA}
\vskip 5pt
$^{f}$ {\small \it Institut de Th\'eorie des Ph\'enom\`enes Physiques, EPFL, CH-1015 Lausanne, Switzerland}
\vskip 5pt

\end{center}

\vskip 13pt 

\begin{abstract}
\vskip 3pt
\noindent
We study the impact of Higgs precision measurements at a high-energy and high-luminosity linear electron positron collider, such as CLIC or the ILC, on the parameter space of a strongly interacting Higgs boson. Some combination of anomalous couplings are already  tightly constrained by current fits to electroweak observables. However, even small deviations in the cross sections of single and double Higgs production, or the mere detection of a triple Higgs final state, can help establish whether it is a composite state and whether or not it emerges as a pseudo-Nambu--Goldstone boson from an underlying broken symmetry. We obtain an estimate of the ILC and CLIC sensitivities on the anomalous Higgs couplings from a study of $WW$ scattering and $hh$ production which can be translated into a sensitivity on the compositeness scale $4\pi f$, or equivalently on the degree of compositeness $\xi=v^2/f^2$. We summarize the current experimental constraints, from electroweak data and direct resonance searches, and the expected reach of the LHC and CLIC on $\xi$ and on the scale of the new resonances.
\end{abstract}

\vskip 13pt
\newpage

\section{Introduction}
\label{sec:introduction}

With the discovery of the Higgs boson~\cite{ATLASdiscovery, CMSdiscovery} and with the absence,  thus far, of any clear evidence for New Physics (NP),  
a basic feature of the dynamics underlying electroweak symmetry breaking (EWSB) has begun to materialize: the new states associated with that dynamics 
do not seem to be as light as naturalness considerations would have recommended. That this might be the case has been suggested for quite some time by 
considerations based on precision electroweak and flavor data, but the LHC results have made this picture more concrete. Of course, we may just sit at the 
edge of New Physics and evidence may plentifully show up in the next run of the LHC. However, once a bit of un-naturalness is accepted, it is {\it natural} to 
expect, or fear, that history may repeat itself at 13~TeV. For instance, keeping the composite Higgs scenario in mind~\cite{compositeHiggs},  a  plausible  
situation  is  one where, at the end of its  program, the LHC  will have only measured inconclusive $O(10\%)$  deviations from the Standard Model (SM) 
in the Higgs couplings. 
In a definitely more optimistic situation  these small deviations would appear along with some new states, but without a clear indication for their role in the  
EWSB dynamics. 
Under the above circumstances, the next  experimental project would be more one of  exploration and discovery than one of refinement and consolidation. 
With this in mind, a high-energy hadron machine like the LHC at $\sqrt{s} = 33\,$TeV~\cite{Todesco:2011np} 
would superficially seem better suited than a cleaner but less powerful  leptonic machine like the ILC~\cite{Baer:2013cma} or CLIC~\cite{CLICCDR}. 
However, given the criticality of the decisions we may face in the coming years, it is important to carefully assess the 
potential of each machine in each plausible NP scenario. It is the goal of this paper to provide one such assessment: by focussing on a high-energy 
lepton machine such as CLIC, we shall explore its potential in the exploration of the composite Higgs scenario.

Single Higgs production at a linear collider, even at 500 GeV,  is known to be a  very sensitive probe of compositeness~\cite{Barger:2003rs}, 
even though an indirect 
one. In the case where resonances are still out of reach, a more direct probe on compositeness is offered by the study of the interactions 
of longitudinally polarized electroweak vector bosons and the Higgs. The relevant processes are $VV\to VV$ and $VV \to hh$ ($V=W,Z$), whose cross sections
grow like~$E^2$. There exist several studies of $VV\to VV$ at hadron 
machines~\cite{bagger,ballestrero,Butterworth:2002tt,Han:2009em,Espriu:2012ih, Freitas:2012uk, Contino:2010mh} but just a few of 
$VV\to hh$~\cite{Contino:2010mh, Giudice:2007fh}. The study of these processes in hadron collisions is not an easy task. $VV\to VV$ offers final states with 
leptons  that stand out well against the QCD background. However, the genuine SM contribution to $VV\to VV$ happens to be numerically so large that  the effects 
of compositeness dominate only at very high energy, where LHC parton luminosities are  small~\cite{Contino:2010mh}. The result is a poor reach on the scale 
of compositeness. In the case of $VV\to hh$ the genuine SM contribution is numerically small, so that in principle this would be a good probe of compositeness.
However the final states and branching ratios of Higgs decay are not favorable in a hadronic environment. The reach on compositeness from this process at the 
LHC is thus also not very good. Instead, as we will show in this paper, a machine like CLIC offers the right combination of a clean environment and 
center-of-mass energy to significantly probe the composite Higgs scenario. The improvement compared to hadron machines is particularly stark for  
$VV\to hh$. Our main result is that $VV\to hh$ at CLIC offers about half the reach on the scale of compositeness as single Higgs production at a $500\,$GeV ILC.
However the observation of a  cross section for $hh$ production that grows with energy would be a more direct and convincing evidence of the strongly coupled 
nature of NP. It also turns out that in the presence of a signal in $VV\to hh$ one may in principle be able to make more refined statements about the 
nature of $h$. First of all,  via a comparison with single $h$ processes,  one  can nicely and directly confirm that $h$ is part of a doublet. Moreover one  can in principle test
 whether $h$ is a pseudo Nambu--Goldstone boson  (PNGB) living in a coset space~\cite{compositeHiggs} or whether it is a generic composite 
scalar.  A way in which this could also be done is by studying triple Higgs production: $VV\to hhh$. 
This process could be marginally observable in the case of a generic 
composite~$h$, while in the case of a PNGB  non-trivial selection rules suppress its rate below observability.

This paper is organized as follows. In Section~\ref{sec:Hcouplings}, we review the general parametrization of the Higgs couplings and the relative importance of 
energy-growing $2 \to 2$ scattering processes.
Current constraints on the couplings, especially from electroweak precision tests, and their consequences on the scale of NP are discussed in detail 
in Section~\ref{sec:Constraints}. In Sections~\ref{sec:SingeDoubleHiggsProd} and~\ref{sec:VVtohhh}, we study the information contained in single, double and 
triple Higgs production on the structure of the underlying theory. A quantitative analysis of the ILC and CLIC sensitivities on the anomalous couplings and their 
reach in the parameter space is given in Section~\ref{VVtoHH}. We present our conclusions and outlook in Section~\ref{sec:conclusions}.

\section{General parametrization of the Higgs couplings}
\label{sec:Hcouplings}

Under the assumption that the mass scale at which new states appear is large, $m_\rho \gg m_h$, 
the recently discovered Higgs  boson can be described by
means of an effective Lagrangian. Motivated by the experimental evidence accumulated both at LEP and recently at the LHC, we will assume
that the dynamics behind electroweak symmetry breaking has an approximate custodial invariance, under which $h$ is a singlet. 
The effective Lagrangian can be organized by expanding in the number of derivatives and classifying
the various terms according to the number of $h$ fields. The expression obtained in this way extends
the EW chiral Lagrangian~\cite{EWchiralL} to include 
the light state $h$. Such a construction does not assume that $h$ is part of an $SU(2)_L$ doublet, nor does it make
hypotheses on  the strength of its interactions, as long as it is weakly coupled at energies of the order of its mass. 
It is thus completely general and applies as well to the case where $h$ is a Higgs-like impostor  not directly involved in  EWSB.
Despite the strong indications from the experimental measurements in favor of a SM-like Higgs boson, hence in favor of
the Higgs being part of a weak doublet, the purpose of this paper is to study processes with multiple Higgs production, 
and the general parametrization adopted can help emphasizing the peculiarity of a linearly-realized EW symmetry.
In the following we will further assume that
$h$ is a CP-even scalar. This choice is both motivated  from the theoretical point of view (it follows for example in minimal composite Higgs theories),
and supported by the preliminary results on the Higgs couplings obtained by the LHC collaborations.

At  $O(p^2)$ in the derivative expansion,
the bosonic part of the effective Lagrangian thus reads~\cite{Contino:2010mh, earlyCL}
\begin{equation} \label{eq:HiggsLag}
{\cal L} = \dst \frac{1}{2} \left(\partial_\mu h\right)^2 - V(h) + \dst \;  \frac{v^2}{4} \text{Tr}\left(D_\mu \Sigma^\dagger D^\mu \Sigma\right)
\left( 1 + 2 a\, \frac{h}{v} + b\, \frac{h^2}{v^2} +  b_3\, \frac{h^3}{v^3} + \dots \right) \, ,
\end{equation}
where the $2\times 2$ matrix $\Sigma$ is defined as
\begin{equation}
\Sigma(x) = \exp\left( i \sigma^a \chi^a(x)/v \right)\, , 
\end{equation}
and $\chi^a(x)$ are the Nambu--Goldstone (NG) bosons of the global coset $SO(4)/SO(3)$ which are eaten in the unitary gauge  to form the longitudinal
polarizations of the $W$ and $Z$. Under the custodial $SO(3)$ the $\chi$'s transform as a triplet.
In eq.~(\ref{eq:HiggsLag}), $V(h)$ is the potential for $h$
\begin{equation}
V(h) = \frac{1}{2} m_h^2 h^2 + d_3 \left( \frac{m_h^2}{2v} \right) h^3 + d_4 \left( \frac{m_h^2}{8v^2} \right) h^4 + \dots\, ,
\end{equation}
and $a,b, b_3,d_3,d_4$ are arbitrary dimensionless parameters. The dots stand for terms of higher order in $h$.
For the SM Higgs $a=b=d_3=d_4=1$ while all higher-order terms vanish.  The dilaton couplings are instead characterized by the relations $a=b^2$, $b_3=0$~\cite{dilaton}.

Any deviation of the couplings $a,b, b_3$ from their SM values implies
the energy growth of some scattering amplitude whose strength can be parametrized in terms of a ``running'' coupling $\bar{g}(\sqrt{s})$ at a given center-of-mass (c.o.m.) energy $\sqrt{s}$.
For example, the couplings $a$ and $b$ control the strength of the interactions in $2\to 2$ processes among $\chi$'s and $h$.
Under the assumption of $SO(3)$ custodial invariance, the scattering amplitudes read
\begin{align}
{\cal A}\!\left(\chi^a \chi^b \to \chi^c \chi^d \right)  & = A(s,t,u) \, \delta^{ab}\delta^{cd} + A(t,s,u)\, \delta^{ac} \delta^{bd} + A(u,t,s)\, \delta^{ad} \delta^{bc} \, , \\[0.3cm]
{\cal A}\!\left(\chi^a \chi^b \to hh \right) & = A_{hh}(s,t,u)\, \delta^{ab} \, , \\[0.3cm]
{\cal A}\!\left(\chi^a \chi^b \to \chi^c h \right) & = A_{h\chi}(s,t,u)\, \eps^{abc} \, , 
\end{align}
where $s,t,u$ are the usual Mandelstam variables. As implied by the equivalence theorem~\cite{equivtheorem, Wulzer:2013mza}, at high energy each of the above amplitudes equals 
one in which each external $\chi$  is replaced by the corresponding longitudinal vector boson ($\chi^\pm \to W_L^\pm$, $\chi^0 \to Z_L$).
From the Lagrangian in eq.~(\ref{eq:HiggsLag}), at leading order in the derivative expansion, it follows $A(s,t,u)  = (1-a^2) s/v^2$ and
$A_{hh}(s,t,u) = (a^2-b) s/v^2$. In both these cases the scattering amplitude  defines a coupling strength
\begin{equation}
\label {eq:twototwo}
{\cal A}( 2 \to 2) = \delta_{hh} \frac{s}{v^2} \equiv (\bar g(\sqrt{s}))^2 \, ,
\end{equation}
where we indicate by $\delta_{hh}$ both $a^2-1$ (for $\chi\chi\to \chi\chi$) and $a^2-b$ (for $\chi\chi\to hh$). 
A measurement of the $V_LV_L\to V_LV_L$ and $V_LV_L\to hh$ ($V=W,Z$) scattering rates at a given center of mass energy $\sqrt s$ thus corresponds to the 
measurement of an effective coupling  $\bar g(\sqrt{s})$, characterizing the strength of the EWSB dynamics.
The effective coupling $\bar g(\sqrt{s})$ grows with energy so that perturbativity, and with it the validity of the effective Lagrangian, would be lost at the scale 
$\sqrt {s_*}$ where  $\bar g(\sqrt{s_*})\sim 4\pi$. A reasonable expectation is then that new states will  UV complete the effective Lagrangian at a scale 
$m_\rho\leq \sqrt {s_*}$. The new states would expectedly saturate the growth of the effective coupling to  $g_\rho \equiv \bar g(m_\rho)\leq 4\pi$.

In the case of the scattering $\chi \chi \to \chi h$, 
Bose and crossing symmetries imply that the function $A_{h\chi}(s,t,u)$ is antisymmetric under the exchange of any two Mandelstam
variables. As a consequence, 
the lowest-order contribution to $A_{h\chi}(s,t,u)$ arises at $O(p^6)$, that is $A_{h\chi} \propto (s-u)(u-t)(t-s)$,
in accordance with the fact that there exists no local operator at the level of two and four derivatives giving a vertex with three $\chi$'s.
The corresponding scattering amplitude, $V_LV_L\to V_L h$, is expected to be suppressed  by a factor $(s/m^2_\rho)^2$ compared to that 
of $V_LV_L\to V_LV_L, hh$, and is thus not a sensitive probe of the  Higgs  interaction strength at energies below the scale $m_\rho$ of New Physics.
In fact, the absence of an energy growth in the $\chi \chi \to \chi h$ amplitude could have been anticipated on the basis of a simple 
symmetry argument. The request of custodial invariance fixes the global coset to be $SO(4)/SO(3)$, which is a symmetric space.
The grading of its algebra, under which all broken generators and thus all NG bosons change sign,
is an accidental symmetry of the $O(p^2)$  Lagrangian~(\ref{eq:HiggsLag})~\footnote{
It coincides with parity up to a spatial inversion: $P = P_0 P_{LR}$, with $P_0: \{ \vec x \to -\vec x, t\to t \}$.}
\begin{equation}
\label{eq:PLR}
P_{LR}: \qquad  \chi^a(x) \to - \chi^a(x)\, , \qquad h(x) \to h(x)\, .
\end{equation}
Any process with an odd number of $\chi$'s, including $\chi \chi \to \chi h$, must thus vanish at leading derivative order. 
Furthermore, although $P_{LR}$ is generically broken at $O(p^4)$, it turns out that none of the $P_{LR}$-odd operators with four derivatives
contributes to $2\to 2$ processes~\cite{Contino:2011np}. 
In absence of custodial symmetry, on the other hand, the global coset is $SU(2)\times U(1)/U(1)$ rather than $SO(4)/SO(3)$.
This is not a symmetric space, and there is no  grading symmetry which forbids  vertices  with three NG bosons at $O(p^2)$.
In particular,  the operator $[\text{Tr}( \Sigma^\dagger D_\mu \Sigma \sigma^3)]^2 h$ contains the term 
$h\, \partial_\mu \chi^3 (\chi^+ i\overleftrightarrow{\partial_\mu} \chi^-)$, which gives ${\cal A}(\chi^+ \chi^- \to \chi^3 h) \propto (t-u)$. 
In practice, the experimental results on the Higgs couplings obtained by the LHC collaborations already set tight limits on
possible custodial breaking effects~\cite{Farina:2012ea, CMS:yva, ATLAS:2013sla} and thus on the energy growth of $V_LV_L\to V_L h$. These new constraints are not surprising given the very strong constraint on custodial symmetry breaking provided by electroweak precision tests at LEP/SLC/Tevatron.

One might ask whether the amplitude of the process $V_T V_L \to V_L h$, with one transversely polarized vector boson, grows
with the energy and thus probes the Higgs interaction strength. By virtue of the equivalence theorem,  at high energy this coincides
with the  amplitude of $V_T \chi \to \chi h$, for which a naive power counting would suggest ${\cal A} \sim g \sqrt{s}/v$. 
A direct calculation, on the other hand, reveals that the energy-growing term cancels after summing all relevant diagrams,
thus implying ${\cal A}(V_T \chi \to \chi h) \sim g^3 (v/\sqrt{s})$.~\footnote{The cancellation follows from the fact that all the diagrams have the same
dependence on the Higgs couplings, namely they are all proportional to $a$. Since in the SM limit $a=1$ the amplitude cannot grow
with the energy, by continuity this implies that the same  holds true for any $a$.}
Eventually, the leading contribution to $VV\to Vh$ comes from the scattering amplitude with \textit{two} transversely polarized vector bosons,
${\cal A}(V_T V_T \to V_L h) \sim g^2$, which makes it clear that this process cannot be used to probe the Higgs interaction strength. 
Incidentally, notice that there is no analog cancellation in the scatterings with zero or two Higgses and one transverse vector boson, that 
is: ${\cal A}(V_T \chi \to hh) \sim (a^2-b) g\sqrt{s}/v$ and ${\cal A}(V_T \chi \to \chi\chi) \sim (a^2-1) g\sqrt{s}/v$.

\vspace{0.5cm}
So far our discussion has been general, since the effective Lagrangian (\ref{eq:HiggsLag}) applies to any scalar $h$ with arbitrary couplings
(provided the custodial symmetry is exact). One might however consider a situation in which future experiments constrain the Higgs couplings to be close 
to their SM value, so that only small deviations are allowed. 
In this case it is convenient to adopt a more specific effective description in which
$h$ is assumed to be part of an $SU(2)_L$ doublet $H$ and the Lagrangian is expanded  in powers of the $H$ field (as well as in the number 
of derivatives). 
The  list of dimension-6 operators of such an effective Lagrangian has been discussed at length in the literature~\cite{DIM6EFFLAG}, for recent 
reviews see Refs.~\cite{Contino:2013kra, Elias-Miro:2013mua}.
The case of a strongly interacting light Higgs (SILH) was addressed in this context in Ref.~\cite{Giudice:2007fh}, where a power counting 
was introduced to estimate the Wilson coefficients. A similar scenario, limited to bosonic operators, was also
studied  earlier in Ref.~\cite{Barger:2003rs}.
A simple yet crucial observation is that any additional power of $H$ costs a factor $g_\rho/m_\rho \equiv 1/f$, where $g_\rho \leq 4\pi$
denotes the coupling strength of the Higgs to New Physics states; any additional derivative instead is suppressed by a factor 
$1/m_\rho$.~\footnote{Extra powers of the gauge fields are also suppressed by $1/m_\rho$ as they can only appear through covariant derivatives.}
If the light Higgs interacts strongly with the new dynamics, $g_\rho \gg 1$, then the  leading corrections to low-energy observables
arise from operators with extra powers of $H$ rather than derivatives. This remark greatly simplifies the list of relevant operators. By concentrating  on the 
bosonic part of the SILH Lagrangian and assuming custodial invariance, there are two such operators involving only the Higgs multiplet  
\begin{equation} \label{SILHcH}
O_H = \frac{c_H}{2 f^2}\partial_\mu|H|^2 \partial^\mu |H|^2\, , \qquad O_6 = - \frac{c_6 \lambda}{f^2} \left( H^\dagger H \right)^3\, ,
\end{equation}
where $\lambda$ is the quartic coupling which appears in front of the marginal operator $(H^\dagger H )^2$. 
The $O(1)$ coefficients $c_H$ and $c_6$ control the Higgs couplings $a,b,b_3,d_3$ at  order $(v/f)^2$.
Under the assumption of $h$ being part of a doublet, the Higgs couplings of eq.~(\ref{eq:HiggsLag})
are thus correlated and functions of a smaller set of parameters. 
Note that there exists only one dimension-6 structure involving fermions: $(H^\dagger H)H\bar f f $. Under the assumption of minimal flavor violation 
there are thus three additional operator coefficients $c_{t}$, $c_{b}$ and $c_{\tau}$, describing the non-linear couplings of the Higgs multiplet to  
up- and down-type quarks and to charged leptons respectively~\cite{Giudice:2007fh}.
Operators of dimension 8  induce corrections of order $(v/f)^4$ to the Higgs couplings.
As described in more detail in Appendix~\ref{app:dim8}, there are only two dimension-8 operators which modify  $a, b, b_3, d_3$
\begin{equation}
\label{eq:OHpO8}
O_H^\prime = \frac{c_H'}{2 f^4}|H|^2\partial_\mu|H|^2 \partial^\mu |H|^2\, , \qquad O_8 = - \frac{c_8 \lambda}{f^4} \left( H^\dagger H \right)^4\, ,
\end{equation}
where the coefficients $c_H'$ and $c_8$ are expected to be of order 1.
The expressions for the couplings at $O(v^4/f^4)$ thus read
\begin{gather}
\label{eq:SILHrelations1}
\begin{gathered}
a = 1 -\frac{c_H}{2} \frac{v^2}{f^2} + \left( \frac{3c_H^2}{8} - \frac{c_H^\prime}{4} \right) \frac{v^4}{f^4}  \,  , \qquad 
b = 1-2 c_H \frac{v^2}{f^2} + \left(  3c_H^2  - \frac{3c_H^\prime}{2}  \right) \frac{v^4}{f^4} \, , \\[0.4cm]
b_3 = - \frac{4c_H}{3}  \frac{v^2}{f^2} + \left(  \frac{14c_H^2}{3} - 2 c_H^\prime \right) \frac{v^4}{f^4}  \, ,  
\end{gathered}
\\[0.7cm]
\label{eq:SILHrelations3}
d_3 = 1+ \left(c_6 - \frac{3c_H}{2}  \right) \frac{v^2}{f^2} 
+ \left( \frac{15c_H^2}{8} -\frac{5c_H^\prime}{4} -\frac{c_6 c_H}{2} -\frac{3c_6^2}{2} + 2 c_8\right) \frac{v^4}{f^4}  \, .
\end{gather}
In the special case in which the Higgs doublet $H$ is a PNGB of a global breaking ${\cal G}/{\cal H}$, all powers $(H/f)^n$ can be resummed exactly by imposing ${\cal G}$ invariance.
In this case $f$ must be identified with the decay constant of the NG bosons.
For example, there are just two custodially symmetric cosets yielding only one complex doublet of Goldstones: $SO(5)/SO(4)$ and $SO(4,1)/SO(4)$. 
They lead to
\begin{equation}
\label{eq:pngbparameters}
a =\sqrt{1-\xi}\, , \qquad b=1-2\xi\, ,  \qquad b_3 = -\frac{4}{3} \xi \sqrt{1-\xi}\, , 
\end{equation}
where $\xi =v^2/f^2>0$ for $SO(5)/SO(4)$ and $\xi =-v^2/f^2< 0$ for $SO(4,1)/SO(4)$. 
By expanding the above relations to $O(\xi^2)$ one re-obtains those of eq.~(\ref{eq:SILHrelations1}) for $c_H =1$,~$c_H^\prime =2$.
The expression for $d_3$ depends instead on the form of the Higgs potential, which is model dependent since it requires some explicit breaking of
the Goldstone symmetry ${\cal G}$. For example, in the minimal models MCHM4 and MCHM5 of  Refs.~\cite{Agashe:2004rs,Contino:2006qr} one has
\begin{equation}\label{eq:pngbparametersd3}
\text{MCHM4}: \quad d_3 = \sqrt{1-\xi}\, , \qquad\qquad \text{MCHM5}: \quad d_3 = \frac{1-2\xi}{\sqrt{1-\xi}}\, ,
\end{equation}
from which it follows $c_6 = 1$, $c_8 = 5/4$ in the MCHM4 and $c_6 = c_8 =0$ in the MCHM5.

The fact that at $O(v^2/f^2)$ the couplings $a, b, b_3$ are affected by
only one operator \cite{Giudice:2007fh}, whose coefficient $c_H$ can be always redefined away by a proper redefinition of $f$  (for example it can be set to 1),
has an important consequence.
Since the predictions of any coset ${\cal G}/{\cal H}$ must match those of the SILH Lagrangian at low energy, this implies that 
the expressions of eq.~(\ref{eq:SILHrelations1})
are universal at first order in $v^2/f^2$, i.e. they are the same for  a PNGB and for a generic scalar.
At order $v^4/f^4$, instead, the couplings $a, b, b_3$ are modified by two operators, whose coefficients are thus related by a specific
relation for any given coset ${\cal G}/{\cal H}$;
for example, the coset $SO(5)/SO(4)$ implies $c_H^\prime = 2 c_H$.
One can thus distinguish the case of a generic SILH, where $c_H^\prime$ can have any value, from that of a PNGB Higgs.

\section{Current constraints on the Higgs couplings}
\label{sec:Constraints}

Past and current experiments set important constraints on the Higgs couplings and on the scale of New Physics $m_\rho$.
In particular, the coupling $a$ is indirectly constrained by the precision tests of the EW observables performed at LEP, SLD and Tevatron,
and directly measured in single Higgs processes studied at the LHC.
However, there is currently no  constraint on the couplings $b$ and $d_3$ as these can be measured only through double Higgs processes.

\subsection{EW precision observables}

The sensitivity of the EW observables on $a$ arises at the 1-loop level only through the Higgs contribution to vector boson self energies.~\footnote{The 
1-loop Higgs contribution to the $b\bar b Z$ vertex is suppressed by $y_{b}^{2}$  and thus  negligible.} 
This is the leading effect, two-loop corrections are small and thus negligible. Compared to a few years ago, the information that comes from the 
EW fit has sharpened considerably~\cite{Baak:2012kk,Erler:2012wz,Batell:2012ca,SilvestriniWP}. 
This is mainly due to the value of the Higgs mass being now precisely known experimentally, so that
 a global fit can be used to extract the Higgs coupling to vector bosons directly, 
but also due to the new and more precise measurement of the $W$ mass from Tevatron.
For example,  compared to the  average of Tevatron and LEP measurements $m_W = (80.425\pm 0.034)\,$GeV used in the 
2006 final report on EW tests at the $Z$ pole~\cite{ALEPH:2005ab}, the current world average $m_W = (80.385 \pm 0.015)\,$GeV~\cite{Group:2012gb} 
has an error smaller by more than a factor 2. 
As a matter of fact, among the various observables sensitive to the Higgs coupling $a$, $m_W$ is the one which leads to the most
precise determination~\cite{SilvestriniWP}. Focusing on the 1-loop Higgs contribution to the vector boson self energies, the dependence on $a$ 
can be straightforwardly derived from that on $m_H$ at $a=1$~\cite{Barbieri:2007bh}: the $b$-quark forward-backward asymmetry $A_{FB}^b$ prefers values $a<1$, 
while the leptonic asymmetries $A_l$ and $m_W$ favor values slightly larger than 1. Overall, the global fit of $a$ is dominated by $m_W$ due to its small 
uncertainty,  with the other observables individually playing a minor role. 
By using the results from the GFitter collaboration~\cite{Baak:2012kk}, we find that in absence of additional NP contributions to the EW fit, the Higgs coupling 
is expected to lie in the interval $0.98 \leq a^2 \leq 1.12$ with 95\% of probability.
A similar result has been recently obtained by Ref.~\cite{SilvestriniWP}.
This is an extremely strong bound which seems to disfavor Higgs compositeness 
as a natural solution of the little hierarchy problem, in particular its realizations through compact cosets where $a$ is always reduced
compared to its SM value (see for example eq.~(\ref{eq:pngbparameters})).

To better understand this result it is useful to perform a two-dimensional fit in terms of the  
Peskin--Takeuchi  $\hat S$ and $\hat T$ parameters~\cite{PT,Barbieri:2004qk}. It is well known that modifying the Higgs coupling to vector bosons 
compared  to its SM prediction leads to a logarithmically divergent shift in these two parameters, with $\Delta \hat S >0$ and 
$\Delta \hat T < 0$~\cite{Barbieri:2007bh}. 
For $m_H=125\,$GeV and $a=1$ the theoretical point lies slightly outside the 68\% contour, and by decreasing $a$ it moves
further outside the experimentally preferred region, following a trajectory almost orthogonal to the probability isocontours.
Thus, small reductions of the  coupling $a$ have dramatic impact on the fit. Values $a>1$ are less
constrained but  also theoretically less motivated, as they require either non-compact cosets (like for example $SO(4,1)/SO(4)$),
or a sizable tree-level contribution from a scalar resonance with isospin $I=2$~\cite{Low:2009di,Falkowski:2012vh}.~\footnote{The latter possibility 
can be directly tested experimentally, since the  $I=2$ multiplet includes doubly-charged scalars which can be produced and observed at the LHC.}
If one excludes these more exotic theoretical scenarios, one concludes that sizable New Physics contributions to the EW observables,
in particular to the vector boson self energies, are required to accommodate $\sim O(10\%)$ shifts in the Higgs coupling $a$.

A negative and large $\Delta \hat S$ can follow from loops of fermion resonances~\cite{Golden:1990ig,Barbieri:2008zt,Grojean:2013qca,AzatovWP}.
A sizable and positive $\Delta \hat T$ could also be generated by the 1-loop 
exchange of composite fermions, in particular
the top partners. For example, if both SM top chiralities 
couple with the same strength to the strong dynamics, one naively expects 
$\Delta \hat T \sim \xi\,  y_t^2/(16\pi^2)$ (see for example Ref.~\cite{Giudice:2007fh}), which shows that it is possible to obtain  
$\Delta \hat T \sim \text{a few}\times 10^{-3}$ for $\xi \sim O(10\%)$.
Corrections of this size would dramatically modify the range of $a$ preferred by the EW fit, especially if accompanied by an additional $\Delta \hat S<0$.
For example, by assuming  $\Delta \hat T = +1.5\times 10^{-3}$ (with no extra $\Delta \hat S$), we find that the 95\% interval on the Higgs coupling becomes 
$0.70 \leq a^2 \leq 0.92$.
It is thus interesting to see under what conditions models can accommodate such corrections while satisfying all other constraints, in particular on the 
$Z\bar b b$ vertex, and investigate what  their predictions for the production of the top partners are at the LHC.
The first analyses that appeared in the literature
seemed to indicate a generic difficulty to obtain positive $\Delta \hat T$~\cite{Carena:2006bn,Barbieri:2007bh, Lodone:2008yy}. However, a more detailed 
exploration of the full  parameter space  in a broader class of  models has shown that there is more freedom in accommodating a positive and sizeable 
$\Delta \hat T$ while keeping  corrections to $Z\bar b b$ under control~\cite{Grojean:2013qca} (see also Ref.~\cite{Pomarol:2008bh}).

One might wonder if a modified value of $a$ can be helpful in relaxing the tension of the $b$-quark observables $A_{FB}^b$ and 
$R_b$.~\footnote{By including the  two-loop calculation of $R_b$ performed by Ref.~\cite{Freitas:2012sy}, 
the pulls of $A_{FB}^b$ and $R_b$ are respectively $+2.7\sigma$ and $-2.1\sigma$~\cite{SilvestriniWP} (see also~\cite{Baak:2012kk} for similar results).} 
In fact,  as we already pointed out, all the observables, including those related to the $b$ quark, depend
on $a$ mainly through the 1-loop  contribution of the Higgs to the vector boson self-energies. As a consequence, any NP correction to $a$
cannot  lead to an effect restricted to the $b$-quark sector, but will propagate to all observables.
Notice also that excluding $A_{FB}^b$ from the fit pushes $a$ towards larger values, which are even more problematic from the theoretical viewpoint,
although the effect is small.
It is thus clear that the existence of a tension in the $b$ observables does not lead to any room for  relaxing the strong
bound on the Higgs coupling $a$.
Yet, the fact that in a fit to the couplings of $b_L$ and $b_R$ to the $Z$ 
the SM point lies outside the 95\% probability contour (see Refs.~\cite{Batell:2012ca, SilvestriniWP, Guadagnoli:2013mru})
might indicate that the contribution from NP states is already at work.

Apart from possible New Physics effects, the fit of $a$ is strongly sensitive to the value of 
the $W$ and top quark masses. We have already stressed that $m_W$ dominates  over the other observables.
Its current experimental measurement is $\sim 1.2\sigma$ larger than the one
preferred by the EW fit, if it goes down in the future also the central value of $a$ will diminish.  The strong dependence
on the top mass originates from the 1-loop correction to the $\rho$ parameter, $\Delta\rho = \Delta \hat T\propto m_t^2 G_F$,   which
we have seen has an important impact on $a$. In this regard one must notice that the error 
reported in the current Tevatron average $m_t = (173.18\pm 0.94)\,$GeV~\cite{Aaltonen:2012ra} does not include the theoretical uncertainty 
on the definition of  the parameter extracted from the event kinematics in terms of the $\overline{MS}$ mass. 
If one instead adopts  the larger error $\sigma_t =2.8\,$GeV that follows from measuring the $\overline{MS}$ mass
directly from the $t\bar t$ cross section~\cite{Alekhin:2012py},
one finds that the uncertainty on $a$ increases by a non-negligible amount~\cite{SilvestriniWP}.
While these issues  have to be considered to make a precise determination of the Higgs coupling,
the overall picture which emerges from the EW fit seems quite robust: 
$O(10\%)$ shifts in $a$ require sizeable NP contributions to the vector-boson self energies.

\subsection{Direct coupling measurements}

The precision currently reached at the LHC on the direct measurement of the Higgs couplings to vector bosons and to fermions 
is, on the other hand, more limited. The exact value depends on the assumptions one makes to extract the couplings.
For example, one can make a two-dimensional fit of $a$ and $c$, where the latter parametrizes a common rescaling of all the Yukawa couplings.
Even neglecting the second solution at $c<0$, the uncertainty  in the official fits of the ATLAS and CMS collaborations is of the
order of $\sim 20\%$ on $a$ and even larger on $c$~\cite{CMS:yva, ATLAS:2013sla}. In particular, while ATLAS prefers values $a>1$, the best fit value of 
CMS is for $a<1$.
A naive combination of these results leads to a smaller uncertainty~\cite{THfits}, but more data are definitely required to form a clearer picture.
Preliminary studies indicate that eventually a precision of $\sim 5\%$ on $a$ should be reached 
at the $14\,$TeV LHC with an integrated luminosity of $300\,\text{fb}^{-1}$~\cite{CMS-NOTE-2012/006,ATLAS-collaboration:2012iza}. 

\subsection{Resonance searches}

Searches for direct production of resonances at the LHC also set important constraints on the mass scale $m_\rho$ of a new strongly-interacting sector. 
Here we consider  the case of a generic $SO(5)/SO(4)$ composite Higgs theory as a benchmark scenario, although the actual bounds will depend on the details 
of the strong dynamics and on how it couples to the SM fermions. For illustrative purposes we focus on the lightest spin-1 resonance of the strong sector,
which we denote by $\rho$, and assume that it transforms as a $(3,1)$ under $SU(2)_L \times SU(2)_R \sim SO(4)$ (for recent studies in this direction see for example \cite{Contino:2011np, Barbieri:2008cc}). 
The dominant production  is via Drell--Yan processes (see for example Ref.~\cite{Falkowski:2011ua}). 
A class of theories motivated both theoretically and experimentally is one in which the spin-1 resonance couples to light fermions
only through its mixing to the SM gauge fields~\cite{Contino:2006nn} (see Ref.~\cite{Redi:2011zi} for alternative possibilities).
In this case the Drell--Yan production cross section scales as $1/g_{\rho}^{2}$, since couplings of the resonances to the SM fermions are suppressed by $1/g_{\rho}$. 
The strongest exclusion limits are currently set by the LHC searches performed at $8$~TeV with $20$ fb$^{-1}$ in final states with one lepton and missing 
transverse energy~\cite{Chatrchyan:2012rva} or dileptons~\cite{Chatrchyan:2012rvb}, looking for charged and neutral spin-1 resonances respectively. 
For values of $\xi$ of order 1, searches for resonances decaying into $WZ$, in particular those with three leptons in the final state~\cite{Chatrchyan:20120114}, give slightly stronger bounds.~\footnote{We find that the more recent searches for spin-1 resonances decaying to pairs of vector bosons with boosted decay products~\cite{Chatrchyan:2012024} give less strong constraints.} Assuming the $\rho$ to be a $({\bf 3},{\bf 1})$ of $SU(2)_{L} \times SU(2)_R$, we translated the bounds on 
$(\sigma \times BR)$ set by the experimental collaborations into a combined exclusion region in the $(\xi, m_{\rho})$ plane. 

\begin{figure}[t]
\begin{center}
\includegraphics[width=0.50\linewidth]{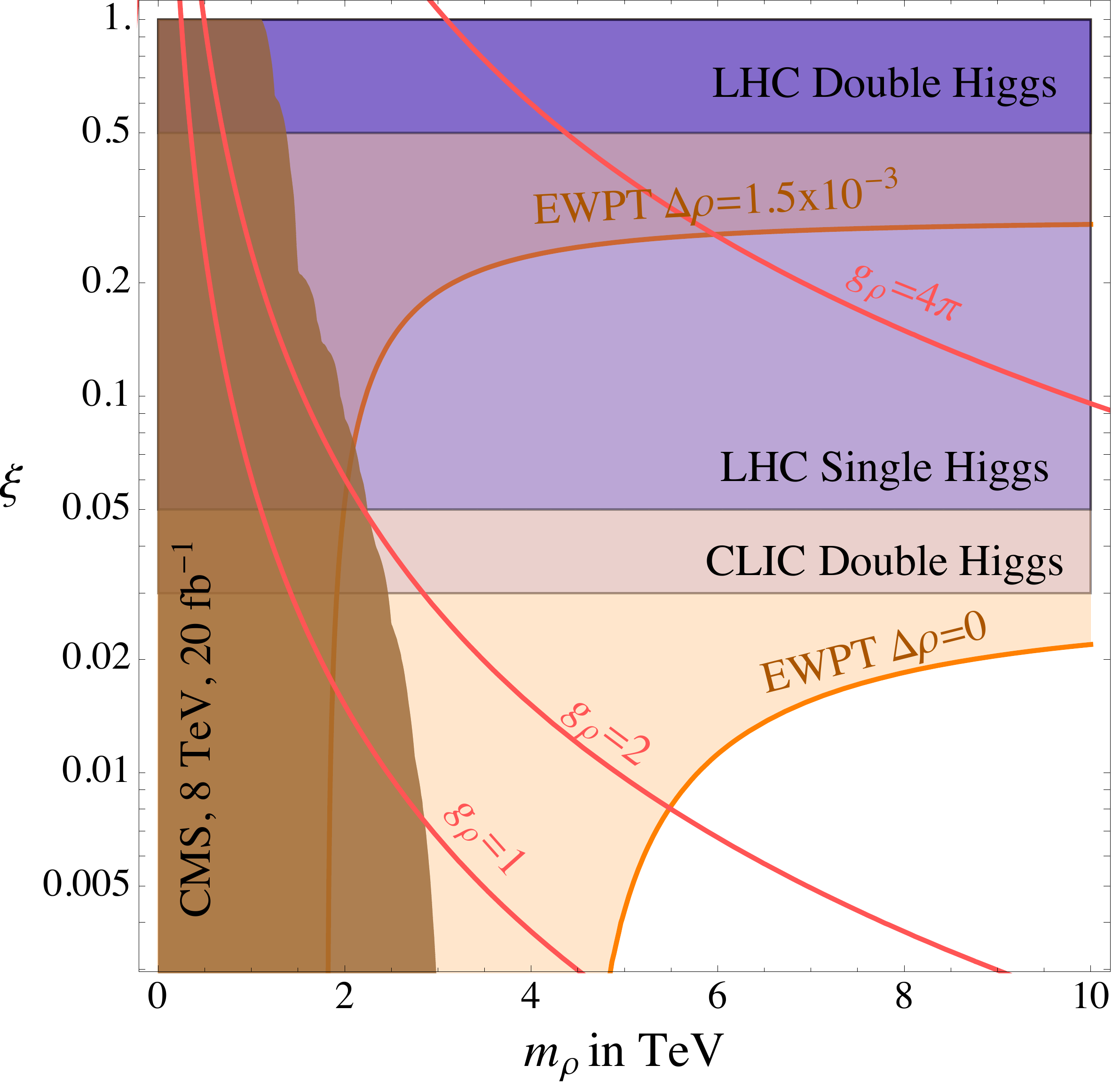}
\\[-0.1cm]
\caption{\small
Summary of current constraints (orange curves and brown region) and expected sensitivities at CLIC and the LHC (horizontal regions) on $\xi = (v/f)^2$ 
and the mass of the lightest spin-1 resonance $m_\rho$  for  $SO(5)/SO(4)$ composite Higgs  theories. See text.
}
\label{fig:summary}
\end{center}
\end{figure}

The situation of direct and also indirect constraints is summarized in Fig.~\ref{fig:summary} 
for the case of a generic $SO(5)/SO(4)$ composite Higgs theory.
%
%
We fix $a_\rho \equiv m_\rho/(g_\rho f)= 1, $  where we define $g_{\rho}$ as the physical coupling strength between three $\rho$ resonances. 
The fundamental free parameters of the new dynamics are then the mass of the spin-1 resonance, $m_\rho$, and the strengths of the Higgs interactions 
parametrized by $\xi = (v^2/f^2)$. 
The dark  brown region on the left shows the current 95\% combined limit from direct production of the charged $\rho^\pm$ at the LHC decaying to 
$l \nu$ and $WZ \to 3l \nu$ final states. A similar exclusion region follows from the limits on the production of the neutral $\rho^0$.
The dark (medium light)  horizontal purple bands of Fig.~\ref{fig:summary} indicate instead the sensitivity on $\xi$ expected at the LHC from double (single)
Higgs production with $300\,\text{fb}^{-1}$ of integrated luminosity (see footnote~\ref{ftn:sensitivity} for the definition of sensitivity adopted in this paper). 
The value shown for the case of double Higgs production is based on a naive (and perhaps
optimistic) extrapolation  of the study of Ref.~\cite{Contino:2010mh}; a more precise determination requires an updated analysis for $m_h=125\,$GeV.
As we will discuss in Section~\ref{VVtoHH},
the study of double Higgs processes alone at CLIC  is expected to lead to a precision on $\xi$ larger than what obtainable at the LHC through single
Higgs studies.
In the plot of Fig.~\ref{fig:summary} this is illustrated by the lowest horizontal band.
The possibility of directly testing such small values of $\xi$  at CLIC has to be compared with the indirect bounds set by the EW precision data.
By including only the tree-level contribution $\Delta \hat S= m_W^2/m_\rho^2$ from the $\rho$ exchange~\cite{Contino:2011np} and the 1-loop IR effect 
from the modified Higgs  couplings, we find that the region above the lower orange band is excluded at $95\%$. For $m_\rho \to \infty$ the upper limit on $\xi$
tends to $\sim 0.02$, as previously reported in the discussion of the EW fit. In the absence of  other contributions to the oblique parameters, masses 
$m_\rho \lesssim 5\,$TeV are already excluded even for very small $\xi$.
Lowering the value of $a_\rho$ makes the bound on $m_\rho$ weaker, but does not change much that on $\xi$.
The  allowed region instead opens up in presence of an additional contribution to $\hat T$:  for $0 < \Delta \hat T \leq + 1.5\times 10^{-3}$
the 95\% exclusion boundary varies between the upper and lower orange lines and masses as low as $m_\rho \sim 2$~TeV can still be viable. 
The domain of validity of our predictions, $g_\rho<4\pi$, is below the upper red line.
%

\section{What can be learned from single and double Higgs production?}
\label{sec:SingeDoubleHiggsProd}

In this Section we discuss what could be learned directly, or indirectly, from a program of precise
Higgs measurements at CLIC. For definiteness we can imagine a  scenario  where the LHC did not measure deviations from the SM larger  than $O(20\%)$ in single Higgs production, and also no clear indications emerged on what the underlying theory may be  (new particles may have been discovered but not with a clear role, i.e.~no supersymmetric particles). There are various broad questions one can in principle address with these measurements. One question is whether the scalar $h$ is elementary or composite. Other questions  concern the nature of $h$, whether or not  it fits into an $SU(2)$ doublet (an explicit, if not well motivated, example of a non-doublet Higgs-like scalar is a dilaton~\cite{dilaton}) or whether or not it is a PNGB. To some extent the information will be indirect, so it is worth illustrating the logic in some detail.

Consider the issue of compositeness first.
 Of course, in order  to directly answer this question, it would be necessary
to explore the energy scale associated with the new states and observe the onset of a novel UV regime, perhaps described by a strongly coupled CFT. 
That would be the analogue of observing the hadron to parton transition in QCD processes. 
However,  the measurement of low-energy quantities can already give an appraisal of the strength of the underlying interactions, thus favoring or disfavoring a 
composite scenario. Indeed for an $SU(2)$ Higgs doublet, a heavy particle with mass $m_\rho$ and coupling  to the Higgs $g_\rho$ modifies the low-energy 
couplings by a relative amount of order $(g_\rho v/m_\rho)^2$.~\footnote{Notice that a light Higgs mass $m_h = 125\,$GeV disfavors maximal values 
$g_\rho \sim 4\pi$ unless some (additional) tuning is present in the Higgs potential. See for example 
Refs.~\cite{Panico:2012uw,DeSimone:2012fs,Pappadopulo:2013vca}.}
For instance, massive fermions with a  vectorlike mass $m_\rho$ and a Yukawa interaction to the Higgs of strength $g_\rho$
affect the coupling of $h$ to two gluons and two photons by a relative amount $\sim (g_\rho v/m_\rho)^2$ (for recent work in the context of composite Higgs models see e.g. Ref.~\cite{Azatov:2011qy}). Similarly, a heavy singlet scalar $S$ coupled to the 
Higgs doublet via a trilinear term $g_\rho m_\rho S|H|^2$ mixes by an angle $\theta\sim g_\rho v/m_\rho$ with $h$, implying shifts in its couplings
of order  $\theta^2\sim (g_\rho v/m_\rho)^2$.~\footnote{One could also consider a potential 
$ V=-m_\rho^2 |S|^2 +g_\rho^2 |S|^2 |H|^2+ g_\rho^2 |S|^4$, by which $\langle S\rangle \sim m_\rho/g_\rho \equiv f$, and reach the same conclusion.}
In the absence of new states below a certain scale $M$, the observation of deviations of order $\delta_h^{exp}$ in single Higgs production  
would then imply a qualitative lower bound on the coupling 
\begin{equation}
g_\rho >\sqrt {\delta_h^{exp}} \,\frac{M}{v}\, .
\label{single}
\end{equation} 
Sizeable deviations $\delta_h^{exp}$ in the absence of new states would  suggest a strong coupling and, indirectly, $h$ compositeness. A more direct 
measurement of the strength of the underlying interaction is obtained by a study of the processes $WW\to WW$ and $WW\to hh$. 
As discussed in Section~\ref{sec:Hcouplings}, a deviation from $a=b=1$ leads to a cross section that grows with~$s$. The $2\to 2$ amplitude can be taken as a 
measure of a  ``running''  coupling $\bar g(\sqrt{s})$, see eq.~(\ref{eq:twototwo}).
The measurement of an enhancement, quantified by $\delta_{hh}^{exp}$, in these processes at an energy $\sqrt{s}$, corresponds directly, though qualitatively, to a 
lower bound on the strength of the interaction
\begin{equation}
g_\rho>\bar g(\sqrt{s}) \sim \sqrt{\delta_{hh}^{exp}} \,\frac{\sqrt{s}}{v}\, .
\label{double}
\end{equation}
Equations~(\ref{single}) and (\ref{double}) look similar, and not by chance. Notice, however, that the second equation corresponds to a direct measurement of the coupling, 
and is thus a more robust estimate. Indeed, at a precise machine such as CLIC a detailed study of $2\to 2$ processes would allow even stronger conclusions. 
The point is that eq.~(\ref{eq:twototwo}) is only the leading term in a derivative expansion, the subleading corrections being of relative size $s/m_\rho^2$:
\begin{equation}
{\cal A}(2\to 2) =\delta_{hh} \frac{s}{v^2}\left (1 +O\left(\frac{s}{m_\rho^2}\right)\right )\, .
\end{equation}
In principle at CLIC one could measure the leading $O(s)$ contribution and set an upper bound $\epsilon_{hh}$ on the relative size of the $O(s^2)$ term. 
That would indirectly suggest that there are no new states below a mass $M\sim \sqrt{s}/\sqrt{\epsilon_{hh}}$ and that the amplitude will keep rising at least 
until that scale. 
That would amount to a stronger indirect bound
\begin{equation}
g_\rho>\bar g(M) \sim \sqrt{\frac{\delta_{hh}^{exp}}{\epsilon_{hh}}} \,\frac{\sqrt{s}}{v}\, .
\end{equation}
We clearly see here the value of being able to measure $2\to 2$ processes with high precision below the threshold of New Physics.
Of course another possibility is that of directly observing, rather than setting limits on, the $O(s^2)$ effects from the tails of heavy resonances.
In this case detailed information on the strong dynamics, such as the quantum numbers of its resonances, can come from the comparison of different 
scattering channels,  see for example Refs.~\cite{chanowitz,Contino:2011np}.~\footnote{The effects from the tails of spin-1 resonances can also be
studied through the process $e^+ e^- \to VV$, see for example Refs.~\cite{eeVV}.}

Consider now the properties of $h$ from the standpoint of symmetries. In the case of the SILH, in which $h$ fits into a doublet of $SU(2)$ arising from some 
unspecified dynamics at the scale $m_\rho$, the bosonic couplings $a, b, b_3$ are predicted in terms of just one parameter at $O(v^2/f^2)$, as illustrated by eq.~(\ref{eq:SILHrelations1}).
In particular, by defining $\Delta a^2\equiv a^2-1$ and $\Delta b\equiv b-1$ one has
\begin{equation}
\Delta b= 2  \Delta a^2 \left (1+ O(\Delta a^2)\right)\,, 
\label{abgeneral}
\end{equation}
where the higher-order corrections are determined by the tower of higher-dimensional operators with two derivatives and $2n$ $H$ fields using the SILH power counting. 
Furthermore, in  the very special case where $H$ is a PNGB the whole tower of operators and  the resulting $WWh^n$ couplings are all fixed 
in terms of a single parameter $\xi$. Equation~(\ref{eq:pngbparameters}) reports for example the predictions of the minimal $SO(5)/SO(4)$ and $SO(4,1)/SO(4)$ theories.
In both cases  eq.~(\ref{abgeneral}) becomes exactly
\begin{equation}
\Delta b= 2 \Delta a^2 \, .
\label{abpngb}
\end{equation}
From single Higgs production one would be  able to measure $\Delta a^2$ with an error $\sim 10^{-2}$, maybe 
of a few per mille~\cite{Baer:2013cma, Abramowicz:2013tzc, Barger:2003rs, Peskin:2012we}.
The measurement of $WW\to hh$, as we will discuss in the next Sections, allows one in principle  to measure $\Delta b$
with an error of order~$10^{-2}$. Equations~(\ref{abgeneral}) and (\ref{abpngb}) can then be tested at the percent level. 
For instance, in the case of a SILH not embedded in a coset one could imagine
  finding $\Delta a^2, \Delta b\gsim 0.1$ and to be compatible with eq.~(\ref{abgeneral}) 
but violating eq.~(\ref{abpngb}) by an amount bigger than the expected percent accuracy. On the other hand, for $\Delta a^2, \Delta b<  0.1$, it would not be possible to distinguish between a SILH and a PNGB. Finally, down to $\Delta a^2, \Delta b \sim 10^{-2}$ one could find that eq.~(\ref{abgeneral}) is not respected, indirectly speaking against the embedding of $h$ in a doublet. It should however be pointed out that such a scenario, normally associated with a fully composite $h$, would more probably imply $\Delta$'s of order 1, which are
already excluded by the current LHC results. It should also be remarked that the only case of this type with some mild motivation is the one of a light dilaton, corresponding to
\begin{equation}
\Delta b=  \Delta a^2 \, ,
\label{abdilaton}
\end{equation}
implying  a vanishing contribution to $WW\to hh$ at leading order in the energy expansion~\cite{Contino:2010mh}. 

We should finally point out the potential role of the rates for $h\to gg$ and $h\to \gamma\gamma$ in distinguishing a pseudo Nambu--Goldstone $h$ from 
a generic composite scalar. The basic remark~\cite{Giudice:2007fh}  is that there are two classes of corrections to these rates. One correction originates from the 
modification of the coupling of $h$ to $WW$ and to $\bar t t$ and affects the on-shell $h\to gg$, $h\to \gamma\gamma$ amplitudes via the $W$ and $t$ 
loop contribution. In a sense this contribution is long distance. A second correction is the genuine short-distance contribution to the Wilson coefficient of the 
operators
\begin{equation}
{\cal O}_{gg} = h G_{\mu\nu}G^{\mu\nu}\, , \qquad\qquad {\cal O}_{\gamma\gamma} = h F_{\mu\nu}F^{\mu\nu} \, ,
\end{equation}
that arises from loops of heavy states. In the case of a PNGB this second class of effects is suppressed with respect to the first by a factor $(g_{\not G}/g_\rho)^2$, 
where by $g_{\not G}$ we indicate a weak spurion coupling which breaks the Goldstone symmetry. This suppression is a consequence of the Goldstone symmetry
selection rules and would be absent in the case of a generic composite scalar, like for instance the dilaton. In the limit where $g_{\not G}/g_\rho\ll 1$, the rates 
$h\to gg$, $h\to \gamma\gamma$ are fully controlled by $a$ and $c_t$ ($c_t$ measures the deviations of the top Yukawa coupling~\cite{Giudice:2007fh}), a result that can in principle be tested. However, one should keep in mind that the measured value of $m_h$ prefers a scenario where the top partners are somewhat lighter than the rest and only moderately strongly coupled~\cite{Matsedonskyi:2012ym,Marzocca:2012zn,Panico:2012uw,Pappadopulo:2013vca}.
In that situation the correlation between  $h\to gg$, $h\to \gamma\gamma$ and the parameters $a, c_t $ may receive important corrections.

\section{What can be learned from triple Higgs production?}
\label{sec:VVtohhh}

In this Section we discuss the relevance of the process $VV \to hhh$ in distinguishing between a generic SILH and a PNGB (for an earlier study of this process at a linear collider, see Ref.~\cite{Ferrera:2007sp}, while a study at the LHC has been recently carried out in Ref.~\cite{Belyaev:2012bm} and triple Higgs production by gluon fusion has also been studied in Ref.~\cite{Binoth:2006ym}).
We will show that this process is suppressed in the PNGB case as a consequence of a 
$Z_2$ invariance of the Lagrangian under which the NG bosons are odd.
A priori, any three-body final state involving the Higgs and gauge bosons could be a further probe of the nature of the Higgs.  In practice, however, $VV \to hhh$  is the only process that adds  new information, thanks to its 
sensitivity to $b_{3}$. 
Studying further  final states like $hhV$, $hVV$ and $VVV$ merely gives a complementary probe of the relation between $a^{2}$ and $b$.

\subsection{Symmetry structure}

In a symmetric coset like $SO(5)/SO(4)$ there exists a  $Z_2$ invariance of the algebra (grading)
under which the broken generators $T^{\hat{a}}$ change sign while the unbroken generators $T^{a}$ do not:
$T^{a} \to + T^{a}$, $T^{\hat{a}} \to - T^{\hat{a}}$.
At the field level this corresponds to a parity $R$ under which all NG bosons are odd
\begin{equation}
R: \qquad \pi^{\hat{a}}(x) \to -\pi^{\hat{a}}(x)\, .
\end{equation}
In general, $R$ is an accidental invariance of the Lagrangian  at the two-derivative level and is violated at higher orders. This is for example the case of  $SO(4)/SO(3)$,
where $R$ coincides with the $P_{LR}$ parity of eq.~(\ref{eq:PLR}).  It may happen however, as for example in the case of $SO(5)/SO(4)$, that $R$ is
an element of ${\cal G}$, in which case it remains unbroken to all orders in the derivative expansion of the strong dynamics. 
In fact, this is true for any coset ${\cal G}/{\cal H}$ involving only doublets under some $SU(2)'\subset {\cal H}$.~\footnote {Such a coset is obviously symmetric, as the commutator of any element in ${\cal G}/{\cal H}$ cannot be a doublet, and must therefore belong to ${\cal H}$. Moreover the residual $SU(2)'$ will forbid odd powers of the NG-bosons at any order in the derivative expansion.} In particular this property is shared by the simplest cosets involving just one scalar doublet, whether custodially symmetric ($SO(5)/SO(4)$ or $SO(4,1)/SO(4)$) or not ($SU(3)/SU(2)\times U(1)$).
Consequently, when the  Higgs doublet is the only PNGB multiplet from some strong dynamics,
processes with an odd number of pseudo-NG bosons are  forbidden to all orders in the strong dynamics and only arise 
as a weak effect of  the SM couplings. 

The above argument implies that, although by a naive counting one would expect the $V_{L} V_{L} \rightarrow hhh$ cross section to grow with 
$\hat{s}^{2}$,~\footnote{From here on we will indicate the partonic c.o.m. energy with $\hat s$, while $s$ will denote the collider energy.}
this does not happen for a PNGB Higgs.
In practice $R$ is weakly broken by the gauging, so that this process is not strictly zero but only suppressed by $g$. The expected energy behavior of the amplitude 
at the parton level can be estimated by power counting and is shown in Table~\ref{NDAPartonic}. 
Longitudinal modes interact with coupling strength $\bar g(\sqrt{\hat s}) \sim \sqrt{\hat s}/f$, while transverse modes  have weak coupling strength $g$.
Measuring the cross section of triple Higgs production can thus give important indications on the nature of the Higgs boson and distinguish the case 
of a PNGB from that of a generic SILH. Indeed, as it will become more clear in a moment, the grading symmetry $R$ is reflected in some non-trivial correlations among the  coefficients of operators of different dimensionality in the expansion of the effective lagrangian in powers of the Higgs doublet $H$.
%
\begin{table}[t]
\begin{center}
\begin{tabular}{c|c|c} 
	\hline
	\hline
	\multirow{2}{*}{Polarisation} 		  & \multicolumn{2}{|c}{Amplitude for } \\[0.02cm]
					  & PNGB & SILH \\ \hline
&& \\[-0.3cm]
	$V_{L} V_{L} \rightarrow hhh$ & $g^{2} v/f^{2}$ & $\hat s v/f^{4}$ \\[0.15cm]
	$V_{L} V_{T} \rightarrow hhh$ & \multicolumn{2}{|c}{ $\sqrt{\hat s} g/f^{2}$ }  \\[0.15cm]
	$V_{T} V_{T} \rightarrow hhh$ & \multicolumn{2}{|c}{ $g^{2} v/f^{2}$ } \\[0.05cm] 
\hline
\hline
\end{tabular}
\\[0.1cm]
\caption{\small 
Naive high-energy and large-angles behavior of partonic $VV\to hhh$ amplitudes for a PNGB Higgs (first column) and a generic SILH scalar (second column).}
\label{NDAPartonic}
\end{center}
\end{table}

\subsection{Quantitative analysis of $VV \rightarrow hhh$}

\begin{figure}[t]	
	\begin{center}
	\includegraphics[width=0.7\textwidth]{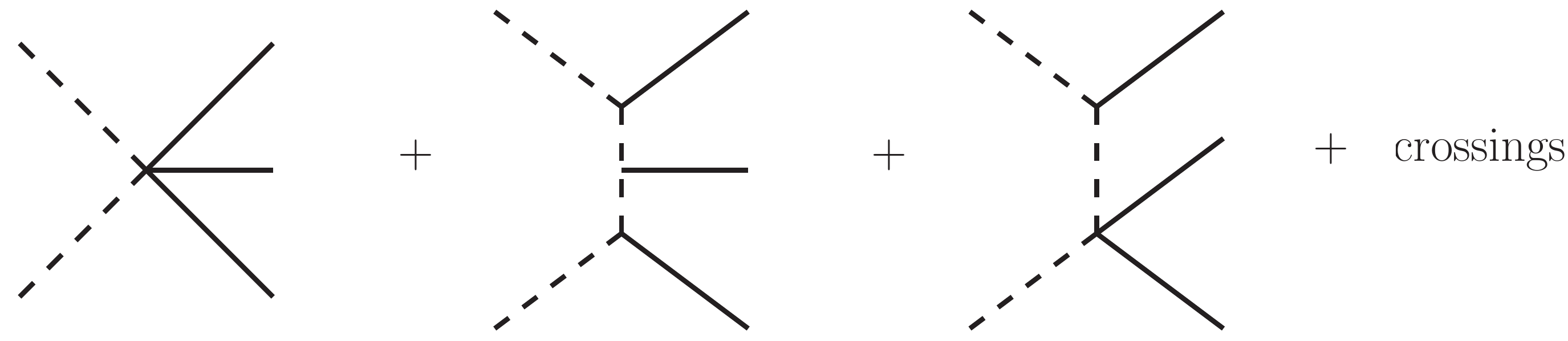}
	\end{center}
	\caption{\small Leading diagrams contributing to the $\chi\chi \rightarrow hhh$ amplitude. Dashed lines represent the NG bosons $\chi$, 
while solid lines denote the Higgs boson $h$. The sum of these diagrams with their crossings cancels out exactly in the gaugeless limit for a symmetric 
coset and at the $O(p^2)$ level for any coset. See text.}
	\label{pipihhhDiagrams}
\end{figure}

We checked that the expected cancellation of the energy-growing term of the $V_L V_L \to hhh$ scattering amplitude 
takes place by performing an explicit computation in the gaugeless limit $g = g^{\prime} =0$. 
By the equivalence theorem, the leading energy behavior of $V_L V_L \to hhh$  is captured by the NG boson scattering
$\chi\chi \to hhh$. From the Lagrangian of eq.~(\ref{eq:HiggsLag})
we find three distinct diagrams, depicted in Fig.~\ref{pipihhhDiagrams}, plus their crossings, which contribute to the amplitude.
%
%
At leading order in $\hat{s}$ we find~\footnote{Note that, similarly to the process $VV\to hh$, anomalous Higgs self-interactions 
parametrized by $d_3$ and~$d_4$ modify the amplitude of triple Higgs production near threshold but do not affect the asymptotic 
behavior at large partonic energy. Their contribution is thus subleading and will be neglected in the following.}
\begin{equation}
\label{pipihhhAmplitude}
\mathcal{A}(\chi \chi \rightarrow hhh) = \frac{i \hat{s}}{v^{3}} \left( 4 a b - 4 a^{3} - 3 b_{3} \right) \, .
\end{equation}
In the case of $SO(5)/SO(4)$ the values of the couplings $a, b$ and $b_{3}$ are given by eq.~(\ref{eq:pngbparameters}) and the coefficient
of the term growing with $\hat s$ in the amplitude vanishes identically.
In the case of a generic Higgs doublet the cancellation works at the $O(v^2/f^2)$ level, as due to the universality of the SILH Lagrangian,
but it fails at higher orders. By substituting the relations of eq.~(\ref{eq:SILHrelations1}) into eq.~(\ref{pipihhhAmplitude}) we find
\begin{equation}
\mathcal{A}(\chi \chi \rightarrow hhh) = 2i \left( c_H^\prime - 2 c_H^2\right) \frac{\hat{s}}{v^3} \left(\frac{v^4}{f^4} \right)\, .
\end{equation}
As expected, the coefficient of the energy-growing term is of order  $v^4/f^4$ and proportional to the linear combination $(c_H^\prime -2 c_H^2)$.
This latter must vanish if the Higgs lives on a symmetric coset ${\cal G}/{\cal H}$.~\footnote{This shows that the relation $c_H^\prime = 2 c_H^2$
holds true in any symmetric coset, and not only in $SO(5)/SO(4)$.}

At CLIC, triple Higgs production proceeds through 
the process $e^{+} e^{-} \rightarrow \nu \bar{\nu}V V\rightarrow \nu \bar{\nu} hhh$, where $V = W^\pm ,Z$. 
Some typical values of the cross section are shown in Table~\ref{tab:xSection_hhh}  for the case of a PNGB and a SILH with $c_{H} = 1$ and $c_{H}^{\prime} = 0$
(and vanishing higher-order operators).
\begin{table}[t]
\begin{center}
{\small
\begin{tabular}{c|ccccccccccc}
\hline
\hline
 $\sigma$ & \multicolumn{7}{c}{$\xi$} \\[0.0cm] 
  [ab]	     & 0 & 0.05 & 0.1 & 0.2  & 0.3 & 0.5 & 0.99 \\
\hline
&&&&&& \\[-0.3cm]
PNGB     & $0.32$ & $0.46$ & $0.71$ & $1.47$  & $2.41$ & $4.13$  & $0.30$ \\[0.02cm]
SILH     & $0.32$ & $0.71$ & $0.87$ & $7.56$  & $42.89$ & $407.9$ & $7808$ \\[0.02cm]
\hline
\hline
\end{tabular}
}
\\[0.1cm]
\caption{\small Cross section for the process $e^{+} e^{-} \rightarrow \nu \bar{\nu} hhh$ for $m_{h} = 125$ GeV at $\sqrt{s} = 3$~TeV. 
The first line shows the cross sections obtained in the symmetric $SO(5)/SO(4)$ coset for various values of $\xi$. The cross sections 
in the second line are for a SILH with
$c_{H} = 1$ and $c_{H}^{\prime} = 0$ and vanishing higher-order operators.}
\label{tab:xSection_hhh}
\end{center}
\end{table}
%
While the cross section for a PNGB  is in the range of a few~ab, in the case of a generic SILH it can be much bigger and grows like $\xi^4$, with the dominant contribution coming
from the subprocess $V_L V_L \to hhh$. A careful analysis of the sensitivity of a linear collider to the anomalous couplings involved in triple Higgs production is beyond the scope of this work. A very conservative approach is to decay every Higgs to $b\bar b$ and to require the identification of at least 5 $b$-jets. The branching ratio of three Higgses into 3 $b\bar b$ pairs is 20\%. Assuming an 80\% $b$-tagging probability, the efficiency to reconstruct at least 5 $b$-jets out of the available 6 in the final state is 66\%.  Including an additional factor 3 reduction due to identification cuts to be performed on the final state jets one obtains an overall efficiency on the signal which is roughly 5\%. Requiring the identification of $O(10)$ triple Higgs events implies the possibility to detect this process with an integrated luminosity of 1~ab${}^{-1}$ as soon as $\xi \gsim 0.3$ for a generic SILH. 

\begin{figure}[t]	
	\begin{center}
	\includegraphics[width=0.6\textwidth]{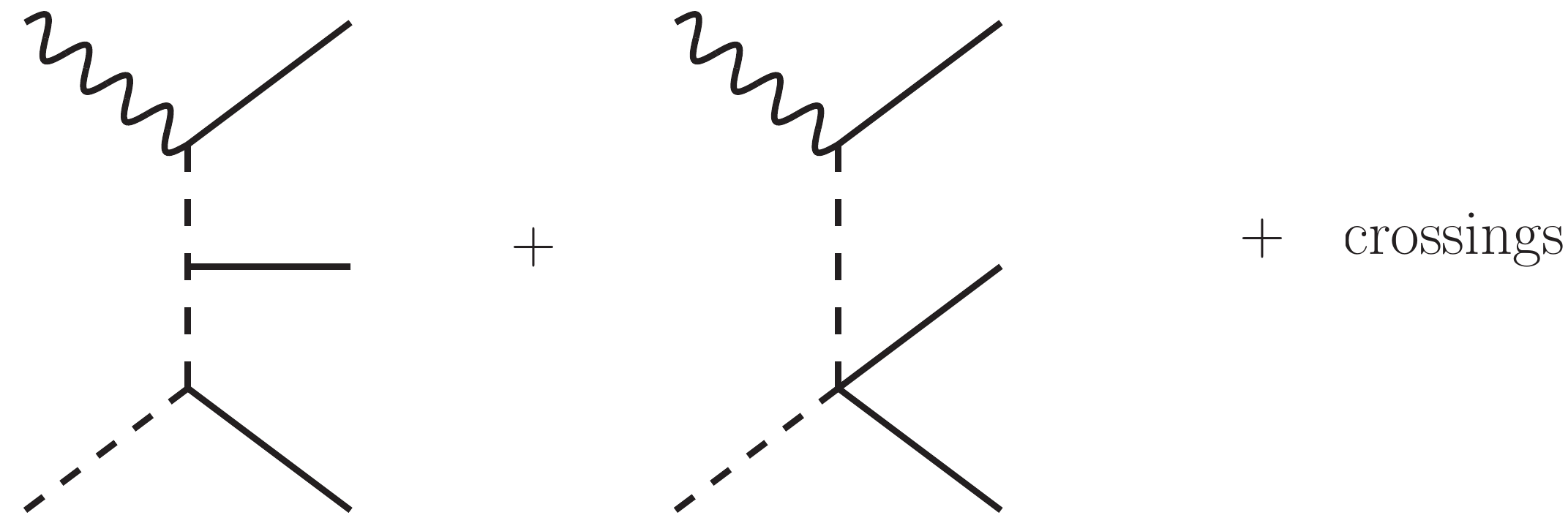}
	\end{center}
	\caption{\small Diagrams giving the dominant contribution to the $V_{T} \chi \rightarrow hhh$ cross-section. Continuous, dashed and wiggly lines 
denote a Higgs boson $h$, the NG bosons $\chi$, and a transverse gauge boson $V_T$ respectively. 
}
	\label{LThhhDiagram}
\end{figure}

The dominant contribution to triple Higgs production in the case of a PNGB Higgs
comes from the subprocess  $W^{\pm}_{L} W^{\mp}_{T} \rightarrow hhh$, whose cross section is expected to grow as $\hat{s}\log\hat s$. 
The leading logarithmic behavior can be extracted by using the equivalence theorem and arises
from the subset of diagrams shown in Fig.~\ref{LThhhDiagram}. 
%
%
In the limit in which the intermediate PNGB line is nearly on-shell, the total cross section factorizes into the product of a collinear $W_T\to \chi h$ splitting times 
the cross section of a hard $\chi^{\pm} \chi^{\mp}\to hh$ scattering
\begin{equation}
\sigma (W^{\pm}_{T} \chi^{\mp}\rightarrow hhh)=
\int \! dx \, d p_T^2 \, f(x,p_T)\, \sigma(\chi^{\pm} \chi^{\mp}\to hh)(x \hat s)\, .
\end{equation}
Here $x$ is the fraction of the $W$ energy carried by the emitted $\chi$,  $p_T$ is its transverse momentum
and $\hat s$ is the total center-of-mass energy of the $W_T \chi \to hhh$ process.
The splitting function $f(x,p_T)$ can be calculated using eq.~(\ref{eq:HiggsLag}) and is given by
\begin{equation}
f(x , p_T) = \frac{1}{p_T^4} \frac{x (1-x)}{8\pi^2} |{\cal A} (W_T\to \chi h)|^2 = \frac{x (1-x)}{p_T^2} \,\frac{a^2 g^2}{32\pi^2}\, ,
\end{equation}
where we have neglected the masses of the gauge and Higgs bosons.
Notice that the amplitude of the splitting ${\cal A}(W_T\to \chi h)$ vanishes in the forward direction as required by angular momentum conservation. 
At leading order in $\hat s$, the cross section of the hard $\chi\chi\to hh$ scattering does not depend on $p_T$, and reads
\begin{equation}
 \sigma(\chi^{\pm} \chi^{\mp}\to hh)(\hat s)=\frac{\hat{s} (b-a^2)^2}{32\pi v^4}\, .
\end{equation}
We thus obtain
\begin{equation}
\label{LThhh}
\sigma (W^{\pm}_{T} \chi^{\mp}\rightarrow hhh)=\frac{g^{2}}{12288\, \pi^{3}} \frac{(ab-a^3)^{2} }{v^4}\hat{s} \log \frac{\hat{s}}{m_{W}^{2}}\, .
\end{equation}
The factor $\log(\hat s/m^2_W)$ originates from the logarithmic divergence of the  integral over $p_T$, which is cut off in the infrared at $p_T^2 \sim m_W^2$
once the $W$ mass dependence is properly taken into account.

Although this calculation captures the exact asymptotic behavior of the $W_T^\pm W_L^\mp \to hhh$ process, it turns out that the subleading contribution 
proportional to $\hat s$ is numerically large, so that
the logarithmically enhanced term starts to dominate only at very high center-of-mass energies.
%
\begin{figure}[t]	
	\begin{center}
	\includegraphics[width=0.6 \textwidth]{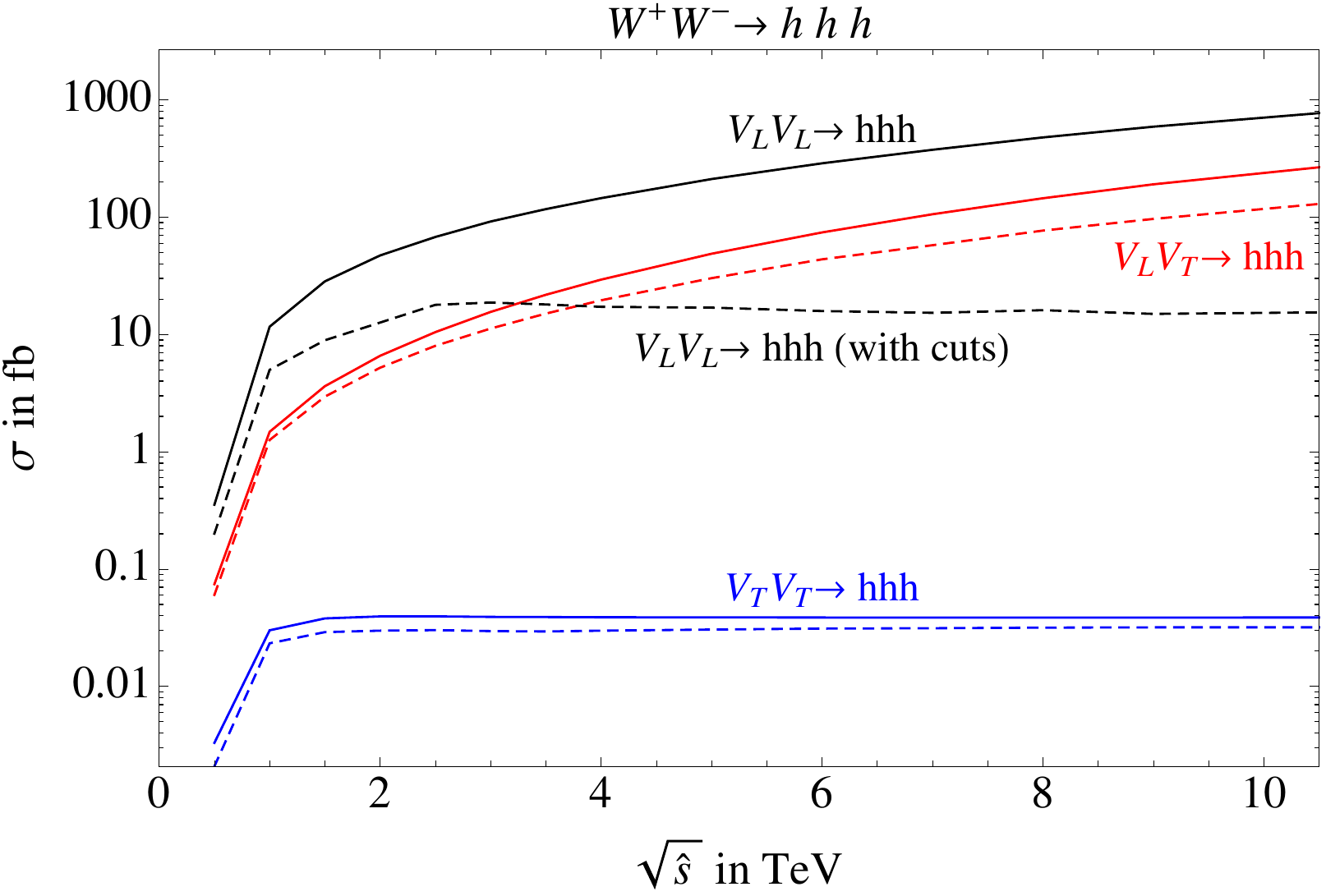}
	\end{center}
	\caption{\small Partonic cross-sections of the processes $V_{L} V_{L} \rightarrow hhh$ (black), $V_{L} V_{T} \rightarrow hhh$ (red) and 
$V_{T} V_{T} \rightarrow hhh$ (blue) as a function of $\sqrt{\hat{s}}$ for a PNGB Higgs with $\xi = 0.1$ and $m_{h} = 125$ GeV. The dashed line shows 
the partonic cross section after applying the cuts $p_{T} > 0.05 \sqrt{\hat{s}}$ for each Higgs and $m_{hh} > 0.1 \sqrt{\hat{s}}$ for all Higgs pairs.}
	\label{EnergyPolarization}
\end{figure}
%
The energy dependence of the process $V V\to hhh$ ($V=W,Z$) in the various polarization channels is shown in Fig.~\ref{EnergyPolarization}. 
We assume that the Higgs boson is a PNGB living on the coset $SO(5)/SO(4)$, so that the selection rules discussed at the beginning of this Section apply. 
The expectation for the various $2\to 3$ amplitudes is reported in Table~\ref{NDAPartonic}. Notice that for a three-body process the product of the flux factor 
and phase space is dimensionless. The naive high-energy behavior of the various cross sections is then obtained by squaring the entries of 
Table~\ref{NDAPartonic}. The total cross section is expected to follow this naive energy behavior only if the phase space integral does not get  any particular 
enhancement from singular kinematic configurations. As explained above, the $LT$ polarization channel indeed gets a logarithmic enhancement from the 
kinematical region where one of the final Higgs bosons is collinear with the incoming transverse vector. As Fig.~\ref{EnergyPolarization} shows, the $TT$ 
channel  cross section is constant at high energy, in agreement with the naive expectation $\sigma(V_TV_T\to hhh)\sim (g^4/(4\pi)^3) (v^2/f^4)$. 
On the other hand the inclusive $V_L V_L\to hhh$ cross section grows like $\hat s$ at high energy, faster than what is expected from Table~\ref{NDAPartonic}. 
This is due to the Coulomb singularity that some of the diagrams 
have in the limit in which one 
of the final Higgs bosons is collinear to the incoming beam.
The amplitude of those diagrams  goes as ${\cal A}(V_L V_L\to hhh)\sim (g^2 v/f^2) (\hat s/\hat t)$, where  $\hat t$ is the squared difference
of  one of the initial momenta and the momentum of the collinear Higgs boson.
The integral over $\hat t$ is  dominated by the singular region $t\simeq t_{min} \sim -m_W^2$, so that the total cross section gets enhanced by a factor
$(\hat s/m_W^2)$. Such an enhancement can be removed by suitable kinematic cuts to avoid all collinear configurations.
For example, the dashed curves of Fig.~\ref{EnergyPolarization} show the energy dependence of the cross sections after 
requiring $p_{T} > 0.05 \sqrt{\hat{s}}$  on each of the Higgs final momenta. As expected, the $TT$ and $LT$ channels are only slightly suppressed by the
cuts, while the Coulomb singularity of the $LL$ channel is removed and its cross section is constant at high energies.

\section{Quantitative analysis of $VV\to VV$ and $VV\to hh$}
\label{VVtoHH}

In this Section we study the sensitivity of a linear $e^+e^-$ collider to the anomalous Higgs couplings $a$, $b$ and $d_3$ through vector boson scattering 
and double Higgs production. In Section~\ref{sec:vvvv_scatt} we discuss $V_LV_L\to V_L V_L$ scattering at low-energy (ILC) and high-energy (CLIC) linear colliders; 
in Sections~\ref{sec:vvhh_scatt} and~\ref{sec:Higgsstrahlung} we focus on double Higgs production. We show that, while CLIC can provide a precise determination 
of the anomalous couplings through the study of the vector boson fusion process $e^+e^-\to hh \nu\bar\nu$, a lower-energy machine like the ILC has to rely on 
the double Higgs-strahlung process $e^+e^-\to h h Z$.  Previous analyses of $VV\to VV$ and $VV\to hh$ at high-energy linear colliders 
appeared in Refs.~\cite{OLDVV} and \cite{Barger:2003rs,Barger:1988kb} respectively.~\footnote{See also Ref.~\cite{gagaVV} for a study of the process
$\gamma\gamma\to VV$  ($V=W,Z$) to extract the anomalous Higgs couplings.}

\subsection{Identification cuts}\label{sec:idcuts}

In the following 
we set $m_h =125\,$GeV and focus  on final states where  $W$, $Z$ and $h$ decay hadronically, with the only exception of double Higgs-strahlung where we 
include leptonic decays of the $Z$. 
Our analysis is at the parton level and does not include corrections due to  QCD radiation. Final-state  partons are passed through a simple algorithm 
to obtain a crude though sufficiently accurate approximation of  jet-reconstruction and detector effects of a real experiment. 
We perform a simple Gaussian smearing of the parton energies assuming a constant resolution $\Delta E/E=5\%$~\cite{CLICCDR}.
Two partons with
\begin{equation} \label{eq:dRjjcut}
\Delta R_{jj} < 0.4 
\end{equation}
are merged together by summing their 4-momenta. The algorithm is applied recursively until it converges to a list of final partons.  A parton thus obtained which satisfies
the  cuts
\begin{equation} \label{idjets}
E_j > \! 20\,{\textrm{GeV}} \, ,  \qquad \quad |\eta_j|<2 
\end{equation}
is identified with a reconstructed jet.
The rapidity cut, in particular, excludes reconstructed jets which fall within $15^\circ$ of the collision axis. 
Leptons (muons and electrons) are identified if they satisfy the following cuts:
\begin{equation}\label{idleptons}
E_\ell > \! 5\,{\textrm{GeV}} \, ,  \qquad\quad |\eta_\ell|<2 \, .
\end{equation}
We furthermore require a separation $\Delta R_{j\ell}>0.4$ between reconstructed jets and leptons.

In events with two Higgs bosons (like those coming from $WW$ fusion or double Higgs-strahlung) we require at least 3 $b$-tags with a $b$-tagging efficiency of 
80\% assumed throughout the analysis. According to Ref.~\cite{CLICCDR}, this is associated with mistag rates of $\sim$10\% and $\sim$1\% for $c$-jets and light jets 
respectively. With this assumption the probability of tagging at least~3~$b$-jets out of 4 is 82\%.
\newline

All our event samples are generated with {\sc{MadGraph5}}~\cite{Alwall:2011uj}, except for the background to $VV\to VV$ scattering which has been
generated with {\sc{Whizard}}~\cite{Whizard}.
We do not include parton showering and hadronization.

\subsection{$VV\to VV$ scattering}\label{sec:vvvv_scatt}

All the channels $e^+e^-\to V V \ell\bar\ell$, where $\ell$ is either an electron or a neutrino, provide a framework for studying vector boson scattering at a linear collider. $VV$ scattering processes with electrons in the final state are initiated by a neutral current splitting $e^\pm\to \gamma e^\pm$ or $e^\pm\to Z e^\pm$. While the first always contributes as a large background to the signal we are interested in, the second splitting is a factor of 2 smaller than the charged current splitting $e\to W\nu$. 
We will focus therefore on $VV\nu\bar\nu$ final states and neglect $VVe^+e^-$ for simplicity. We will consider hadronically decaying vector bosons and, to avoid possible experimental issues related to energy resolution and $W/Z$ separation, we will be inclusive and sum over final states with $W$'s or $Z$'s.

The cross section for the process $e^+e^-\to VV\nu\bar\nu$ can be parametrized in terms of the coupling shift $\Delta a^2$ as
\begin{equation} \label{xsectionxiExp}
	\sigma(\Delta a^2)= \sigma_{SM} \left(1 + A \, \Delta a^2  + B \, (\Delta a^2)^{2} \right) \, ,
\end{equation} 
where $\sigma_{SM}$ is the SM cross section and $A,B$ are two dimensionless coefficients.
Notice that $\Delta a^2=\xi$ in the MCHM. For $\sqrt{s}=3$~TeV we find, before any cut,
\begin{equation}
\label{xsectionxifit}
\sigma_{SM}= 184\,\text{fb}\,, \qquad  \{ A, B \}  = \{ 0.01, 0.15 \} 
\qquad 
{\small
\left[
\begin{array}{c}
e^+ e^- \to VV\nu\bar\nu \qquad \sqrt{s}=3\,\textrm{TeV}\\
\text{before any cut}
\end{array}
\right]
}
.
\end{equation}
It is clear that at this level the cross section is largely dominated by the SM term.
One reason is the ``accidental'' numerical enhancement, discussed in Ref.~\cite{Contino:2010mh}, of the partonic $V_TV_T\to V_TV_T$ cross section compared to 
$V_LV_L\to V_LV_L$.  Another reason is that the total cross section displayed here is dominated by threshold production and does not really probe the highest energies.
 
The situation is worsened by the presence of backgrounds. The largest contribution arises from the process $e^+e^-\to W^+W^-e^+e^-$,
 which goes through a  $\gamma\gamma\to W^+W^-$ hard scattering, where the final electron and positron escape  the detector.
A similar though smaller background comes from $e^+e^-\to W^\pm Ze^\mp \nu$. 
Before cuts, the cross section of these background processes is of the order of hundreds of picobarns. 

We focus on hadronic decays of the $W$ and $Z$ bosons and select events with at least four reconstructed jets,
where jet reconstruction is done  according to the procedure discussed in Section~\ref{sec:idcuts}.
The two $V$ candidates are defined by considering the four most energetic jets in each event, $j_{1,\dots 4}$, and 
by identifying the pairing $(j_1j_2,j_3j_4)$ which minimizes the $\chi^2$ function 
\begin{equation}\label{chi2pairingV}
(m_{j_1j_2}-m_V)^2+(m_{j_3j_4}-m_V)^2 \, ,
\end{equation}
where $m_V \equiv (m_W +m_Z)/2$. We use the average mass $m_V$ in the $\chi^2$ function since we do not know a priori if the 
$V$ candidate is a $W$ or a $Z$. The algorithm is however quite effective to identify real vector bosons, and the percentage of fake  pairings is negligible.
After their reconstruction, we  impose the following cut on the invariant mass of each of the two $V$ candidates:
\begin{equation} \label{deltaminvWZ}
 |m_{jj}-m_{V}|<15\,{\textrm{GeV}} \, .
 \end{equation}
Events where such requirement is not satisfied are discarded.
The overall efficiency of the identification cuts in eqs.~(\ref{idjets}) and (\ref{deltaminvWZ})  is roughly 30\% for both signal and background.
After imposing the identification cuts and including the hadronic branching ratios of the 
$W$ and $Z$ bosons, we find that the signal rate $r = \sigma(e^+ e^- \to VV \nu\bar\nu) \times BR(VV \to 4j)$ is parametrized by
\begin{equation} \label{fitrateVV}
r(\Delta a^2)= r_{SM} \left(1 + A_r \, \Delta a^2  + B_r \, (\Delta a^2)^{2} \right) \, ,
\end{equation} 
with
\begin{equation}
r_{SM}= 28.7\,\text{fb}\,, \qquad  \{ A_r, B_r \}  = \{ 0.04, 0.16 \} \qquad 
{\small
\left[
\begin{array}{c}
e^+ e^- \to 4j\nu\bar\nu \qquad \sqrt{s}=3\,\textrm{TeV}\\
\text{after identification cuts}
\end{array}
\right]
}
.
\end{equation}
In order to enhance the signal and reduce the backgrounds we  apply the following additional set of cuts 
\begin{equation}\label{cutWW1}
\begin{split}
m_{V_1 V_2}&> 500\, {\textrm{GeV}}\, , \\[0.1cm]
\min p_{T}(V_i)&> 100\, {\textrm{GeV}}\, , \\[0.1cm]
\max |\eta_{V_i}|&< 1.1\, , \\[0.1cm]
m_{\nu\nu}&> 150\,{\textrm{GeV}} \, ,
\end{split}
\end{equation}
where $V_{1,2}$ denote the two $V$ candidates. The cut on the invariant mass of the two neutrinos, in particular, eliminates those backgrounds,  
like $e^+e^-\to VVZ$ (with $Z \to\nu\bar\nu$), where the missing energy 
arises from the invisible decay of an on-shell $Z$ boson.
Finally, we require
\begin{equation}\label{cutWW2}
p_T(V_1 V_2)>75\, {\textrm{GeV}}.
\end{equation}
This latter cut on the  transverse momentum of the $VV$ system is applied to further reduce the $e^+e^-\to W^+W^-e^+e^-$ and $e^+e^-\to W^\pm Ze^\mp \nu$ backgrounds: 
if the final electrons are so forward to be lost in the beam-pipe it is reasonable to expect the total $p_T$ of the recoiling vectors to be small. After all these cuts, the signal 
rate  is parametrized by
\begin{equation}
r_{SM} = 1.7\,\text{fb}\,, \qquad  \{ A_r, B_r\}  = \{ 0.04, 0.7 \} 
\qquad 
{\small
\left[
\begin{array}{c}
e^+ e^- \to 4j\nu\bar\nu \qquad \sqrt{s}=3\,\textrm{TeV}\\
\text{after analysis cuts}
\end{array}
\right]
}
.
\end{equation}
The background rate from $e^+e^-\to W^+W^-e^+e^-$ and $e^+e^-\to W^\pm Ze^\mp \nu$ processes amounts to roughly $r_b = 2.5\,$fb after the cuts. 
The calculation has been performed using {\sc{Whizard}}~\cite{Whizard} by requiring the electrons in the final state to be undetected 
($\eta(e^\pm)>2.5$).~\footnote{We found significant numerical instabilities in the MC computation of the cross section, with variations in the final result up 
to $30-50\%$. As explained below, we took into account such uncertainty in our analysis by  rescaling the final background rate by a factor 1.5.}

We thus proceed to estimate the expected sensitivity on $\Delta a^2$. We follow a Bayesian approach and 
construct a posterior probability for the total event rate $r_{tot}$ 
\begin{equation}
\label{eq:posterior}
p(r_{tot}|N_{obs}) \propto {\cal L}(N_{obs} | r_{tot} L ) \pi(r_{tot})\, , 
\end{equation}
where $N_{obs}$ is the assumed number of observed events and $L$ is the integrated luminosity.
We denote with $\pi(r_{tot})$ the prior distribution and with ${\cal L}(N_{obs} | r_{tot} L )$ the likelihood function,
which we take to be a Poisson distribution
\begin{equation}
\label{eq:poisson}
{\cal L}(N_{obs}| r_{tot} L ) = \frac{e^{-r_{tot} L} \, (r_{tot} L)^{N_{obs}}}{N_{obs}!}\, .
\end{equation}
For a given true value~$\Delta \bar a^2$ of the coupling shift, we assume the number of observed events to be $N_{obs} = ( r(\Delta \bar a^2) +r_b ) L$,
while the total rate 
is $r_{tot} = r(\Delta a^2) +r_b$. As we do not explicitly introduce additional uncertainties 
(theoretical or systematic) on the estimate of the background in our statistical analysis,~\footnote{This would require introducing one or more 
corresponding nuisance parameters in the likelihood function, which is beyond the scope of our simple statistical analysis.} we have conservatively rescaled the 
background rate $r_b$ by a factor 1.5  compared to the MC prediction.

By assuming a flat prior on $\Delta a^2$ and setting the integrated
luminosity to $L=1\,{\textrm{ab}}^{-1}$, we obtain the 68\% probability intervals shown in Table~\ref{tab:xierrors_WW} (second column)
for different true values $\Delta \bar a^2$. We find that for large $\Delta \bar a^2$, the term proportional to $(\Delta a^2)^2$ dominates the rate and
a second peak of the likelihood appears at negative values of the coupling shift. The 68\% interval in these cases consists of two disconnected parts.
%
\begin{table}[t]
\begin{center}
{\small
\begin{tabular}{c|cc}
\hline \hline
&& \\[-0.3cm]
$\Delta \bar a^2 = \bar \xi$  & $\Delta a^2$ & $\xi$ \\[0.1cm]
\hline 
&& \\[-0.3cm]
0        & $(-0.21,0.17)$                              &$(0,0.17)$ \\[0.15cm]
0.05   & $(-0.22,0.17)$                              & $(0,0.18)$ \\[0.15cm]
0.1    &  $(-0.23,0.18)$                              & $(0,0.19)$ \\[0.15cm]
0.2    &  $(-0.34,-0.1) \cup (0.04,0.28)$   & $(0.06,0.28)$ \\[0.15cm]
0.3    &  $(-0.45,-0.22) \cup (0.17,0.39)$  & $(0.17,0.39)$ \\[0.15cm]
0.5    &  $(-0.62,-0.49) \cup (0.45,0.56)$  &$(0.43,0.56)$ \\[0.15cm]
\hline
\hline
\end{tabular}
}
\\[0.1cm]
\caption{\small  Expected 68\% probability intervals on $\Delta a^2$  (second column) and $\xi$ (third column) for different true values
$\Delta \bar a^2 = \bar \xi$ measured at CLIC $3\,$TeV through $VV\to VV$ scattering. 
The limits on $\xi$ have been derived by taking into account that
only values in the range $\xi \in [0,1]$ are theoretically allowed.
See the text for details on the statistical analysis.
}
\label{tab:xierrors_WW}
\end{center}
\end{table}
%
%
We also considered the $SO(5)/SO(4)$ models MCHM where the coupling shift is
$\Delta a^2 =\xi$, see eq.~(\ref{eq:pngbparameters}).  In this case we have imposed a prior on $\xi$ which is flat in the theoretically allowed range $[0,1]$
and vanishing outside. The corresponding 68\% probability intervals  on  $\xi$ are reported in the third column of Table~\ref{tab:xierrors_WW} 
for different true values $\bar \xi$.
With our set of cuts, a 3\,TeV linear collider is sensitive to values of $\Delta a^2$ ($\xi$) bigger than $\sim 0.2$ through 
$WW$ scattering.~\footnote{\label{ftn:sensitivity}Here and in the following, by sensitivity/precision on some anomalous Higgs coupling 
we mean the $68\%$ error  on its measured value for injected SM signal.}
We do not find any significant gain in resolution by applying a harder cut on the $VV$ invariant mass.

A similar analysis can be carried out for a lower energy machine. We considered for example the case of a $500\,$GeV linear collider.
Parametrizing the signal cross section as in  eq.~(\ref{xsectionxiExp}), before cuts we find
\begin{equation} \label{xsectionxifitILC}
\sigma_{SM}= 5.12\,\text{fb}\,, \qquad  \{ A, B \}  = \{ -0.03, 0.06 \} 
\qquad 
{\small
\left[
\begin{array}{c}
e^+ e^- \to VV\nu\bar\nu \qquad \sqrt{s}=500\,\textrm{GeV}\\
\text{before any cut}
\end{array}
\right]
}
.
\end{equation}
At this stage the background processes have very large cross sections, of the order of 100\,fb. 
However, all backgrounds can  be reduced to a negligible level by applying the identification cuts of eqs.~(\ref{idjets}) and (\ref{deltaminvWZ}), the additional cuts
$p_T(V_1 V_2)>40$\,GeV, $m_{\nu\nu}>100$\,GeV and requiring all electrons in the final states to escape detection, $\eta(e^\pm)>2.4$.
After these cuts, and including the hadronic branching ratio of $W$ and $Z$, the signal rate is parametrized as in eq.~(\ref{fitrateVV}) with:
\begin{equation}\label{xsectionxifitILCCut}
r_{SM} =0.5\,\text{fb}\,, \qquad  \{ A_r, B_r \}  = \{ -0.03, 0.15 \} 
\qquad 
{\small
\left[
\begin{array}{c}
e^+ e^- \to 4j\nu\bar\nu \qquad \sqrt{s}=500\,\textrm{GeV}\\
\text{after analysis cuts}
\end{array}
\right]
}
.
\end{equation}
By repeating the previous statistical analysis, we find that  with an integrated luminosity $L=1\,{\textrm{ab}}^{-1}$
the effect of a non-vanishing $hVV$ anomalous  coupling can be  resolved in $e^+e^-\to VV \nu\bar\nu$ only for large values of $\Delta a^2$ ($\xi$), of the order $0.5-0.6$.

\subsection{$VV\to hh$ scattering}
\label{sec:vvhh_scatt}

The scattering amplitude for $V_L V_L\to hh$ depends on $a$, $b$ and $d_3$ and can be conveniently written as 
${\cal A} = a^2 \left( {\cal A}_{SM} + {\cal A}_1 \, \delta_b + {\cal A}_2 \, \delta_{d_3} \right)$, where ${\cal A}_{SM}$ is the value predicted by the SM 
and~\footnote{In the MCHM4 $\delta_b = \xi/(1-\xi)$, $\delta_{d_3}=0$, while in the MCHM5 $\delta_b = \delta_{d_3}= \xi/(1-\xi)$. 
See eqs.~(\ref{eq:pngbparameters}),(\ref{eq:pngbparametersd3})}
\begin{equation} \label{dbdd3}
\delta_b \equiv 1 - \frac{b}{a^2} \, , \qquad \delta_{d_3} \equiv 1 - \frac{d_3}{a}\, .
\end{equation}
At large partonic center-of-mass energies, $E\gg m_V$, ${\cal A}_1$ grows like $E^2$, while ${\cal A}_2$ and ${\cal A}_{SM}$ are constant. 
The parameter $\delta_b$ thus controls the high-energy behavior of the amplitude and gives a genuine ``strong coupling'' signature.
On the contrary, $\delta_{d_3}$  determines the value of the cross section at threshold~\cite{Contino:2010mh}.
In an $e^+e^-$ collider, $V_L V_L\to hh$ scatterings can be studied via the processes $e^+e^- \to \nu\bar\nu hh$ and $e^+e^- \to e^+e^- hh$. 
The latter, initiated by a partonic $ZZ$ state, has a cross section which is roughly one order of magnitude smaller than the former. This is due in particular to the 
fact that the $e^\pm\to Z e^\pm$ splitting function is roughly a factor of~2 smaller than $e^\pm \to W^\pm \nu$.
For this reason we neglect $e^+e^- \to e^+e^- hh$ in the following.
The $e^+e^- \to \nu\bar\nu hh$ cross section can be written as
\begin{equation}\label{sigmatot}
\sigma = a^4 \, \sigma_{SM} \left( 1 + A \, \delta_b + B \, \delta_{d_3} + C\,  \delta_b \delta_{d_3} + D\,  \delta_b^2 + E\,  \delta_{d_3}^2 \right)\, ,
\end{equation}
where $\sigma_{SM}$ denotes its SM value.
Notice that $a$ enters only as an overall factor.
Without applying any kinematic cut on the Higgs decay products (nor including the branching fraction of Higgs decays) we find, for $\sqrt{s}=3$~TeV,
\begin{equation}\label{sigmahhdeltas}
\sigma_{SM} = 0.83\,\text{fb}\,, \quad  \{ A, B, C, D, E \}  = \{ 3.83, 0.64, 3.41, 15.6, 0.48 \} 
\quad 
{\small
\left[
\!\begin{array}{c}
 \sqrt{s}=3\,\textrm{TeV}\\
\text{before any cut}
\end{array}
\!\right]
}
.
\end{equation}
Notice that although the SM cross section $\sigma_{SM}$ of the processes $VV\to VV$ and $VV\to hh$ differs by more than two orders of magnitude,
the energy-growing contributions (given by $\sigma_{SM} B = 27.6\,$fb for $VV\to VV$, see eqs.~(\ref{xsectionxiExp}) and~(\ref{xsectionxifit}),  
and $\sigma_{SM} D=12.9\,$fb for $VV\to hh$,   see eqs.~(\ref{sigmatot}) and~(\ref{sigmahhdeltas})) are of the same size, as required by the $SO(4)$ invariance.

In contrast to $VV\to VV$ scattering, in the case of double Higgs production simple acceptance 
and reconstruction cuts keep the background at a negligible level.
For our analysis we focus on events where both Higgs bosons in the signal decay to $b\bar b$, and select events with four or more jets and at least
three $b$-tags.
The most important processes which can fake the signal are then $e^+e^-\to \nu\bar\nu h Z$, $e^+e^-\to \nu\bar\nu ZZ$ and  $e^+e^-\to e^+e^-ZZ$.  
In all cases the $Z$ boson must decay to a $b\bar b$ pair, and in the latter process both electrons have to be missed in the beam pipe. 
Before cuts we find 
\begin{equation}
\label{hh_backgrounds}
\begin{split}
\sigma(e^+e^-\to hZ\nu\bar\nu \to b\bar b b\bar b \, \nu\bar\nu)&= 0.88 \,{\rm fb}\, ,\\[0.2cm]
\sigma(e^+e^-\to ZZ\nu\bar\nu \to  b\bar b b\bar b \, \nu\bar\nu)&= 1.26 \,{\rm fb}\, ,\\[0.2cm]
\sigma(e^+e^-\to ZZ e^+e^- \to b\bar b b\bar b \, e^+e^-)&= 0.58 \,{\rm fb}\, , 
\end{split}
\end{equation}
which can be compared to the signal cross section in eq.~(\ref{sigmahhdeltas}) after multiplying this latter by the Higgs pair branching fraction 
$BR(hh\to b\bar b b\bar b) \simeq BR(hh\to b\bar b b\bar b)_{SM} = 0.34$.
Further backgrounds, like for example $t\bar t\to b \bar b W^+ W^- \to b \bar b jj l\nu$, can fake our signal only if one or more light jets are mistagged as $b$-jets and if extra
charged leptons escape into the beam pipe. This is enough suppression to safely ignore them.
The backgrounds in eq.~(\ref{hh_backgrounds}), on the other hand, are largely suppressed, and thus negligible, if the jet energy resolution of the detector is sufficiently 
good to accurately distinguish a $Z$  from a Higgs boson. This seems to be a valid assumption according to Ref.~\cite{CLICCDR}, and in the following we will consequently assume the
backgrounds to be negligible.

A simple-minded approach to the extraction of the two parameters $\delta_b$ and $\delta_{d_3}$ is the following. Let us consider a kinematical variable 
$\mathcal O$ whose value increases with the c.o.m.~energy of the $W^+W^-\to hh$ subprocess.  The invariant mass of the two Higgses, $m_{hh}$, and the 
sum of their transverse momenta, $H_T$, are two valid examples for $\mathcal O$. We can divide the set of $e^+e^- \to \nu\bar\nu hh$ events into two categories
according to whether $\mathcal O <\bar{\mathcal O}$ or $\mathcal O >\bar{\mathcal O}$, where~$\bar{\mathcal O}$ is some fixed value. The number of observed 
events in these two categories can be fitted to $\sigma_<(\delta_b,\delta_{d_3})$ and $\sigma_>(\delta_b,\delta_{d_3})$.
Notice that thanks to the cut on $\mathcal O$, $\sigma_>(\delta_b,\delta_{d_3})$ will have an enhanced sensitivity to $\delta_b$ while 
$\sigma_<(\delta_b,\delta_{d_3})$ is more sensitive to $\delta_{d_3}$. 

We thus adopt the above strategy and proceed as follows.  We start by selecting events with four or more reconstructed jets.
The Higgs candidates are identified from the list of the four most energetic jets, $j_{1, \dots 4}$, by selecting 
the pairing $(j_1 j_2, j_3 j_4)$  which minimizes the $\chi^2$ function
\begin{equation}
\label{chi2pairingh}
(m_{j_1j_2}-m_h)^2+(m_{j_3j_4}-m_h)^2 \, .
\end{equation}
We impose the following cut on the invariant mass of each of the two Higgs candidates
\begin{equation}\label{deltaminvh}
|m_{jj}-m_h|<15\,{\textrm{GeV}} \, ,
\end{equation}
and require that at least three of the jets $j_{1, \dots 4}$ are $b$-tagged.
Events where these requirements are not fulfilled are discarded.
We find that the overall efficiency of the identification cuts of eqs.~(\ref{eq:dRjjcut}),~(\ref{idjets}) and (\ref{deltaminvh})
varies from 20\% to roughly 35\% when $\delta_b$  ranges in the interval $0-0.5$, while it is only marginally sensitive to $\delta_{d_3}$.
In particular, the energy cut on the jets has an  almost constant efficiency (roughly 80\%)  over the whole parameter space. 
The  variation in the total efficiency comes mainly from the cuts on pseudorapidity and on $\Delta R$. The  cut on $\eta$ disfavors small values of 
$\delta_b$, since these typically lead to more forward Higgses and consequently more forward $b$-jets, which in turn have a smaller probability to pass 
the $\eta$ cut. The cut on minimum $\Delta R$, eq.~(\ref{eq:dRjjcut}), on the other hand, disfavors large values of $\delta_b$, since these lead to more 
boosted Higgses and thus more 
collimated decay products. Finally, the cut in eq.~(\ref{deltaminvh}) has an almost unit efficiency in our parton-level analysis with our assumed energy resolution.

\begin{figure}[t]
\begin{center}
\includegraphics[width=0.46\linewidth]{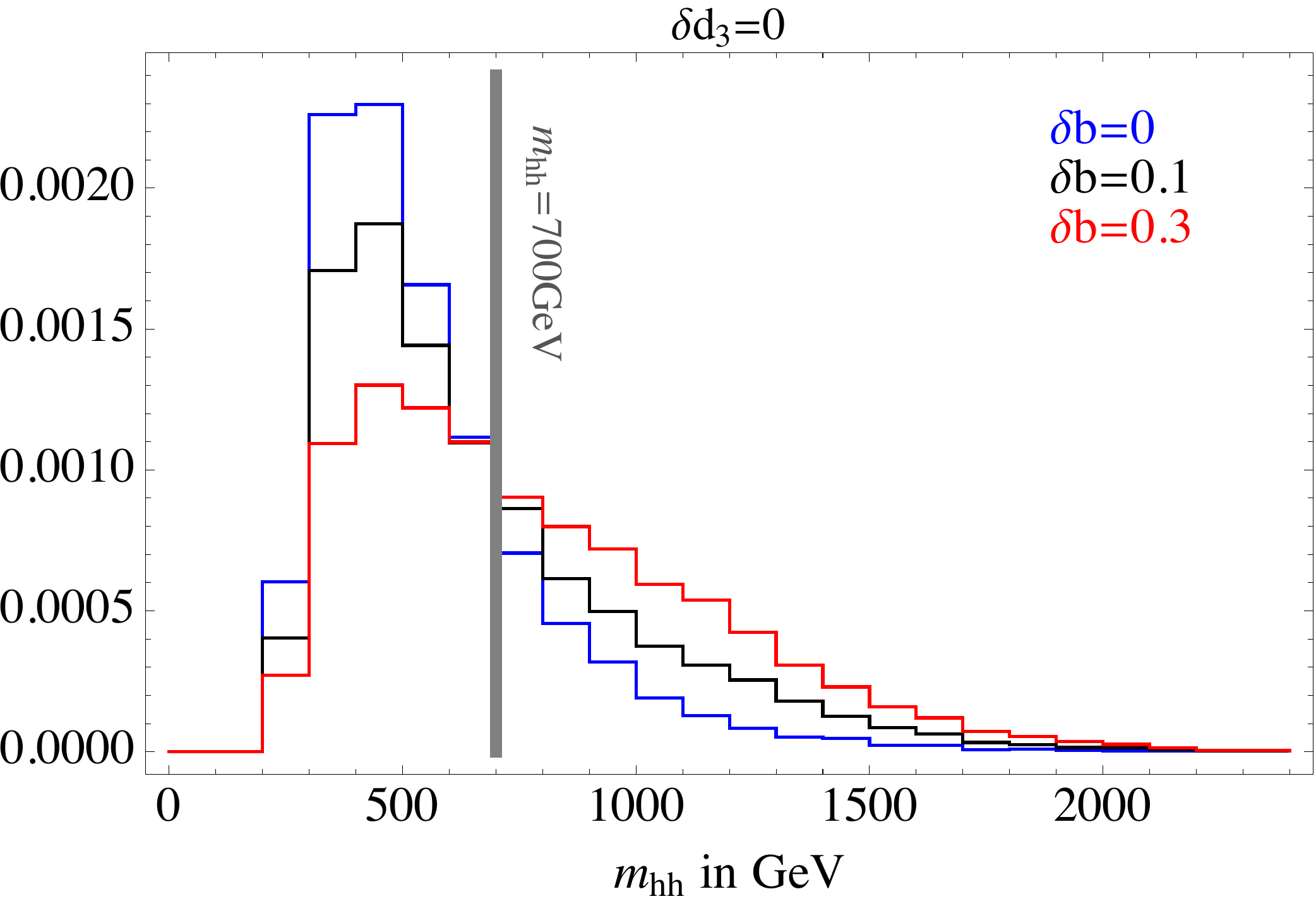} \hspace{0.75cm}
\includegraphics[width=0.46\linewidth]{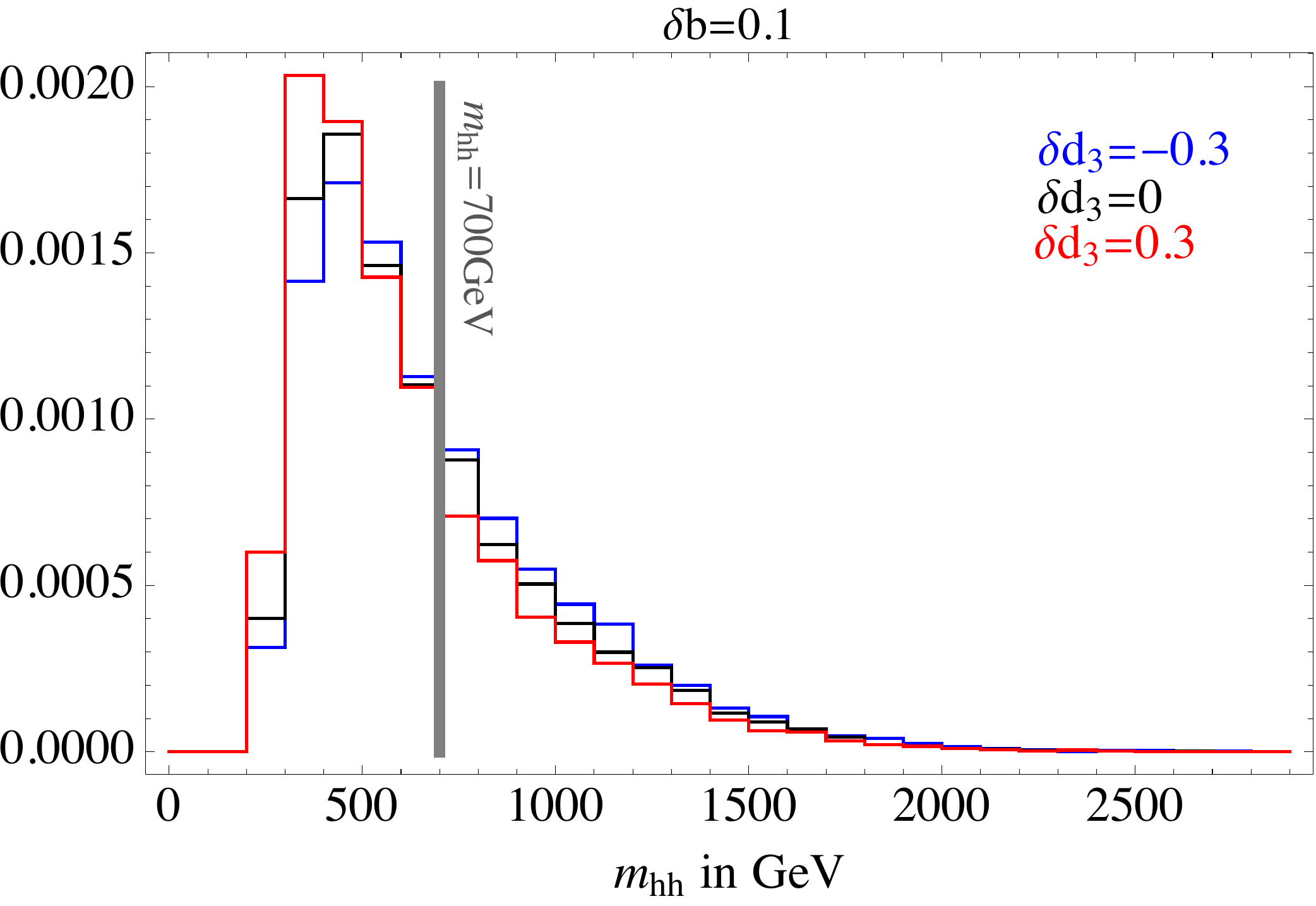}\\ \vspace{0.5cm}
\includegraphics[width=0.46\linewidth]{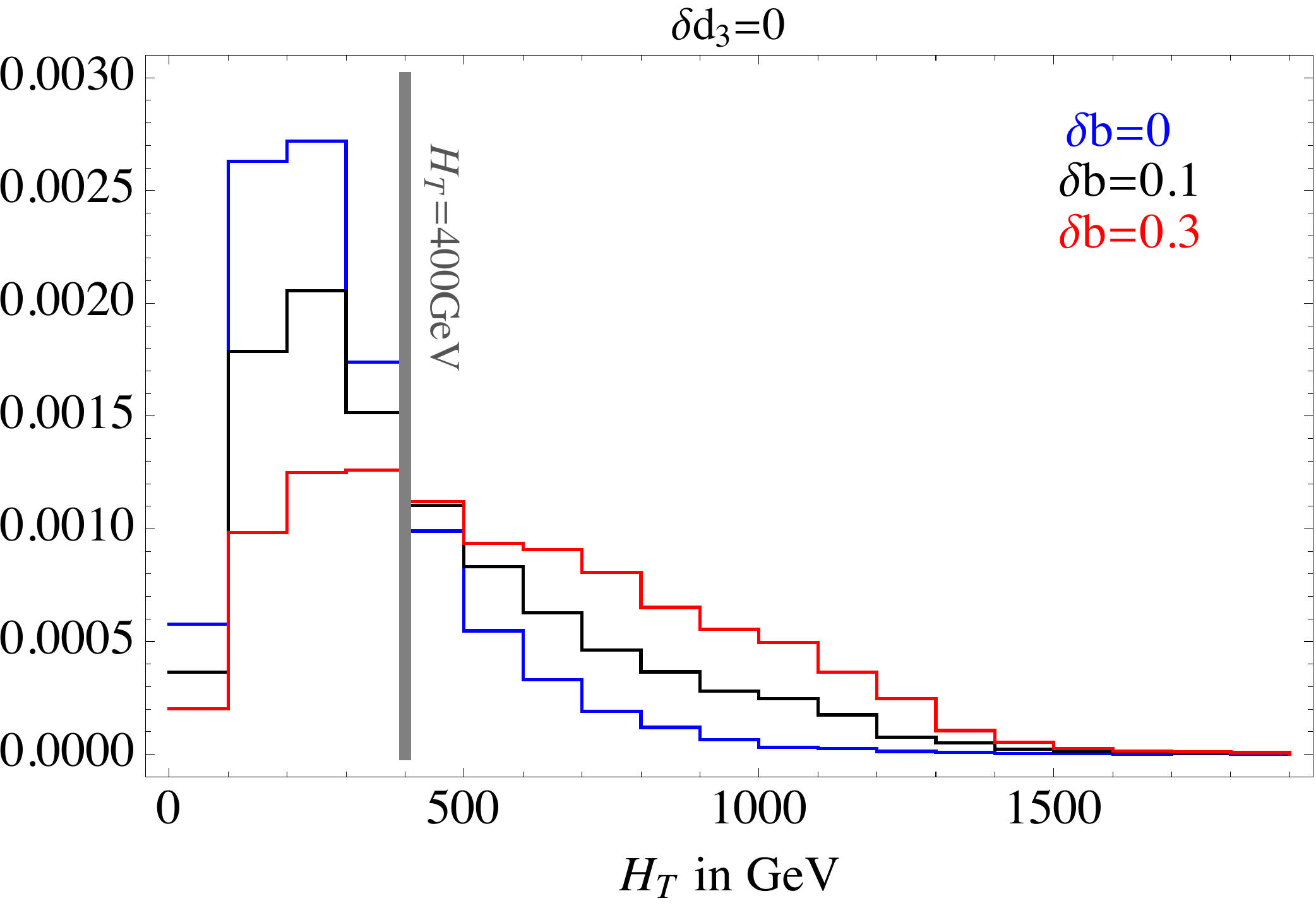}\hspace{0.75cm}
\includegraphics[width=0.46\linewidth]{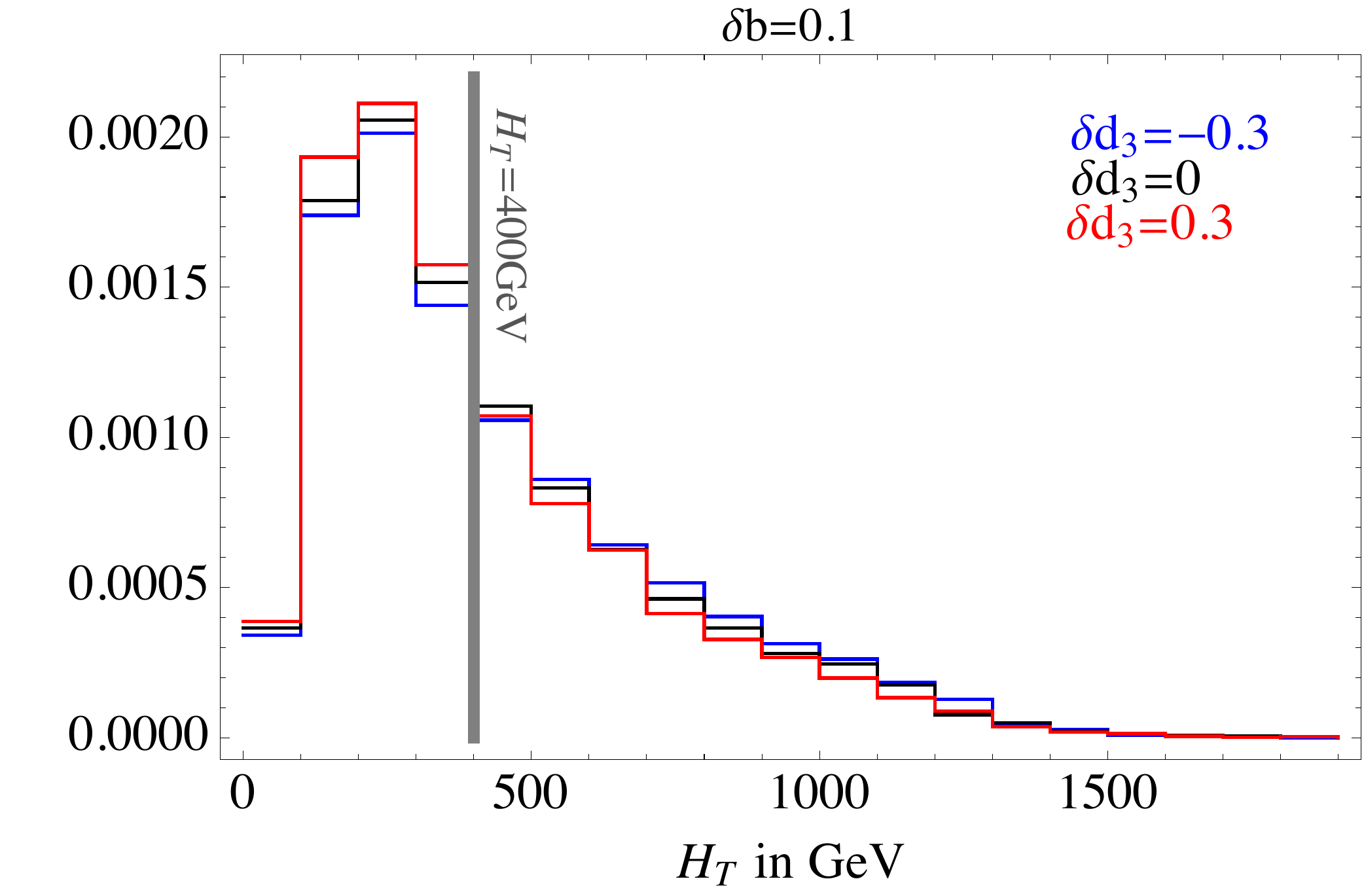}
\\[-0.1cm]
\caption{\small 
Normalized differential cross sections $d\sigma/dm_{hh}$ and $d\sigma/d H_T$ for $e^+e^- \to  \nu\bar\nu hh$  at CLIC with
$\sqrt{s} = 3$~TeV after the identification cuts of eqs.~(\ref{idjets}) and (\ref{deltaminvh}), 
for several values of $\delta_b$ and $\delta_{d_3}$.
}
\label{distributions}
\end{center}
\end{figure}

\begin{figure}[tp]
\begin{center}
\includegraphics[width=0.32\linewidth]{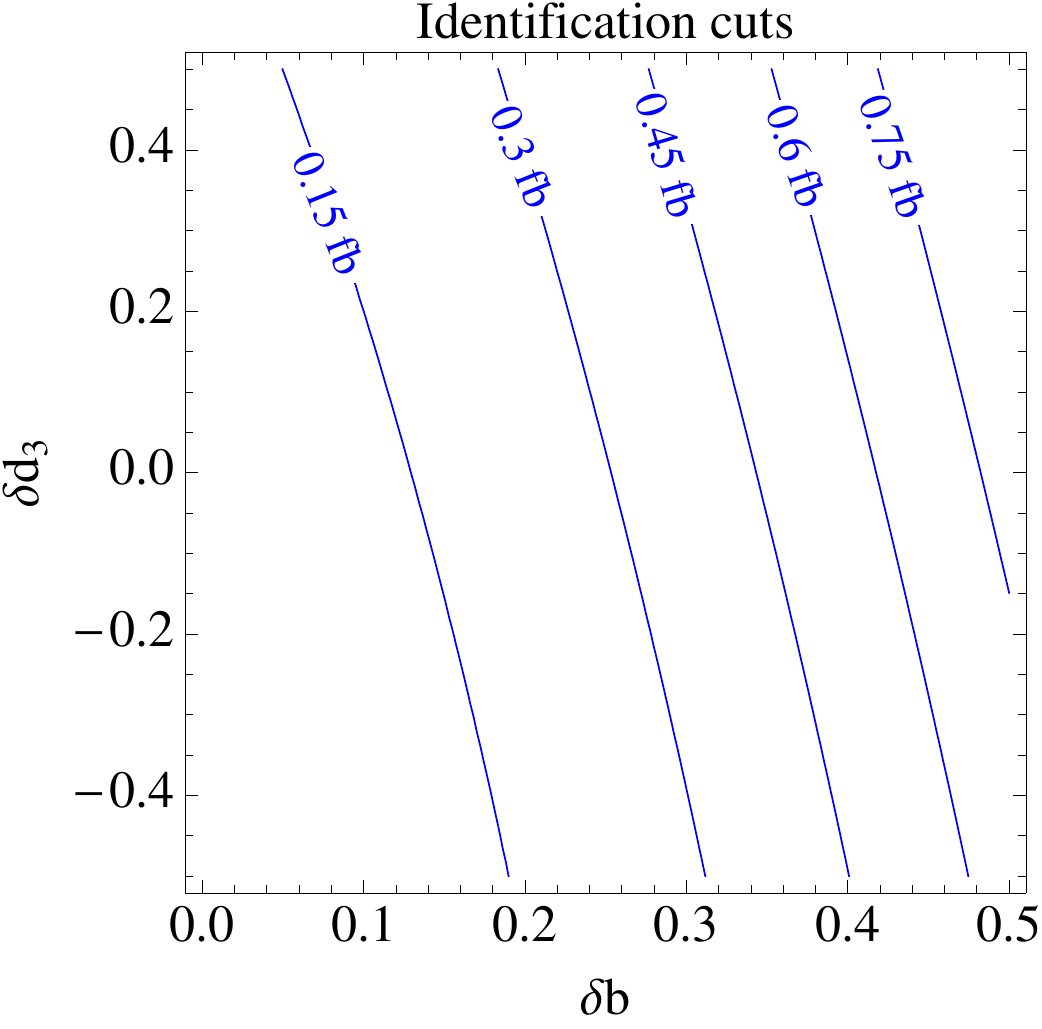} \hspace{0.05cm}
\includegraphics[width=0.32\linewidth]{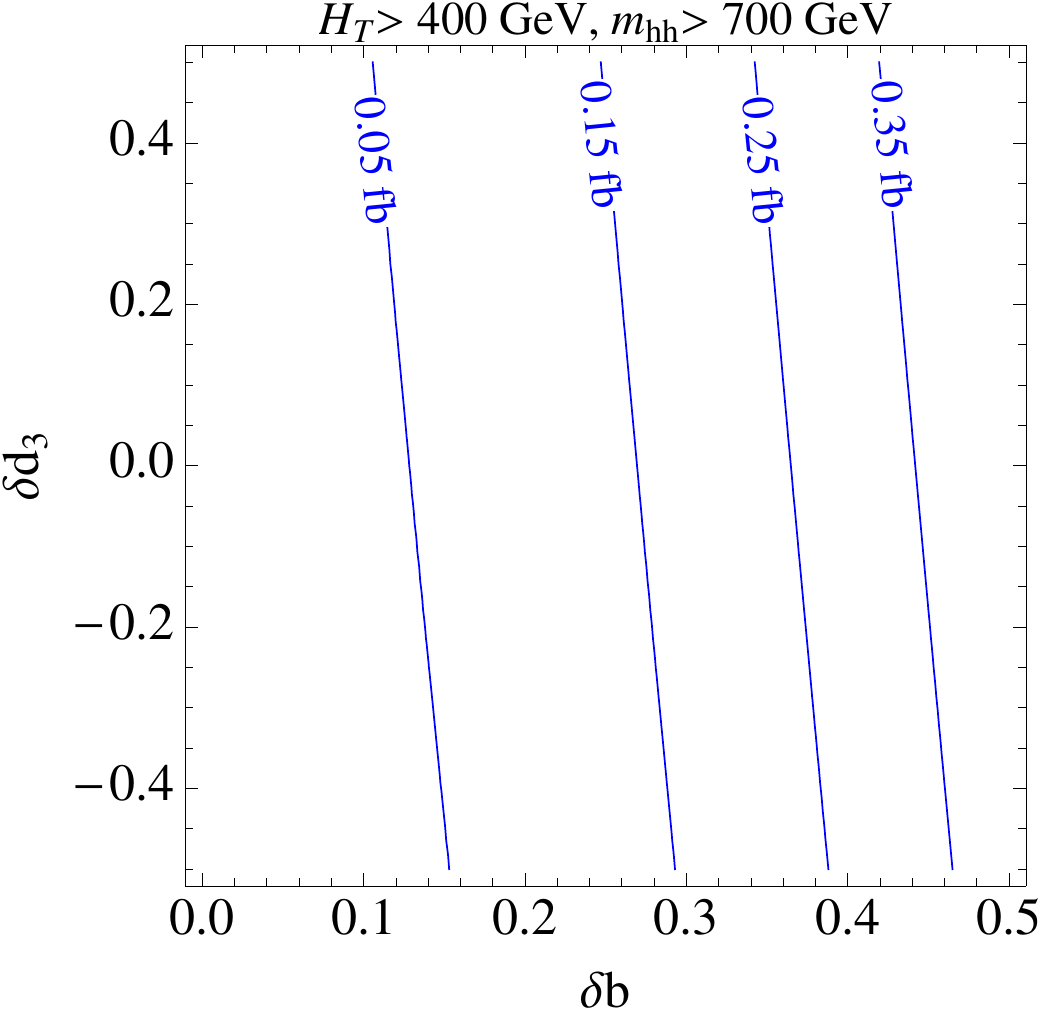} \hspace{0.05cm}
\includegraphics[width=0.32\linewidth]{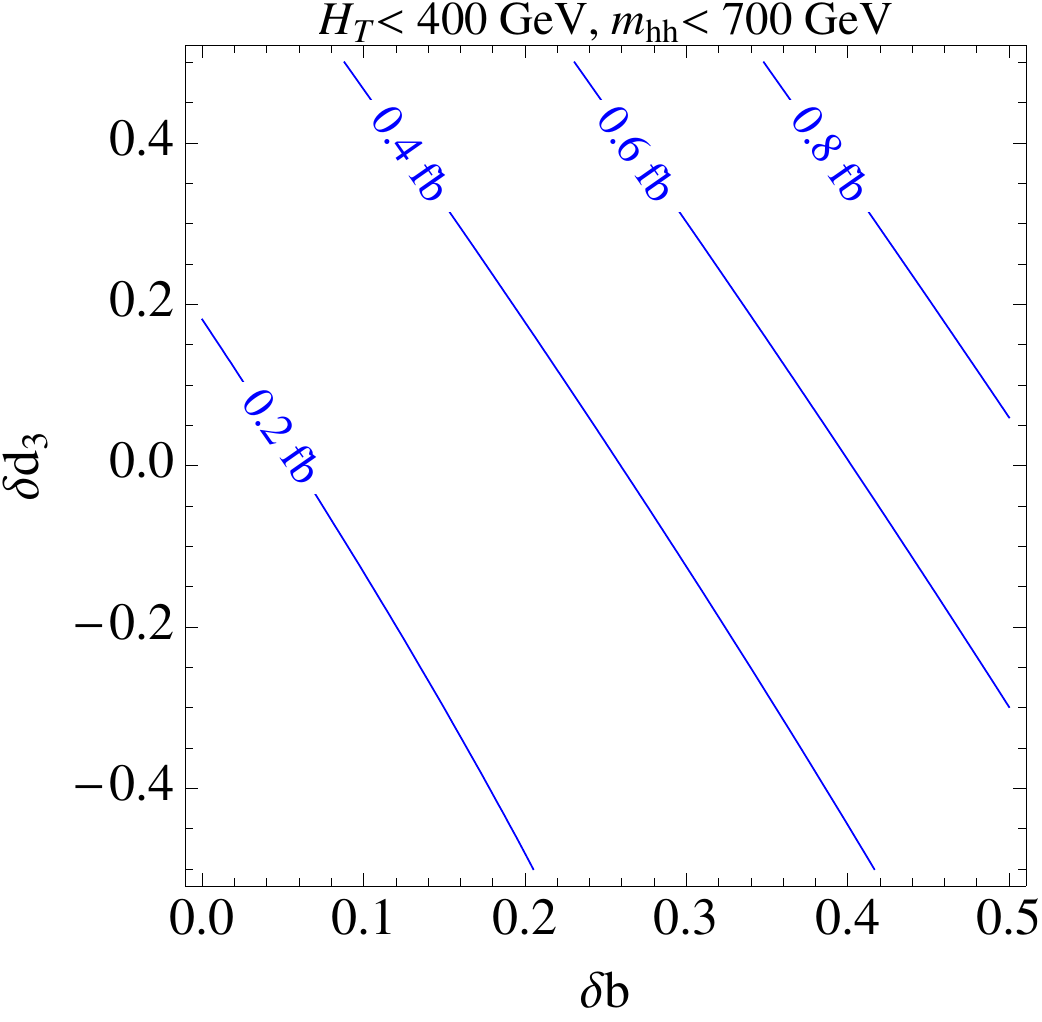}
\\[-0.1cm]
\caption{\small Contours of constant $e^+e^-\to hh (\to b\bar b b\bar b)  \nu\bar\nu$ rate (see eq.~(\ref{fitratehh}))
 for $\sqrt{s}=3\,$TeV in the plane $(\delta_b,\delta_{d_3})$.
We set $a=1$ and $BR(hh\to bb\bar b\bar b) = BR(hh\to bb\bar b\bar b)_{SM}$.
}
\label{fig:hhnunuxsec}
\end{center}
\end{figure}

Figure~\ref{distributions} shows the distributions of $m_{hh}$ and $H_T$ for some fixed values of the parameters~$\delta_b$ and $\delta_{d_3}$ after the identification cuts. 
%
%
%
While a single cut on either of these two kinematic variables is sufficient to extract the dependence on~$\delta_b$ and $\delta_{d_3}$, we found that
using both $m_{hh}$ and $H_T$ gives a slightly better sensitivity. 
We thus consider the four independent kinematical regions
\begin{equation}
\label{eq:+cut}
\begin{aligned}
{\textrm{I}}:\hspace{0.3cm} & m_{hh}>700\,{\textrm{GeV}}~ {\textrm{and}}~ H_{T}>400\,{\textrm{GeV}} \, ,\\[0.25cm]
{\textrm{II}}:\hspace{0.3cm} & m_{hh}>700\,{\textrm{GeV}}~ {\textrm{and}}~ H_{T}<400\,{\textrm{GeV}} \, ,\\[0.25cm]
{\textrm{III}}:\hspace{0.3cm} & m_{hh}<700\,{\textrm{GeV}}~ {\textrm{and}}~ H_{T}>400\,{\textrm{GeV}} \, ,\\[0.25cm]
{\textrm{IV}}:\hspace{0.3cm} & m_{hh}<700\,{\textrm{GeV}}~ {\textrm{and}}~ H_{T}<400\,{\textrm{GeV}}.
\end{aligned}
\end{equation}
Our final results do not crucially depend on the specific choice of the cuts on $m_{hh}$ and $H_T$.
One could in principle optimize them to obtain the best sensitivity on the parameters. We checked, however, that reasonable variations around the values 
 adopted in eq.~(\ref{eq:+cut}) result in small variations of the final results.
For each of  the kinematic regions (\ref{eq:+cut}), the signal rate $r\equiv \sigma(e^+e^- \to \nu\bar\nu hh)\times BR(hh\to bb\bar b\bar b)$ can be parametrized as
follows
\begin{equation}\label{fitratehh}
r =  r_{SM}\, a^4 \, \left(\frac{BR(b\bar b)}{BR(b\bar b)_{SM}}\right)^2 
\left( 1 + A_r \, \delta_b + B_r \, \delta_{d_3} + C_r\,  \delta_b \delta_{d_3} + D_r\,  \delta_b^2 + E_r\,  \delta_{d_3}^2 \right)\, ,
\end{equation}
where $r_{SM}$ is the SM rate and $BR(b\bar b)$ is the Higgs branching fraction to $b\bar b$.
The  values of the coefficients $A_r,B_r,C_r,D_r,E_r$ and of $r_{SM}$ are reported in Table~\ref{tab:fitregions}.
%
\begin{table}[tp]
\begin{center}
{\small
\begin{tabular}{c|ccccccc}
\hline
\hline
& $r_{SM}$ [ab]& $A_r$ & $B_r$ & $C_r$ & $D_r$ & $E_r$\\
\hline
&&&&&& \\[-0.3cm]
I& 8.8    &15.6 &0.88 & 14.5& 164&0.07\\
II&  4.5   &3.87& 0.30&0.92&4.44& -0.08\\
III&  6.5  &9.89& 1.25&17.1&55.4&1.54\\
IV&  44   & 3.95& 1.23&5.09&7.3&1.10\\
\hline
\hline
\end{tabular}
}
\\[0.1cm]
\caption{\small Fit of the $e^+e^-\to hh (\to b\bar b b\bar b)  \nu\bar\nu$ rate (see eq.~(\ref{fitratehh})) at CLIC with $\sqrt{s}=3\,$TeV
in the various kinematical regions defined in eq.~(\ref{eq:+cut}). 
}
\label{tab:fitregions}
\end{center}
\end{table}
%
Figure~\ref{fig:hhnunuxsec} shows the curves of constant rate in the plane ($\delta_b$, $\delta_{d_3}$) for three choices of cuts:  
only the identification cuts of eqs.~(\ref{idjets}) and (\ref{deltaminvh}), identification cuts + region I, identification cuts + region IV.
%

In order to derive the expected sensitivity on $\delta_b$ and $\delta_{d_3}$, we construct a Poisson likelihood function (see eq.~(\ref{eq:poisson}))
for each of the  kinematical regions of eq.~(\ref{eq:+cut}), and a global likelihood as the product of the individual ones.
We assumed a flat prior on $\delta_b$ and $\delta_{d_3}$. 
Since $a$ and the branching ratio $BR(b\bar b)$ appear in eq.~(\ref{fitratehh}) as overall factors, they can be conveniently absorbed by rescaling 
the integrated luminosity $L$ (note that, by the time the study of $VV\to hh$ will be feasible, both $a$ and $BR(b\bar b)$ will be known precisely enough through single Higgs processes).
The sensitivity on $\delta_b\,$($\delta_{d_3}$) is obtained by marginalizing the posterior probability  over $\delta_{d_3}\,$($\delta_b$) 
and using the  resulting single-parameter function to find the 68\% probability interval on $\delta_b\,$($\delta_{d_3}$).

The results for a 3\,TeV linear collider with 
$L = 1 \,\text{ab}^{-1}/(a^2 BR(b\bar b)/BR(b\bar b)_{SM})^2$ are shown in 
Tables~\ref{tab:errors1}~and~\ref{tab:errors2}.~\footnote{Notice that for small $\delta_b$ and 
large and negative $\delta_{d_3}$,  the central value of the measured parameter sometimes does not coincide with the true value.
This is because in this limit, for our choice of integrated luminosity, the 2D likelihood can be largely non-gaussian and its marginalization over one parameter can
lead to a shift of the central value of the second one.}
%
\begin{table}[tp]
\begin{center}
{\footnotesize
\begin{tabular}{cr|rrrrrrr}
\hline \hline 
 &&&&&&&& \\[-0.3cm]
\multicolumn{2}{c|}{measured}  & \multicolumn{7}{c}{$\bar\delta_{d_3}$} \\[0.05cm]
\multicolumn{2}{c|}{$\delta_{b}$} & -0.5  & -0.3 & -0.1 & 0 & 0.1 &  0.3 &  0.5 \\
 \hline 
&&&&&&& \\[-0.2cm]
\multirow{8}{*}{$\bar\delta_{b}$}  
 &0                  & $-0.045_{-0.025}^{+0.060}$&$0.015_{-0.040}^{+0.020}$&$0.010_{-0.045}^{+0.070}$&$0.00_{-0.05}^{+0.05}$&$0.00_{-0.03}^{+0.03}$&$0.00_{-0.03}^{+0.03}$&$0.00_{-0.03}^{+0.03}$\\[0.25cm]
 &0.01             & $-0.055_{-0.020}^{+0.070}$&$0.030_{-0.045}^{+0.030}$&$0.020_{-0.035}^{+0.080}$&$0.015_{-0.035}^{+0.030}$&$0.010_{-0.030}^{+0.020}$&$0.010_{-0.025}^{+0.025}$&$0.010_{-0.025}^{+0.025}$\\[0.25cm]
& 0.02              & $0.02_{-0.035}^{+0.030}$&$0.040_{-0.050}^{+0.040}$&$0.025_{-0.020}^{+0.075}$&$0.020_{-0.035}^{+0.030}$&$0.020_{-0.025}^{+0.025}$&$0.020_{-0.025}^{+0.025}$&$0.020_{-0.025}^{+0.025}$\\[0.25cm]
& 0.03              & $0.03_{-0.035}^{+0.030}$&$0.050_{-0.050}^{+0.040}$&$0.035_{-0.020}^{+0.030}$&$0.030_{-0.025}^{+0.025}$&$0.030_{-0.025}^{+0.025}$&$0.030_{-0.025}^{+0.025}$&$0.030_{-0.020}^{+0.020}$ \\[0.25cm]
& 0.05              & $0.05_{-0.035}^{+0.030}$&$0.080_{-0.040}^{+0.020}$&$0.055_{-0.020}^{+0.025}$&$0.050_{-0.020}^{+0.025}$&$0.050_{-0.025}^{+0.025}$&$0.050_{-0.025}^{+0.025}$&$0.050_{-0.020}^{+0.020}$ \\[0.25cm]
& 0.1               & $0.12_{-0.030}^{+0.025}$&$0.10_{-0.02}^{+0.03}$&$0.10_{-0.03}^{+0.03}$&$0.10_{-0.03}^{+0.02}$&$0.10_{-0.02}^{+0.02}$&$0.10_{-0.02}^{+0.02}$&$0.10_{-0.02}^{+0.02}$\\[0.15cm]
& 0.3             & $0.30_{-0.02}^{+0.02}$&$0.30_{-0.02}^{+0.02}$&$0.30_{-0.02}^{+0.02}$&$0.30_{-0.02}^{+0.02}$&$0.30_{-0.02}^{+0.02}$&$0.30_{-0.02}^{+0.02}$&$0.30_{-0.02}^{+0.02}$\\[0.25cm]
& 0.5             & $0.50_{-0.02}^{+0.02}$&$0.50_{-0.02}^{+0.02}$&$0.50_{-0.02}^{+0.02}$&$0.50_{-0.02}^{+0.02}$&$0.50_{-0.02}^{+0.02}$&$0.50_{-0.02}^{+0.02}$&$0.50_{-0.02}^{+0.02}$\\[0.25cm]
\hline \hline
\end{tabular}
} \\[0.1cm]
\caption{\small 
Expected precision
 on $\delta_b$ for different true values $\bar\delta_b$ and $\bar\delta_{d_3}$  obtained at CLIC with $\sqrt{s}=3\,$TeV 
and $L = 1 \,\text{ab}^{-1}/(a^2 BR(b\bar b)/BR(b\bar b)_{SM})^2$ through $VV\to hh$ scattering.
}
\label{tab:errors1}
\end{center}
\end{table}
\begin{table}[tp]
\begin{center}
{\footnotesize
\begin{tabular}{cr|rrrrrrr}
\hline \hline
 &&&&&&&& \\[-0.3cm]
\multicolumn{2}{c|}{measured} & \multicolumn{7}{c}{$\bar\delta_{d_3}$} \\[0.05cm]
 \multicolumn{2}{c|}{$\delta_{d_3}$} & -0.5  & -0.3 & -0.1 & 0 & 0.1 &  0.3 &  0.5 \\
 \hline 
&&&&&&& \\[-0.2cm]
\multirow{8}{*}{$\bar\delta_{b}$} 
& 0                  & $-0.50_{-0.25}^{+0.35}$&$-0.25_{-0.50}^{+0.20}$&$0.00_{-0.40}^{+0.25}$&$0.05_{-0.30}^{+0.30}$&$0.10_{-0.20}^{+0.25}$&$0.30_{-0.15}^{+0.20}$&$0.50_{-0.15}^{+0.15}$\\[0.25cm]
& 0.01             & $-0.45_{-0.30}^{+0.35}$&$-0.20_{-0.55}^{+0.30}$&$-0.05_{-0.30}^{+0.30}$&$0.00_{-0.25}^{+0.25}$&$0.10_{-0.20}^{+0.20}$&$0.30_{-0.15}^{+0.15}$&$0.50_{-0.15}^{+0.15}$\\[0.25cm]
& 0.02              & $-0.35_{-0.35}^{+0.30}$&$-0.25_{-0.60}^{+0.25}$&$-0.10_{-0.30}^{+0.25}$&$0.00_{-0.25}^{+0.20}$&$0.10_{-0.20}^{+0.15}$&$0.30_{-0.15}^{+0.15}$&$0.50_{-0.15}^{+0.15}$\\[0.25cm]
& 0.03              & $-0.40_{-0.35}^{+0.30}$&$-0.25_{-0.70}^{+0.20}$&$-0.10_{-0.25}^{+0.20}$&$0.00_{-0.20}^{+0.15}$&$0.10_{-0.20}^{+0.15}$&$0.30_{-0.15}^{+0.15}$&$0.50_{-0.15}^{+0.15}$\\[0.25cm]
& 0.05              & $-0.55_{-0.40}^{+0.30}$&$-0.30_{-0.30}^{+0.20}$&$-0.10_{-0.20}^{+0.20}$&$0.00_{-0.20}^{+0.15}$&$0.10_{-0.15}^{+0.15}$&$0.30_{-0.15}^{+0.15}$&$0.50_{-0.10}^{+0.10}$  \\[0.25cm]
& 0.1               & $-0.50_{-0.25}^{+0.15}$&$-0.30_{-0.20}^{+0.15}$&$-0.10_{-0.20}^{+0.20}$&$0.00_{-0.15}^{+0.15}$&$0.10_{-0.15}^{+0.15}$&$0.30_{-0.10}^{+0.10}$&$0.50_{-0.10}^{+0.10}$  \\[0.25cm]
& 0.3             & $-0.50_{-0.15}^{+0.15}$&$-0.30_{-0.15}^{+0.15}$&$-0.10_{-0.10}^{+0.10}$&$0.00_{-0.10}^{+0.10}$&$0.10_{-0.10}^{+0.10}$&$0.30_{-0.10}^{+0.10}$&$0.50_{-0.10}^{+0.10}$\\[0.25cm]
& 0.5             &$-0.50_{-0.10}^{+0.15}$&$-0.30_{-0.10}^{+0.10}$&$-0.10_{-0.10}^{+0.10}$&$0.00_{-0.10}^{+0.10}$&$0.10_{-0.10}^{+0.10}$&$0.30_{-0.10}^{+0.10}$&$0.50_{-0.10}^{+0.10}$\\[0.25cm]
\hline \hline
\end{tabular}
} \\[0.1cm]
\caption{\small 
Expected precision
on $\delta_{d_3}$ for different true values $\bar\delta_b$ and $\bar\delta_{d_3}$  obtained at CLIC with $\sqrt{s}=3\,$TeV 
and $L = 1 \,\text{ab}^{-1}/(a^2 BR(b\bar b)/BR(b\bar b)_{SM})^2$ through $VV\to hh$ scattering.
}
\label{tab:errors2}
\end{center}
\end{table}
%
For injected (true) values $(\bar\delta_b, \bar\delta_{d_3})=(0,0)$ we find that the 68\% error on $\delta_{d_3}$  is equal to $\sim 0.3$ (see Table~\ref{tab:errors2}),
which means that a measurement of the Higgs trilinear coupling in the SM should be possible with a precision of $\sim 30\%$ with $L=1\,\text{ab}^{-1}$. 
This has to be compared with the 16\% and 20\% precisions reported respectively in Ref.~\cite{Abramowicz:2013tzc} and in the third paper of Ref.~\cite{CLICCDR}
for $2\,\text{ab}^{-1}$ of integrated luminosity and unpolarized beams.
For injected $(\bar\delta_b, \bar\delta_{d_3})=(0,0)$ we also find that the precision attainable on $\delta_{b}$ with $L=1\,\text{ab}^{-1}$ is $\sim 5\%$ (see Table~\ref{tab:errors1}),
which is compatible with the 3\% recently reported for $L=2\,\text{ab}^{-1}$ by Ref.~\cite{Abramowicz:2013tzc}.

The results of Tables~\ref{tab:errors1} and~\ref{tab:errors2} have been obtained by considering $a$, $b$ and $d_3$ as independent parameters.
Alternatively, by assuming them 
to be related as in eqs.~(\ref{eq:pngbparameters})~and~(\ref{eq:pngbparametersd3}) 
for the $SO(5)/SO(4)$ model  MCHM4 (where $BR(b\bar b) = BR(b\bar b)_{SM}$), one can optimize the analysis to extract $\xi$.
We do so by applying, besides the identification cuts of eqs.~(\ref{idjets}) and~(\ref{deltaminvh}),
a single cut on $H_T$ to isolate the energy growing behavior.
Since we need to fit a single parameter, we  select  events with $H_T >400\,$GeV. 
The corresponding efficiencies are reported in Table~\ref{tab:eff}. Larger values of $\xi$ give 
larger efficiencies for the identification cuts, as mainly due to the stronger boost of the Higgses, as previously discussed. 
%
\begin{table}[tp]
\begin{center}
{\small
\begin{tabular}{l|ccc}
\hline \hline
&&& \\[-0.35cm]
& All$\times [H_T>400$\,GeV]& No $\eta$$\times [H_T>400$\,GeV]& No $\Delta R$$\times [H_T>400$\,GeV]\\[0.05cm]
\hline
&&& \\[-0.3cm]
$\xi=0$ & 0.07=0.28$\times$0.24& 0.10=0.90$\times$0.11 & 0.08=0.30$\times$0.26    \\[0.05cm]
$\xi=0.1$ & 0.15=0.35$\times$0.44& 0.20=0.89$\times$0.23 &  0.18=0.39$\times$0.46  \\[0.05cm]
$\xi=0.5$ & 0.42=0.55$\times$0.77 & 0.50=0.81$\times$0.62&  0.54=0.65$\times$0.83  \\[0.05cm]
\hline \hline
\end{tabular}
}
\\[0.1cm]
\caption{\small Efficiencies of the  kinematic cuts imposed on the $e^+e^- \to \nu\bar\nu hh$ signal events to extract the parameter $\xi$
at CLIC with $\sqrt{s}=3\,$TeV. 
The format is $A=B\times C$, where $B$ is the efficiency for the identification cuts of eqs.~(\ref{idjets}) and (\ref{deltaminvh}), 
and $C$ is the efficiency of the cut $H_T>400$\,GeV on the reconstructed Higgses.}
\label{tab:eff}
\end{center}
\end{table}
%
The signal rate can be parametrized in this case as follows
\begin{equation}
\label{eq:fithhxi}
r(\xi) = r_{SM} \left( 1 + A_r \,\xi + B_r \,\xi^2 \right)\, . 
\end{equation}
The SM rate $r_{SM}$ and the coefficients $A_r$, $B_r$ are reported in Table~\ref{tab:fithhxi}.
%
\begin{table}[t]
\begin{center}
{\small
\begin{tabular}{r|ccc}
\hline \hline
&&& \\[-0.3cm]
&$r_{SM}$ [ab]& $A_r$ & $B_r$   \\[0.1cm]
\hline
&&& \\[-0.25cm]
All cuts& 15 & 11 &106 \\[0.1cm]
No $H_T$& 63 &4.1& 28.3 \\[0.1cm]
No $\eta$& 23&10.5&76.9 \\[0.1cm]
No $\Delta R$& 17&11&118\\[0.1cm]
 \hline
\hline
\end{tabular}
}
\\[0.1cm]
\caption{\small Fit of the $e^+ e^- \to hh(\to b\bar b b\bar b)\nu\bar\nu$ in the MCHM4 (see eq.~(\ref{eq:fithhxi}))  at CLIC with $\sqrt{s}=3\,$TeV.
The numbers in the second row have been obtained by applying the whole set of kinematic cuts described in the text (eqs.~(\ref{idjets}),(\ref{deltaminvh}) and 
the cut on $H_T$),  while each of the last three rows is obtained by removing one the cuts.
}
\label{tab:fithhxi}
\end{center}
\end{table}
%
In order to estimate the sensitivity on $\xi$ that can be reached at CLIC, for any given true value $\bar\xi$ 
we construct a posterior probability (see eqs.~(\ref{eq:posterior}) and (\ref{eq:poisson})) by assuming a prior on $\xi$ which is flat in the theoretically allowed range $[0,1]$
and vanishing outside.
The results are shown in Table~\ref{tab:xierrors} for $L=1\,$ab$^{-1}$.
%
\begin{table}[t]
\begin{center}
{\small
\begin{tabular}{r|cccccc}
\hline \hline
&&&&& \\[-0.3cm]
& \multicolumn{6}{c}{$\bar \xi$}\\[0.05cm]
&0  & 0.02 & 0.05 & 0.1 & 0.2 & 0.5 \\
\hline
&&&&& \\[-0.2cm]
All cuts&$0_{-0}^{+0.020}$  & $0.02_{-0.015}^{+0.015}$ & $0.05_{-0.015}^{+0.015}$ & $0.1_{-0.015}^{+0.015}$ & $0.2_{-0.015}^{+0.015} $ & $0.5_{-0.015}^{+0.010} $ \\[0.3cm]
No $H_T$& $0_{-0}^{+0.025}$ & $0.02_{-0.020}^{+0.015}$ &  $0.05_{-0.020}^{+0.020}$ & $0.1_{-0.015}^{+0.015}$ & $0.2_{-0.015}^{+0.015}$ & $0.5_{-0.015}^{+0.010} $ \\[0.3cm]
No $\eta$&$0_{-0}^{+0.015}$  & $0.02_{-0.015}^{+0.015} $ & $0.05_{-0.015}^{+0.015}$ & $0.1_{-0.015}^{+0.015}$ & $0.2_{-0.015}^{+0.010}$  & $0.5_{-0.010}^{+0.010}$\\[0.3cm]
No $\Delta R$&$0_{-0}^{+0.020}$  & $0.02_{-0.015}^{+0.015} $ & $0.05_{-0.015}^{+0.015}$ & $0.1_{-0.015}^{+0.015}$ & $0.2_{-0.010}^{+0.010}$  & $0.5_{-0.010}^{+0.010}$\\[0.3cm]
 \hline
\hline
\end{tabular}
}
\\[0.1cm]
\caption{\small Expected 68\% probability intervals on $\xi$ for different  true values $\bar\xi$  
obtained at CLIC with $\sqrt{s}=3\,$TeV and $L = 1 \,\text{ab}^{-1}$  through $VV\to hh$ scattering.
}
\label{tab:xierrors}
\end{center}
\end{table}

The results obtained in this section can be translated into an estimate of the sensitivity of CLIC on the scale of compositeness. 
In the presence of a shift in the Higgs couplings, $\delta_h  \sim (v/f)^2$, the low-energy theory 
becomes strongly coupled at the scale $\Lambda = 4\pi f \sim 4\pi  v/\sqrt{\delta_h}$ unless New Physics states set in at a scale $m_\rho < \Lambda$, 
expectedly freezing the growth of the coupling at $g_\rho \sim m_\rho / f < 4 \pi$.
From Tables~\ref{tab:errors1} and~\ref{tab:xierrors} we conclude that the study of double-Higgs  production at
CLIC with 3\,TeV can lead to a sensitivity on
$\Lambda$ of the order of~$\sim 15-20\, \text{TeV}$ with an accumulated luminosity of $1\, \text{ab}^{-1}$.
This has to be compared with sensitivities of the order of $\sim 10\,$TeV and~$\sim 30-40\,$TeV 
expected from the study of single-Higgs processes respectively at the
14\,TeV LHC with $300\,$fb$^{-1}$~\cite{CMS-NOTE-2012/006,ATLAS-collaboration:2012iza} and at the ILC with $250\,\text{fb}^{-1}$ of luminosity
accumulated at $\sqrt{s}=250\,$GeV plus another $500\,\text{fb}^{-1}$  at $\sqrt{s}=500\,$GeV~\cite{Baer:2013cma,Peskin:2012we}.
Table~\ref{tab:reachoncomp} summarizes
the reach on the compositeness scale at various experiments from the study of single and double Higgs processes.

\subsection{Double Higgs-strahlung}
\label{sec:Higgsstrahlung}

The cross section for double Higgs production through $WW$ fusion drops as the energy of the collider is lowered: in the SM it goes from 1\,fb for $\sqrt{s} =3\,$TeV 
down to 0.01~fb for $\sqrt{s} =500\,$GeV. At such low energies one has to resort to other processes in order to measure the anomalous Higgs couplings $\delta_b$ and 
$\delta_{d_3}$. One possibility is double Higgs-strahlung (DHS), $e^+e^-\to hhZ$~\cite{Barger:1988jk, Djouadi:1996ah,Djouadi:1999gv, Biswal:2008tg}. 
The relevant Feynman diagrams are shown in Fig.~\ref{fig:dhs},
and the analytic expression of the differential cross section is known (see Appendix~\ref{app2}). 
Figure~\ref{fig:dhs2} shows the value of the total cross section as a function of the $e^+e^-$ center-of-mass energy for some  values of  $\delta_b$ and $\delta_{d_3}$. 
For $\delta_b=0$ the cross section drops as $1/s$ at high energy, while it asymptotically approaches a constant value for $\delta_b\neq 0$. This  different high-energy 
behavior is due to the $e^+e^-\to hhZ_L$ amplitude and can be easily derived by using the equivalence theorem (see Appendix~\ref{app2}). 
Notice that for $m_h=125$\,GeV  the cross section is maximal between 500\,GeV and 1\,TeV.
%
\begin{figure}[t] 
\begin{center}
\includegraphics[width=0.65\linewidth]{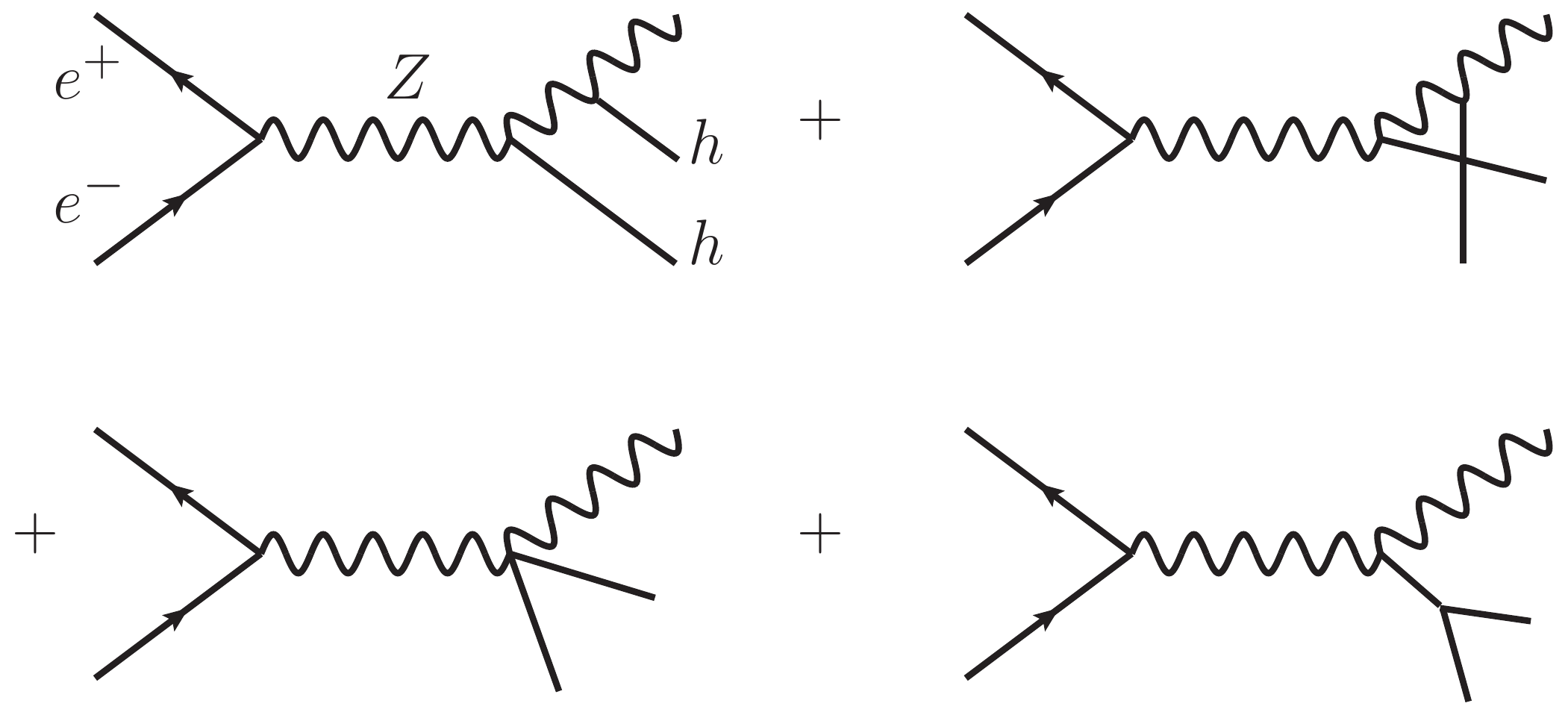}
\\[-0.1cm]
\caption{\small Diagrams contributing to double Higgs-strahlung at an $e^+e^-$ collider.
The two diagrams in the upper row are proportional to $a^2$, while the first and second diagrams in the lower row are proportional
respectively to $b$ and $a d_3$.}
\label{fig:dhs}
\end{center}
\end{figure}
%
%
\begin{figure}[t]
\begin{center}
\includegraphics[width=0.48\linewidth]{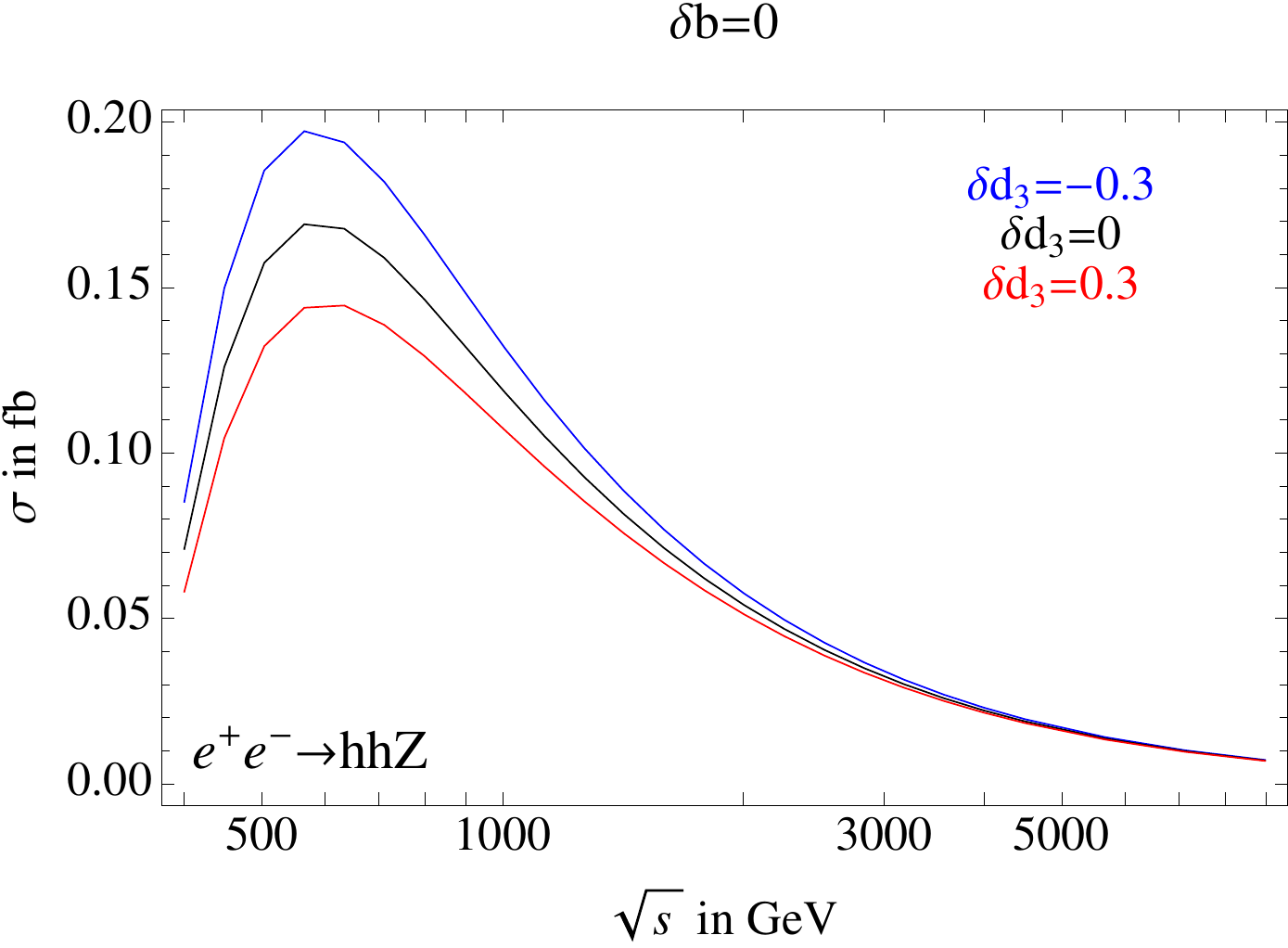}\hspace{0.6cm}\includegraphics[width=0.48\linewidth]{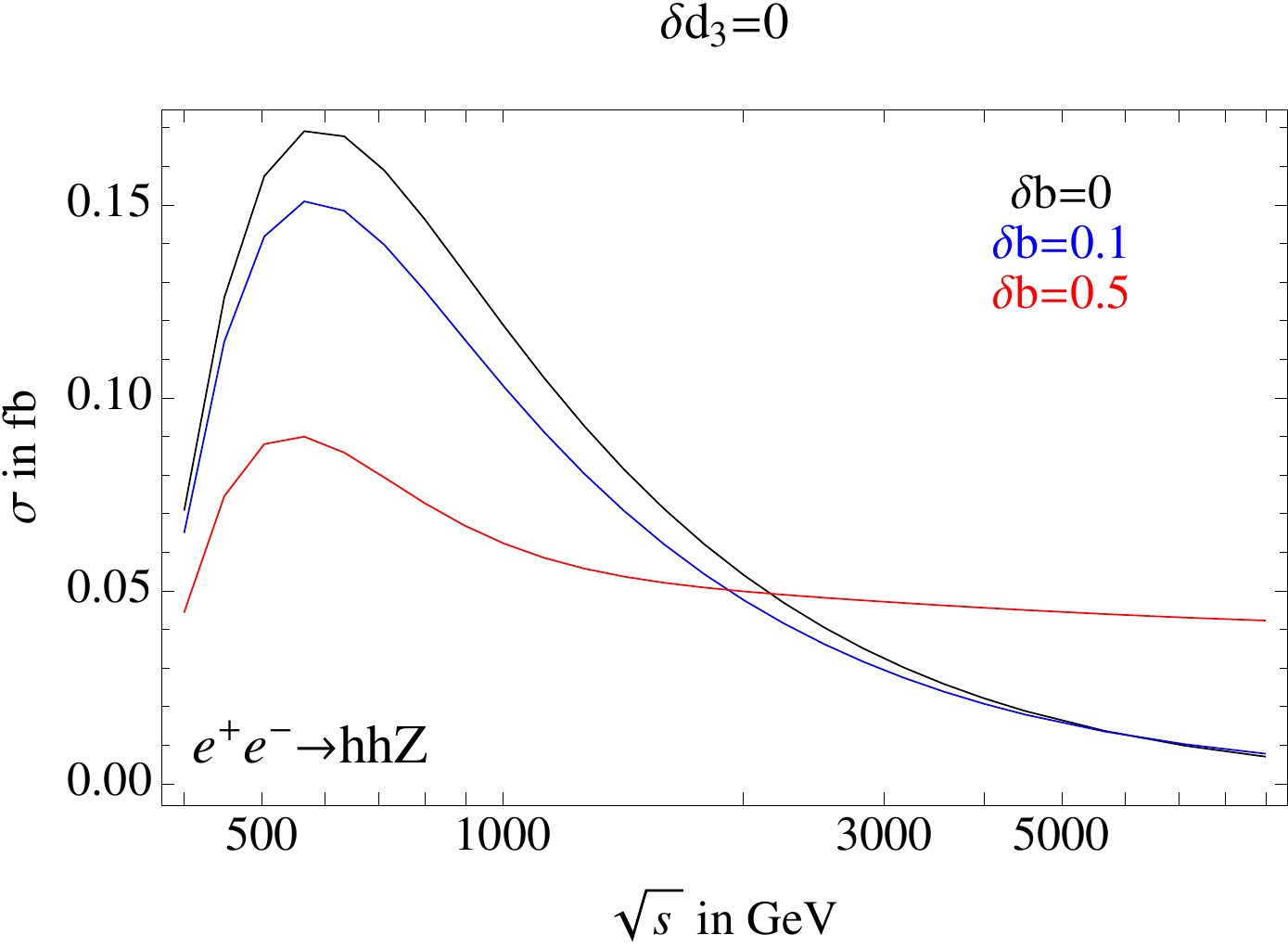}
\\[-0.1cm]
\caption{\small Total cross section of double Higgs-strahlung, $e^+e^-\to hhZ$,  as a function of the c.o.m energy for several values of the parameters $\delta_b$ 
and $\delta_{d_3}$.
}
\label{fig:dhs2}
\end{center}
\end{figure}
%

Before decaying the Higgs bosons, the DHS total cross section 
can  also be parametrized as in eq.~(\ref{sigmatot}). 
By using the analytic expressions given in Appendix~\ref{app2}, we find the coefficients reported in Table~\ref{tab:fitdhs}.
%
\begin{table}[tp]
\begin{center}
{\small
\begin{tabular}{c|ccccccc}
\hline
\hline
&&&&&& \\[-0.32cm]
$e^+e^-\to hhZ$& $\sigma_{SM}$ [fb]& $A$ & $B$ & $C$ & $D$ & $E$\\[0.1cm]
\hline
&&&&&& \\[-0.32cm]
500\,GeV& 0.16    & -1.02& -0.56& 0.31& 0.28& 0.10 \\[0.1cm]
1\,TeV&  0.12   &-1.42& -0.35& 0.48& 0.93& 0.91\\[0.1cm]
1\,TeV\,($m_{hh}<500$)&  0.03  &-2.45& -1.02& 1.42& 1.85 & 0.33\\[0.1cm]
1\,TeV\,($m_{hh}>500$)&  0.09  &-3.15& -0.36& 0.48& 1.83& 0.03\\[0.1cm]
\hline
\hline
\end{tabular}
}
\\[0.1cm]
\caption{\small 
Parametrization of the double Higgsstrahlung $e^+e^-\to hh Z$ cross section (see eq.~(\ref{sigmatot})) for various center-of-mass energies and cuts 
on the invariant mass of the two Higgses. The coefficients in the Table have been computed by using the analytic expressions given in Appendix~\ref{app2}. 
Decay branching fractions and reconstruction efficiencies are not included.
}
\label{tab:fitdhs}
\end{center}
\end{table}
%
%
\begin{figure}[t]
\begin{center}
\includegraphics[width=0.485\linewidth]{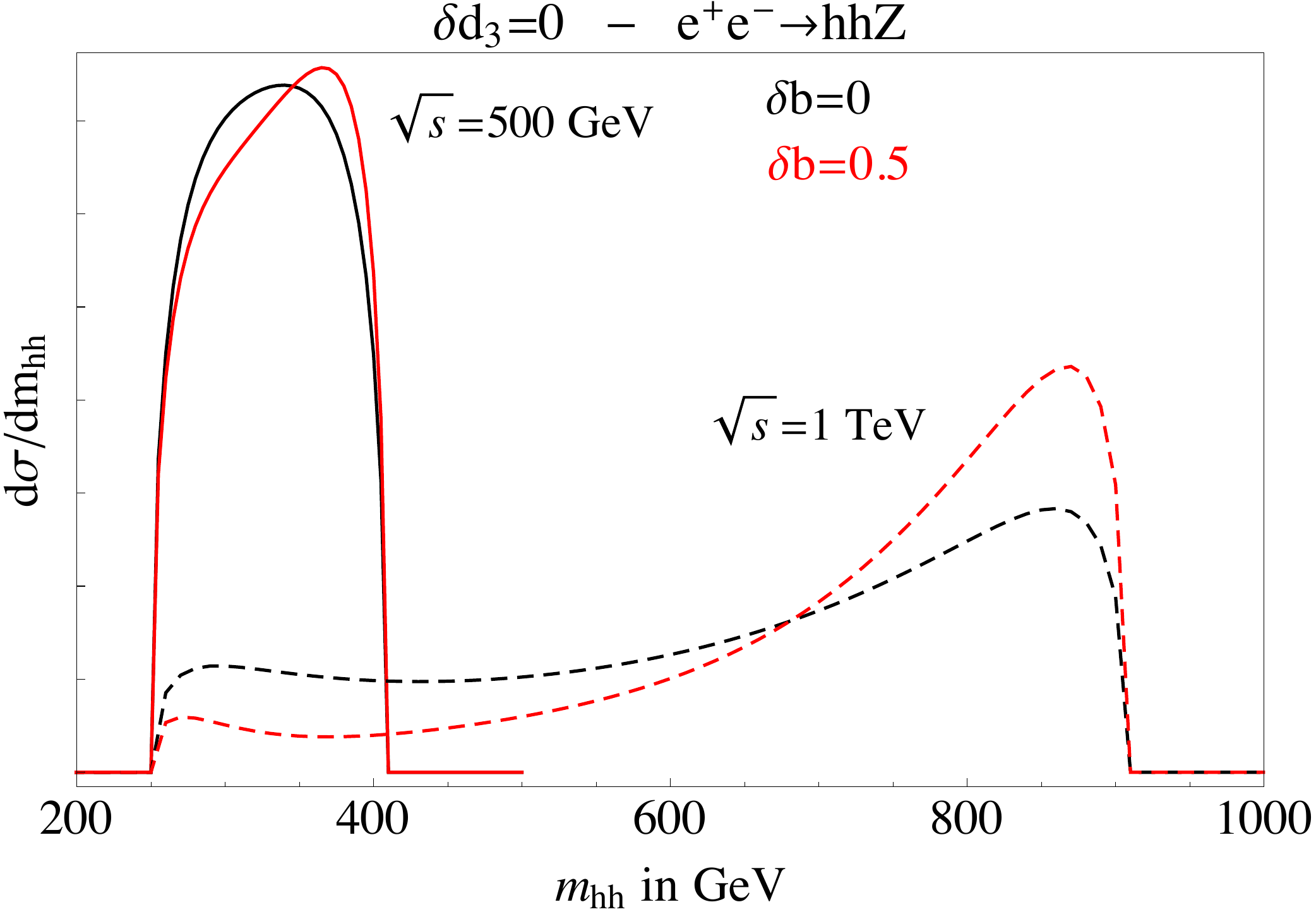}
\hspace{0.2cm}
\includegraphics[width=0.485\linewidth]{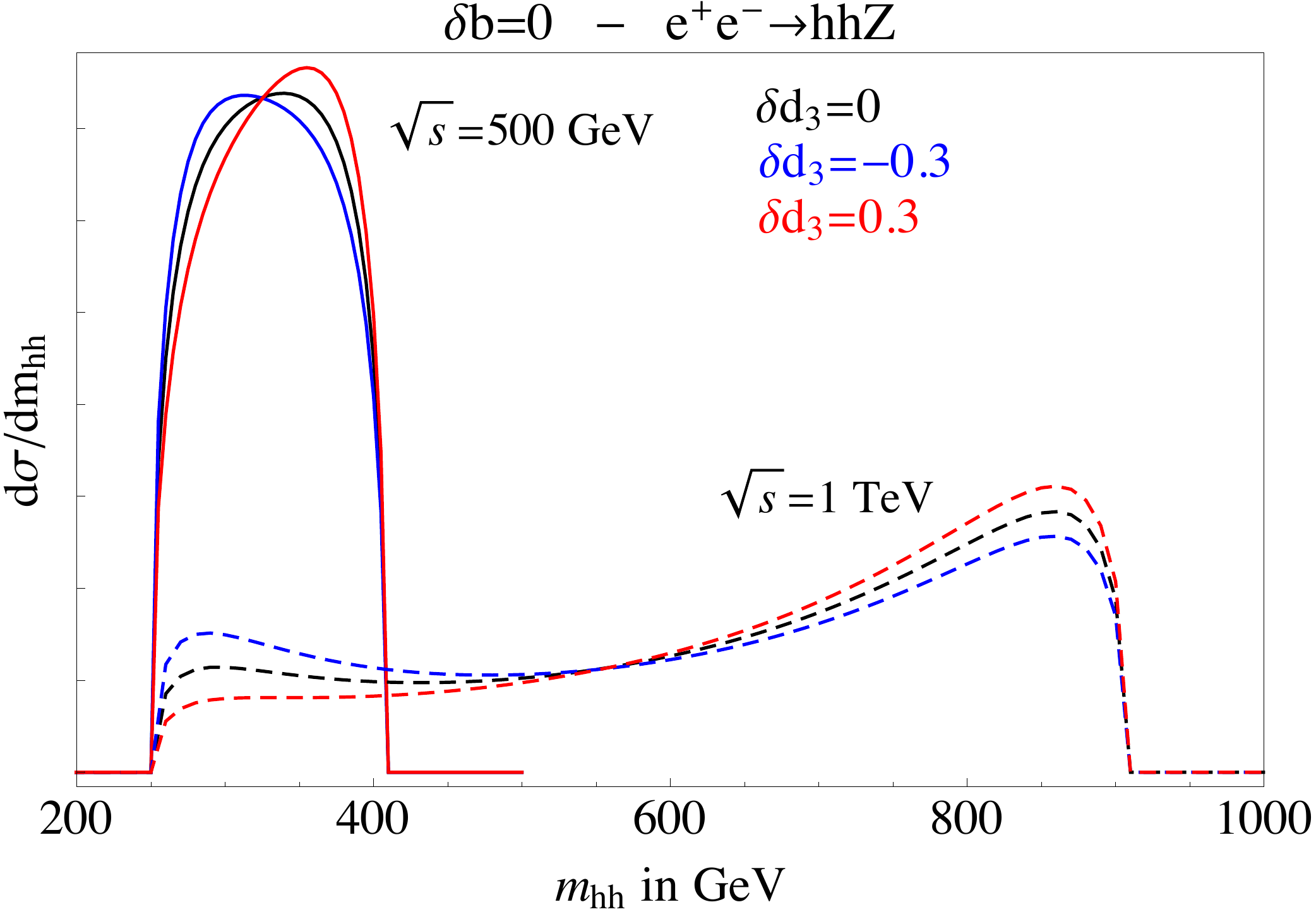}
\\[-0.1cm]
\caption{\small Differential cross section $d\sigma/dm_{hh}$ of double Higgsstrahlung 
at a linear collider with $\sqrt{s}=500\,$GeV (solid lines) and $\sqrt{s}=1\,$TeV (dashed lines),   for several values of 
$\delta_b$ and $\delta_{d_3}$.
All distributions have been normalized to unit area.}
\label{dhsmhhinv}
\end{center}
\end{figure}
In Fig.~\ref{dhsmhhinv} we compare the invariant mass distributions of the two Higgses 
at $\sqrt{s} = 500$ GeV and  $\sqrt{s} = 1$~TeV for various values of the parameters.~\footnote{\label{ftn:softsingularity} The enhancement of the cross section at 
$m_{hh} \sim \sqrt{s}$ is due to the infrared singularity associated with the soft emission of a transversely-polarized  $Z$ in the diagrams in the first row of Fig.~\ref{fig:dhs}.
The energy of the  $Z$ boson, $E_3$, is related to the invariant mass of the two Higgses by the formula $m_{hh}^2/s = 1- 2E_3/\sqrt{s} + m_Z^2/s$.
In the limit $E_3 \to 0$ it then follows $m_{hh}\to \sqrt{s} \, (1 + O(m_Z^2/s))$.
See Appendix~\ref{app2} for more details.}

Our strategy to extract the anomalous couplings in this Section differs in part from the one employed to analyze $WW$ scattering. 
In the case of DHS, at the energies we are interested in, the efficiency of the  identification and reconstruction cuts is practically insensitive
to the value of the Higgs couplings. The final rate can then be obtained by
starting from the analytical expression of the cross section in terms of the parameters $\delta_b$ and $\delta_{d_3}$ 
given in eq.~(\ref{sigmatot}) and Table~\ref{tab:fitdhs}, and  rescaling the value of $\sigma_{SM}$ by an overall efficiency factor to include the decay 
branching fractions and the effect of 
 kinematic cuts.  We extracted such efficiency factor by generating a single sample of events corresponding to the SM choice of parameters.
Such simplified approach  fully exploits the analytic expression of the DHS cross section and greatly reduces the complexity of the Monte Carlo simulation.

The analysis of DHS turns out to be more difficult than the one of double Higgs production via $WW$ fusion due to the presence of non-negligible background processes.
We focus on final states where both Higgses decay to $b\bar b$ and the $Z$ decays either hadronically or to a pair of charged leptons.
We thus select events with  6 or more  jets (and no lepton), or with 4 or more jets plus 2 opposite-charge  leptons (electrons or muons). 
Jets and leptons are reconstructed according to
the criteria defined in eqs.~(\ref{idjets}) and (\ref{idleptons}). Our selection ensures a full reconstruction of the  momentum of the $Z$ boson 
in signal events and consequently a substantial reduction of background contamination. As a final discrimination we  require at least 3 of the jets in the event to be $b$-tagged.

The reconstruction of the Higgs and $Z$ candidates proceeds as follows. 
In the case of  events with 4 or more jets and 2 leptons, the $Z$ is reconstructed from the lepton pair, while the two Higgs candidates are identified 
as done in the case of $WW\to hh$, see Section~\ref{sec:vvhh_scatt}.
We impose the following cut on the invariant mass of the lepton pair
\begin{equation}
|m_{ll}-m_Z|<10\,{\textrm{GeV}} \, , 
\end{equation}
while the invariant masses of the Higgs candidates are required to satisfy eq.~(\ref{deltaminvh}).
Events which do not satisfy these cuts are rejected. 
In the case of fully hadronic events, the Higgs and $Z$ candidates are reconstructed from the six most energetic  jets, $j_{1, \dots 6}$, by identifying 
the pairing $(j_1j_2,j_3j_4,j_5j_6)$ that minimizes the $\chi^2$ function
\begin{equation}\label{chi2dhs}
(m_{j_1j_2}-m_h)^2+(m_{j_3j_4}-m_h)^2+(m_{j_5j_6}-m_Z)^2 \, .
\end{equation}
We focus only on pairings where at least three among the four jets $j_{1\ldots 4}$ are  $b$-tagged, discarding the other pairings.
This implies the presence of at least 3 $b$-tags in the decay products of the two reconstructed Higgs bosons. 
After the  Higgs ($j_1j_2$ and $j_3j_4$) and $Z$ ($j_5j_6$) candidates have been reconstructed, we impose a cut
\begin{equation}
|m_{j_5 j_6}-m_Z|<10\,{\textrm{GeV}} 
\end{equation}
on the invariant mass of the $Z$ candidate, and the cut of eq.~(\ref{deltaminvh}) on the invariant mass of each of the two Higgs candidates. 
Events where these requirements are not fulfilled are rejected.
This algorithm has a fake rate (i.e. the rate at which it reconstructs fake Higgs or $Z$ candidates)  always below 5\% in the case of fully-hadronic events, and even
smaller in the case of events with two leptons.
The  identification efficiencies on the signal at $\sqrt{s} =500\,$GeV and 1\,TeV 
are given in Table~\ref{tab:effDHS} for each of the two final states under consideration. 
%
\begin{table}[tp]
\begin{center}
{\small
\begin{tabular}{r|ccc}
\hline \hline
&&& \\[-0.35cm]
&Energy & Efficiency &$r_{SM}\,$[ab] \\[0.05cm]
\hline
&&& \\[-0.3cm]
\multirow{3}{*}{$hhZ\to b\bar bb\bar bjj$} &500\,GeV & 0.56 & 21   \\[0.05cm]
 &1\,TeV-I & 0.50 & 3.5       \\[0.05cm]
 &1\,TeV-II & 0.50 & 10.2       \\[0.4cm]
\multirow{3}{*}{$hhZ\to b\bar bb\bar b\ell\ell$} &500\,GeV & 0.58 & 2.2  \\[0.05cm]
 &1\,TeV-I & 0.56 & 0.34      \\[0.05cm]
 &1\,TeV-II & 0.56 & 1.0     \\[0.05cm]
\hline \hline
\end{tabular}
}
\\[0.1cm]
\caption{\small
Efficiencies for the identification of the Higgs and $Z$ candidates in the DHS signal at 500\,GeV and 1\,TeV (for
both regions I and II of eq.~(\ref{eq:cut_dhs})) for the two final states discussed in the analysis. 
The variation of the efficiency with the parameters $\delta_b$ and $\delta_{d_3}$ is negligible. 
The last column reports the SM rate  after the identification cuts, $r_{SM}$, as defined in eq.~(\ref{fitratehh}).
}
\label{tab:effDHS}
\end{center}
\end{table}
%
They are to a large  extent constant upon variations of $\delta_b$ and $\delta_{d_3}$. The signal rate  $r(\delta_b,\delta_{d_3})$ can be thus parametrized as in eq.~(\ref{fitratehh})
with coefficients $A_r,B_r,C_r,D_r,E_r$ equal to the $A, B, C, D, E$ given in Table~\ref{tab:fitdhs}, and an overall factor $r_{SM}$ fully subsuming the reconstruction
efficiency and the decay branching fraction. 

At a center-of-mass energy of 500\,GeV, the signal rate is mostly dominated by  events at the kinematical threshold.
Disentangling the effect of $\delta_b$ from that of~$\delta_{d_3}$ by means of kinematic cuts does not seem possible 
(at least for reasonable  values of  integrated luminosity).  A measurement of the total cross section gives 
nevertheless the possibility to constrain a combination of the two relevant parameters $\delta_b$ and~$\delta_{d_3}$. 
At the higher center-of-mass energy $\sqrt{s} =1\,$TeV,
a better determination of both parameters is possible by  cutting on the invariant mass of the two Higgs bosons.
We define the two kinematical regions
\begin{equation}
\label{eq:cut_dhs}
\begin{split}
{\textrm{I}}:  \quad m_{hh} & > 500\,{\textrm{GeV}}\, ,\\[0.15cm]
{\textrm{II}}: \quad m_{hh} & < 500\,{\textrm{GeV}}\, .
\end{split}
\end{equation}
The signal rate in each  region and for each of the two final states (leptonic and fully hadronic)
is  obtained by multiplying the cross section by the decay branching fraction and an overall reconstruction efficiency.~\footnote{Notice that in order for 
this procedure to be accurate it is crucial that the fake-rate of our  algorithm for the reconstruction of the Higgs and $Z$ candidates is very small. 
If this was not the case, the measured invariant mass distribution of the Higgs pair could be affected by the  reconstruction and this would invalidate our  procedure.}
The value of the SM rate  $r_{SM}$ for each of the event categories at $\sqrt{s}=500\,$GeV and 1\,TeV is reported  in  Table~\ref{tab:effDHS}.

As  previously mentioned, a crucial difference between DHS and double Higgs production via vector boson fusion is the presence of  backgrounds processes that cannot 
be neglected. 
They  can be classified according to their scaling with the parameter $a$, which sets the strength of the $hVV$ coupling. The powers of $a$ thus control the number 
of external Higgs boson legs. Notice that up to effects of order $\Gamma/m$ the interference between amplitudes with a different number of on-shell Higgs boson legs 
is negligible. Under this assumption, each factor $a^2$ is accompanied by one power of the Higgs decay branching ratio to $b\bar b$. The total  rate
can thus be parametrized as follows:
\begin{equation} 
\label{dhstotalrate}
r_{tot}(\delta_b,\delta_{d_3}) = r_b^{(0)}+a^2 \, \frac{BR(b\bar b)}{BR(b\bar b)_{SM}} r_b^{(1)}+a^4 \, \left(\frac{BR(b\bar b)}{BR(b\bar b)_{SM}}\right)^2 r(\delta_b,\delta_{d_3}) \, ,
\end{equation}
where $r^{(0)}$ and $r^{(1)}$ are the rates for background processes respectively with 0 and 1 Higgs boson.
Simple inspection of eq.~(\ref{dhstotalrate}) shows that in this case
our  results  will depend in a non-trivial way on three quantities: $a^2 (BR(b\bar b)/BR(b\bar b)_{SM})$, $\delta_b$ and $\delta_{d_3}$.
The backgrounds included in our analysis are listed in Table~\ref{tab:bkgDHS}, together with their rate after  applying the same reconstruction algorithm 
adopted for the signal.
%
\begin{table}[tp]
\begin{center}
{\small
\begin{tabular}{r|clll}
\hline \hline
&&&&\\[-0.35cm]
&  & \multicolumn{3}{c}{Rate}  \\[0.05cm]
& scaling with $a$ & 500 GeV & 1 TeV-I & 1 TeV-II \\[0.05cm]
\hline
&&&&\\[-0.3cm]
$b\bar bb\bar bjj$          & $a^0$ & $r_b^{(0)}=20\,\text{ab}$  & $r_b^{(0)}=1.4\,\text{ab}$   &  $r_b^{(0)}=3.1\,\text{ab}$ \\[0.05cm]
$hb\bar bjj$                    & $a^2$ & $r_b^{(1)}=5.5\,\text{ab}$ & $r_b^{(1)}=0.4\,\text{ab}$   &  $r_b^{(1)}=0.3\,\text{ab}$ \\[0.35cm]
$b\bar bb\bar b\ell\ell$ & $a^0$ & $r_b^{(0)}=0.2\,\text{ab}$ & $r_b^{(0)}=0.05\,\text{ab}$ &  $r_b^{(0)}=0.01\,\text{ab}$  \\[0.05cm]
\hline \hline
\end{tabular}
}
\\[0.1cm]
\caption{\small
Rates after identification and reconstruction cuts for the  backgrounds included in our double Higgs-strahlung analysis at $\sqrt{s}=500\,$GeV and 1\,TeV (for
both regions I and II of eq.~(\ref{eq:cut_dhs})).
The leptonic background  $hbb\bar b\ell\ell$ is negligible.}
\label{tab:bkgDHS}
\end{center}
\end{table}
%

We derive the expected sensitivity on $\delta_b$ and $\delta_{d_3}$ by assuming a flat prior on these coupling shifts and
constructing a Poissonian likelihood function for each 
event category: two categories for a $\sqrt{s}=500\,$GeV collider (the leptonic and fully hadronic final states); four categories 
for a $\sqrt{s}=1\,$TeV collider (two kinematic regions for each of the two final states). In each case, the total
likelihood is obtained by taking the product of the individual ones. 
In general, we find that the fully hadronic final states lead to a better sensitivity on the couplings than the leptonic ones.
As a way to effectively take into account the systematic and theoretical uncertainties on the estimate of the background in our statistical analysis, 
we have rescaled all the background rates by a factor 1.5  compared to the MC predictions reported in Table~\ref{tab:bkgDHS}.

\begin{figure}[t]
\begin{center}
\includegraphics[width=0.4\linewidth]{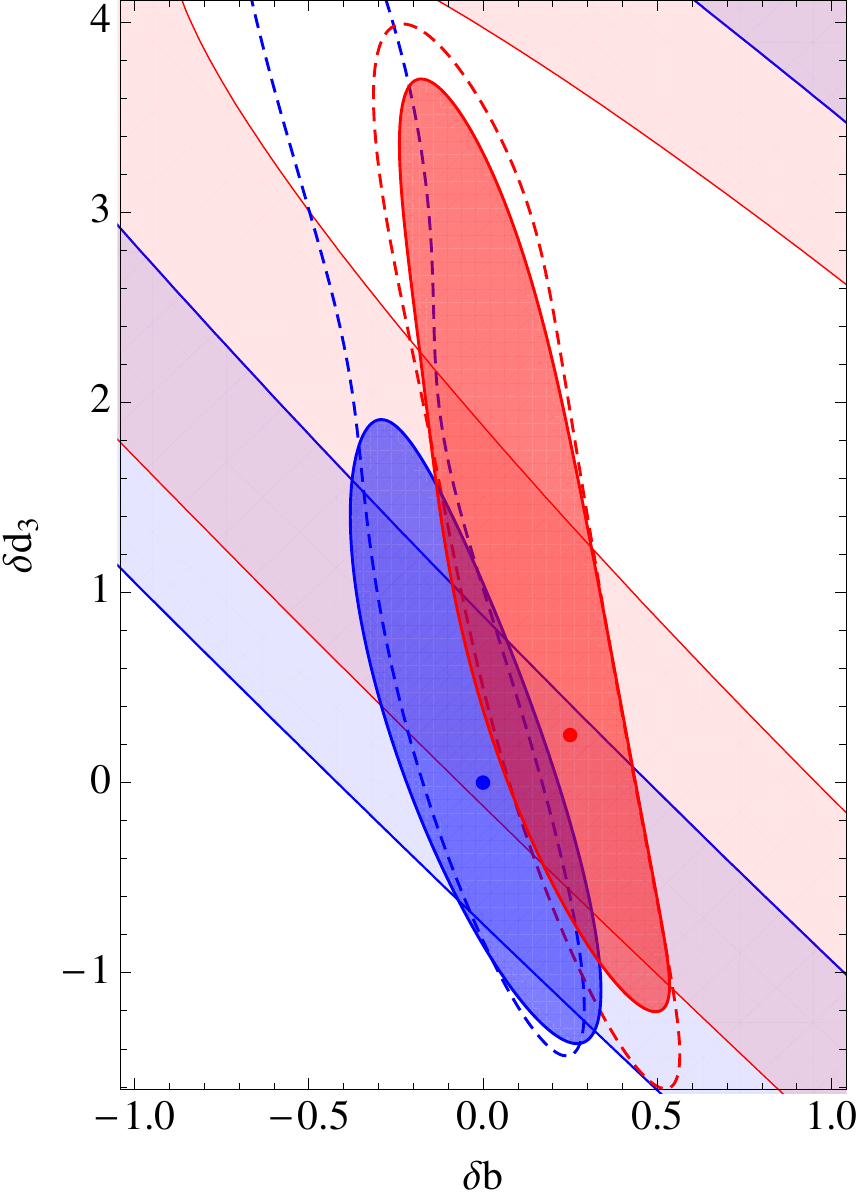}
\hspace{1.5cm}
\includegraphics[width=0.4\linewidth]{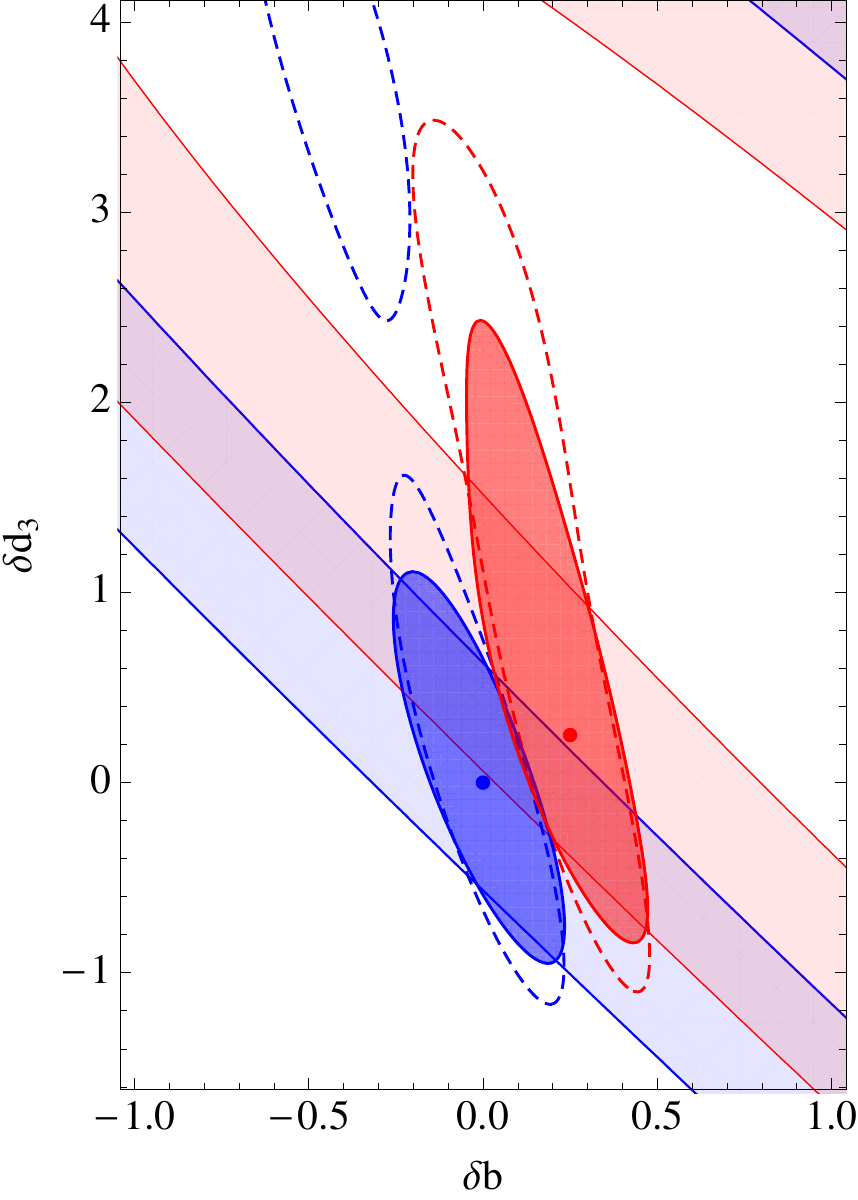}
\\[-0.1cm]
\caption{\small 
Regions of 68\% probability in the plane $(\delta_b,\delta_{d_3})$ obtained from the  analysis of double Higgs-strahlung at various collider energies.
Blue~(red) shapes and contours are relative to the case of injected  values $\bar\delta_b=0$, $\bar\delta_{d_3}=0$ ($\bar\delta_b=0.25$, $\bar\delta_{d_3}=0.25$).   
Lighter shaded bands: 500\,GeV; Dashed contours: $1$\,TeV; Darker shaded regions: 500\,GeV + 1\,TeV.
The plots have been obtained by assuming an integrated luminosity $L = 1\,$ab$^{-1}$ and setting 
$a^2 (BR(b\bar b)/BR(b\bar b)_{SM})=0.81$ (left plot) and $a^2 (BR(b\bar b)/BR(b\bar b)_{SM})=1$ (right plot).
}
\label{dhsbands}
\end{center}
\end{figure}

Figure~\ref{dhsbands}  shows the regions of 68\% probability obtained in the plane $(\delta_b, \delta_{d_3})$ with $L=1\,$ab$^{-1}$ by setting
$a^2 (BR(b\bar b)/BR(b\bar b)_{SM}) =0.81$ (left plot) and   $a^2 (BR(b\bar b)/BR(b\bar b)_{SM}) =1$  (right plot).
The various contours are relative to the following two benchmark points:  $(\delta_b, \delta_{d_3}) =(0,0)$ (in blue), and $(\delta_b, \delta_{d_3}) =(0.25,0.25)$
(in red). The latter point is obtained in the $SO(4)/SO(5)$ MCHM5 for $\xi=0.2$, see eqs.~(\ref{eq:pngbparameters}),(\ref{eq:pngbparametersd3}).
%
%
The bands in light (red and blue) color indicate the result obtainable by measuring just the total cross section at a 500\,GeV collider.
The red and blue dashed curves show instead the precision achievable at a linear collider with 1\,TeV c.o.m. energy by exploiting the cut on $m_{hh}$.
If  measurements at both 500\,GeV and 1\,TeV c.o.m. energies are possible 
(each with an integrated luminosity $L=1\,\text{ab}^{-1}$), an even more accurate precision 
on the couplings can be reached.
The corresponding 68\% regions are shown in darker (red and blue) color in Fig.~\ref{dhsbands}.

An estimate of the precision attainable on the Higgs trilinear coupling at the ILC has been recently derived in Ref.~\cite{Baer:2013cma} assuming $a=1$, $b=1$. 
It  is found that  for $\sqrt{s}=500\,$GeV  and with an integrated luminosity $L=500\,\text{fb}^{-1}$, $d_3$ can be measured with a precision of $104\%$ through DHS. 
At $\sqrt{s}=1\,$TeV with $L=1\,\text{ab}^{-1}$,  Ref.~\cite{Baer:2013cma} cites a precision of $28\%$ through $VV\to hh$ scattering, while DHS is found to be less powerful.
This suggests that a substantial improvement of our results can be obtained by including double Higgs production via vector boson fusion at $\sqrt{s}=1\,$TeV into the analysis.
Another recent study  appeared  in Ref.~\cite{Killick:2013mya} whose approach is more similar to the one presented in this paper.
The couplings $b$ and $d_3$ are extracted through the
measurement of the total DHS cross section at 500\,GeV and 1\,TeV
(with no cut on $m_{hh}$ applied), and on double Higgs production through $VV$ fusion at 1\,TeV. 
In the case of DHS, all the final states $Z\to \ell\ell,\nu\bar\nu, jj$ were included.
We find that, although we did not include $VV \to hh$ and the $\nu\bar\nu$ final state in DHS, 
our analysis gives a  better precision on the couplings $d_3$ and $b$ for those  benchmark points where a comparison can be performed.

\section{Discussion}
\label{sec:conclusions}

The observation of a resonance with a mass around 125~GeV and properties  remarkably compatible 
with  those of the Standard Model Higgs boson  makes the questions about the dynamics at the origin of  electroweak symmetry breaking more pressing.
The most relevant and urgent issue now facing us  concerns the structure of the newly discovered (Higgs) scalar. Are there additional states accompanying it?
Is it elementary or is it composite? Could this really be the first elementary scalar observed in Nature, or could it just be a bound state arising from some novel strong dynamics,  like a $\pi$ or $\eta$ in QCD? The answer to these questions
will have profound implications on our picture of fundamental physics. That is because of the hierarchy problem. Establishing, to the best of our experimental capability, that the Higgs boson is elementary, weakly coupled and solitary, would surely be shocking, but  it  may well start a revolution in the basic concepts of quantum mechanics and space-time.
If instead deviations from the SM will emerge in the dynamics of the Higgs, we will have to use them as a diagnostic tool of the underlying dynamics. A crucial  part of this program is the identification of the smoking guns of compositeness in  Higgs dynamics. Moreover, along this basic question there are more specific ones we can ask, related to the symmetry properties of the new state.
For instance, it will be essential to establish whether the new scalar is indeed ``a Higgs" fitting into an $SU(2)$ doublet and  not some exotic impostor, like for instance a pseudo-dilaton. Although there is really no strong theoretical motivation for such an alternative, and so far the data disfavor it,  it remains a logical possibility that can be tested and possibly ruled out. A perhaps more interesting question is whether the Higgs particle is just an ordinary composite, like a $\sigma$, or whether it is a pseudo-Nambu--Goldstone boson, like the~$\pi$. The answer to this question will give us important clues on the UV completion of the electroweak breaking dynamics.

It is well known that in a fully natural theory of electroweak symmetry breaking the Higgs couplings must deviate, even in a significant way, from the predictions of the Standard Model. Note, however, that past and current experiments already put stringent constraints on these deviations, specifically on the single-Higgs $hVV$  coupling $a$, while the quadratic $h^2VV$  coupling $b$ and the Higgs cubic self-coupling $d_{3}$ so far remain unconstrained.  Thanks to finally precise knowledge of the Higgs mass and to a more accurate
measurement of the $W$ mass, the Higgs coupling $a$ is now constrained
to lie with $95\%$ probability in the  interval $0.98 \le a^{2} \le 1.12$ under the assumption of no further contribution from New Physics. The tension with the theoretically motivated range $a < 1$ can be slightly lifted by including an extra positive contribution to $\Delta \hat{T}$ in the electroweak fit, as it can arise in explicit models. Assuming $\Delta \hat{T} = +1.5 \times 10^{-3}$ leads to an interval $0.70 \le a^{2} \le 0.92$. 
Important constraints are also set by direct searches for spin-1 resonances at the LHC, which start to exclude interesting portions of the parameter space. Direct Higgs coupling measurements at the LHC, on the other hand, still have a limited precision but they are expected to reach a $\sim 5\%$ resolution on $a$ at the $14$~TeV LHC.

In this paper we  laid down a strategy to infer information about the scale and the nature of the dynamics behind EWSB  through a precise measurement of the Higgs couplings. 
Observing a shift in the Higgs couplings of order $\delta_h$ in single-Higgs processes,
together with the absence of any other new degree of freedom below a scale $M$, puts a qualitative lower bound on the strength of the interaction within the New Physics sector, $g_\rho > \sqrt{\delta_h} \times M/v$. This could provide a first indirect evidence for strong dynamics at the origin of EWSB and a hint towards a composite nature of the Higgs boson. For instance, $O(10 \%)$ deviations without any new states below $2\,$TeV, would already correspond to a coupling exceeding  all the SM interactions in strength. Qualitatively, an analogous lower bound on $g_\rho$ can be obtained from the observation of an enhanced amplitude for the scattering of massive gauge bosons. Indeed, as we point out in Section~\ref{sec:SingeDoubleHiggsProd}, at an electron-positron collider, the precision of the measurements could in principle allow us to estimate the size of subleading  terms
in the growth of the amplitude, thus providing a stronger indirect bound on the strength of the coupling.

Multi-Higgs production can bring additional valuable information to characterize the strong sector, even if it does not provide stronger bounds on its coupling 
strength or its scale. Double Higgs production by vector boson fusion gives access to the linear and quadratic couplings of the Higgs to the electroweak gauge bosons. 
We established a {\it universal} relation among these couplings, valid at order $v^2/f^2$ 
(where the scale $f$ is defined by $f = m_\rho/g_\rho$ for a generic composite state and corresponds to the decay constant for a PNGB),  
that follows when the Higgs boson is part of an electroweak doublet. 
This is because a single operator of dimension-6  controls the leading corrections to both  scattering amplitudes and single Higgs couplings. 
Furthermore,  we studied the corrections to this relation that arise at 
order $v^4/f^4$ from dimension-8 operators and we demonstrated that they can distinguish scenarios with a PNGB Higgs from those where the discovered boson is
a generic light scalar resonance of the strong dynamics.  
The reason for this non-trivial result, is that, in the case of a PNGB Higgs, the  non-linearly realized symmetry relates operators of different dimension.

We also emphasized the importance of a precise and energetic lepton collider such as CLIC to study the rare process of triple Higgs production through vector boson fusion, 
$VV\to hhh$. For  a generic composite Higgs, the leading expected growing behaviour of the cross section below the scale of the resonances is  $\sim v^2s^2/f^8$ and could 
in principle be observed provided $v^2/f^2\sim 0.1$. However we pointed out that for a PNGB Higgs based on cosets involving only doublets, in particular in the simplest cases  
of $SO(5)/SO(4)$ or $SO(4,1)/SO(4)$, this leading term exactly cancels. This cancellation is a simple consequence of the homogeneity and of the grading symmetry of such cosets,
 but in the effective lagrangian it corresponds to a more obscure correlation among the coefficients of operators of dimension~6 and dimension~8. This is the same correlation we 
mentioned before. The observation, or lack thereof, of a visible rate for $VV\to hhh$ could then play a relevant role in the reconstruction of the underlying theory.

We presented a quantitative analysis of vector boson scattering and double Higgs production, both through vector boson fusion and double Higgs-strahlung, 
at the ILC and CLIC for two different center-of-mass energies. Focusing on $VV\to VV$ scattering processes and using a simple cut-and-count analysis, 
we found that a $\sqrt{s} = 500\,$GeV linear collider with an integrated luminosity of $1\,$ab$^{-1}$ 
is only sensitive to large deviations of the coupling $a$ from its SM value, $\Delta a^2 \sim 0.5$
(see footnote~\ref{ftn:sensitivity} for the definition of sensitivity used in this paper).
A $3\,$TeV linear collider with  the same 
luminosity, on the other hand, is sensitive to shifts in $a^2$ larger than $\sim 0.2$.
Double Higgs production depends both on $a$ and on the couplings $b$ and $d_{3}$. Its cross section can be conveniently expressed in terms of the two shifts 
$\delta_{b}\equiv 1-b/a^2$ and $\delta_{d_{3}}\equiv 1-d_3/a$, while the parameter $a$ enters as a simple overall rescaling which can be absorbed in the value of the luminosity 
(note that, at the time of the studies we are proposing, $a$ will be known with good accuracy thanks to single Higgs processes, hence $\delta_b$ and $\delta_{d_3}$ will really measure 
the deviations in the $b$ and $d_3$ couplings).
As it emerges clearly throughout our study, $\delta_b$ offers a more sensitive probe into the Higgs structure than the trilinear $\delta_{d_{3}}$. In the case of a $3\,$TeV 
CLIC machine with $L=1\,\text{ab}^{-1}$, the study of $VV\to hh$ offers a sensitivity of about $0.05$ on $\delta_b$ while that on $\delta_{d_{3}}$ is hardly better than $0.3$. 
In a specific model like the Minimal Composite Higgs model the couplings $a$, $b$ and $d_3$ depend on the single parameter $\xi = (v/f)^2$. 
Through the study of $e^+e^-\to hh\nu\bar\nu$, a machine like CLIC with $\sqrt{s}=3\,$TeV and $L=1\,\text{ab}^{-1}$ can reach a sensitivity as small as $0.02$ on $\xi$. 
These sensitivities can be translated into an indirect reach  on the cutoff scale $\Lambda \equiv 4\pi f$, that is the mass scale of the resonances for  the case where the 
underlying dynamics is maximally strong. We find a reach $\Lambda \sim 15-20\,$TeV,  which should be compared with the reach $\Lambda \sim 30-40\,$TeV 
expected through single-Higgs processes at the ILC with $250\,\text{fb}^{-1}$ of luminosity
accumulated at $\sqrt{s}=250\,$GeV plus another $500\,\text{fb}^{-1}$  at $\sqrt{s}=500\,$GeV~\cite{Baer:2013cma,Peskin:2012we}.
Table~\ref{tab:reachoncomp} summarizes the values of $\Lambda$ which can be probed  at various experiments through the study of single and double Higgs processes.
%
\begin{table}[tp]
\begin{center}
{\small
\begin{tabular}{lll} 
 & $\xi = (v/f)^2$ & $\Lambda = 4\pi f$  \\[0.2cm]
\hline
\\
LHC \ $14\,$TeV \ $L=300\,\text{fb}^{-1}$ & 0.5  (double Higgs~\cite{Giudice:2007fh,Contino:2010mh}) & 4.5 TeV \\[0.15cm]
& 0.1  (single Higgs~\cite{CMS-NOTE-2012/006,ATLAS-collaboration:2012iza}) & 10 TeV \\[0.7cm]
ILC \ \phantom{+} $250\,$GeV \ $L = 250\,\text{fb}^{-1}$  
                                                                     & \multirow{2}{*}{0.6-1.2$\, \times 10^{-2}$ (single Higgs~\cite{Baer:2013cma,Peskin:2012we})} & \multirow{2}{*}{30-40\,TeV} \\[0.05cm]
\phantom{ILC} \ + $500\,$GeV \ $L = 500\,\text{fb}^{-1}$ & & \\[0.7cm]
CLIC \ $3\,$TeV \ $L=1\,\text{ab}^{-1}$ & 2-5$\, \times 10^{-2}$ (double Higgs~[this work]) & 15-20 TeV \\[0.7cm]
CLIC \ \phantom{+} $350\,$GeV \ $L = 500\,\text{fb}^{-1}$  
                                                                     & \multirow{3}{*}{1.1-2.4$\, \times 10^{-3}$ (single Higgs~\cite{Abramowicz:2013tzc})} & \multirow{3}{*}{60-90\,TeV} \\[0.05cm]
\phantom{CLIC} \ + $1.4\,$TeV \hspace{0.1cm} $L = 1.5\,\text{ab}^{-1}$ & & \\[0.05cm]
\phantom{CLIC} \ + $3.0\,$TeV \hspace{0.1cm} $L = 2\,\text{ab}^{-1}$ & & \\[0.5cm]
 \hline
\end{tabular}
}
\\[0.1cm]
\caption{\small 
Summary of the precision on $\xi$ (as defined in footnote~\ref{ftn:sensitivity})
and the corresponding reach on the compositeness scale at various experiments  from the study of single and double Higgs processes.
}
\label{tab:reachoncomp}
\end{center}
\end{table}
%
Though the reach on $\Lambda$  seems remarkable, one should not forget that the measured value of the Higgs mass disfavors a maximally strong coupling ~\cite{Pappadopulo:2013vca}:  new states are therefore expected significantly below $\Lambda$ with a mass around $m_\rho \sim \Lambda \times g_\rho/(4\pi)$. Still, even in the case of a moderately strong sector $g_\rho\sim 3$, direct production  of resonances at a high-energy hadron collider like the LHC with $\sqrt{s} =33\,$TeV may not  become competitive. Of course one must beware of these qualitative arguments, as the model's details often matter. 

At lower energies the $e^+e^-\to hh\nu\bar\nu$ process is not effective to measure the couplings $b$, $d_3$, and one has to resort to double Higgs-strahlung, 
$e^+e^-\to hhZ$.  At $500\,$GeV center-of-mass energy, only a linear combination of the two couplings $\delta_b$ and $\delta_{d_3}$ can be extracted from a measurement
of  the total cross section. For $\sqrt{s} = 1\,$TeV, on the other hand, it is possible to exploit the  
kinematical distribution of the final state to extract both couplings independently, even though with large uncertainties. 
The combined measurement of double Higgs-strahlung at both 500\,GeV and 1\,TeV allows us to obtain the sensitivity contours shown in 
Figure~\ref{dhsbands}, which again indicate that $\delta_b$ can be measured more precisely than $\delta_{d_3}$.
For all those interested in the structure of the Higgs the message is then very clear. The parameter $b$
not only encodes more robust information than $d_3$ about the nature of $h$, whether an impostor, a composite or a PNGB, but it also affords better sensitivity. 

In the absence of direct production of new particles at the LHC, precision measurements in the sector of the newly discovered Higgs boson
can play a key role in the search for New Physics. The time has come to establish a clear strategy to extract the
information on the origin of electroweak symmetry breaking  encoded in the Higgs measurements and to pave the way for a future experimental program.

\section*{Acknowledgments}

We would like to thank 
Daniele Del Re,
Enrico Franco,
Gian Giudice,
Lucie Linssen,
Markus Luty,
Barbara Mele,
Paolo Meridiani,
Satoshi Mishima,
Alexandra Oliveira,
Rogerio Rosenfeld,
Luca Silvestrini,
Alessandro Strumia,
Riccardo Torre,
James Wells and
Andrea Wulzer
for useful discussions and comments.
We are grateful to Philipp Roloff and Marc Thomson for discussions and clarifications about the CLIC detector simulation.
We thank Ian Low for pointing out a typo in the first version of the paper. 
The work of C.G. and A.T. has been partly supported by the European Commission under the contract  ERC advanced grant 226371 `MassTeV' and the 
contract PITN-GA-2009-237920 `UNILHC'. The work of D.P. is supported by the NSF Grant PHY-0855653.
The work of R.R. and A.T. is supported by the Swiss National Science Foundation
under contracts No. 200020-138131 and No. 200022-126941. 
R.C. was partly supported by the ERC Advanced Grant No.~267985 
{\em Electroweak Symmetry Breaking, Flavour and Dark Matter: One Solution for Three Mysteries (DaMeSyFla)}.

\appendix
\section{Dimension-8 operators  for strong scatterings}
\label{app:dim8}

At the dimension-8 level, the following three operators can be constructed with two derivatives and six Higgs fields  
\begin{equation}
\label{eq:operators8}
\begin{aligned}
c'_r \mathcal O'_r  &= \frac{c'_r}{f^2}\, |H|^4|D_\mu H|^2                                         &&= |H|^2 c'_r\mathcal O_r \, , \\[0.1cm]
c'_H \mathcal O'_H &= \frac{c'_H}{2 f^2}\, |H|^2\partial_\mu|H|^2\partial^\mu|H|^2 &&= |H|^2 c'_H\mathcal O_H \, ,\\[0.1cm]
c'_T \mathcal O'_T &= \frac{c'_T}{2f^2}\, |H|^2\left(H^\dagger \overleftrightarrow D_\mu H\right)\left(H^\dagger \overleftrightarrow  D^\mu H\right) \hspace{-0.35cm}
                                                                                                                             &&= |H|^2 c'_T\mathcal O_T  \, ,
\end{aligned}
\end{equation}
which can be found by constructing all possible $SU(2)_{L}$-invariant structures and 
using integration by parts and the identities
\begin{equation}
\begin{split}
\sigma^A_{ij}\sigma^A_{hk}&=2 \delta_{ik}\delta_{jh}-\delta_{ij}\delta_{hk} \, ,\\[0.2cm]
\sigma^2_{ij}\sigma^2_{hk}&=-\delta_{ih}\delta_{jk}+\delta_{jh}\delta_{ik} \, .
\end{split}
\end{equation}
The operators of eq.~(\ref{eq:operators8}) consist of the dimension-6 structures discussed in Ref.~\cite{Giudice:2007fh} extended by two Higgs fields. Note that 
$\mathcal O'_T$ violates custodial symmetry, in analogy to $\mathcal O_T$. 
At the dimension-8 level there is also the operator $O_8 = -c_{8}(H^{\dagger} H)^{4}/f^{2}$ defined in eq.~(\ref{eq:OHpO8}), which involves no derivative.
The operator $\mathcal O_r$ can be redefined away by the following field redefinition
\begin{equation}
H\rightarrow H+ \alpha H\frac{|H|^2}{f^2}+ \beta H\frac{|H|^4}{f^4}\, ,
\end{equation}
under which the kinetic term transforms as follows
\begin{equation}
|D_\mu H|^2\rightarrow |D_\mu H|^2 + \frac{2 \alpha}{f^2} \mathcal O_r+\frac{\alpha}{f^2}\mathcal O_H + \frac{\alpha^2+2\beta}{f^4}(\mathcal O'_r+ 2 \mathcal O'_H) \, ,
\end{equation}
while the dimension-6 operators give
\begin{equation}
\mathcal O_T\rightarrow\mathcal O_T+\frac{5\alpha}{f^2}O'_T+\ldots ,
\quad O_H\rightarrow \mathcal O_H+\frac{8\alpha}{f^2}\mathcal O'_H+\ldots,
\quad  \mathcal O_r\rightarrow O_r + \frac{8 \alpha}{f^2} \mathcal O'_r+\frac{2 \alpha}{f^2}\mathcal O'_H+\ldots
\end{equation}
Hence by choosing $\alpha$ and $\beta$ appropriately both $\mathcal O_r$ and $\mathcal O'_r$ can be redefined away. We thus remain with only one custodially-invariant 
operator: $\mathcal O'_H$.

\section{Double Higgs-strahlung}\label{app2}

The differential cross section for double Higgsstrahlung can be expressed in term of the Dalitz variables $x_i\equiv2 E_i/\sqrt{s}$ where $E_{1,2}$ are the energies of the two Higgses and $x_3\equiv 2 E_3/\sqrt{s}$ for the $Z$ boson~\cite{Djouadi:1996ah,Djouadi:1999gv}:
\begin{equation}
\label{eq:DHSdiffxsec}
\frac{d\sigma}{d x_1 dx_2}=\frac{G_F^3 m_Z^6}{384 \sqrt 2 \pi^3 s}\frac{1+(1-4 s_W^2)^2}{(1-\mu_Z)^2} \mathcal A\, .
\end{equation}
Here $\sqrt{s}$ is the collider center-of-mass energy. We define $\mu_i\equiv m_i^2/s$, $\mu_{ij}\equiv\mu_i-\mu_j$, $y_i\equiv1-x_i$, so that $x_3=2-x_1-x_2$ follows by energy conservation. 
We have
\begin{equation}
\mathcal A=\mathcal A_0^2 f_0+ \frac{a^2}{4\mu_Z(y_1+\mu_{hZ})}\left(\frac{a^2 f_1}{y_1+\mu_{hZ}}+\frac{a^2 f_2}{y_2+\mu_{hZ}}+2\mu_Z \mathcal A_0 f_3\right)+(y_1\leftrightarrow y_2)
\end{equation}
with 
\begin{equation}
\mathcal A_0=3\frac{m_h^2}{m_Z^2}\frac{a d_3}{y_3-\mu_{hZ}}+\frac{2 a^2}{y_1+\mu_{hZ}}+\frac{2 a^2}{y_2+\mu_{hZ}}+\frac{b}{\mu_Z}
\end{equation}
and
\begin{equation}
\begin{split}
f_0 = & \, \frac{1}{8}\mu_Z((y_1+y_2)^2+8\mu_Z),\\[0.15cm]
f_1=& \, (y_1-1)^2(\mu_Z-y_1)^2-4\mu_h y_1(y_1+y_1\mu_Z-4\mu_Z) + \mu_Z(\mu_Z-4\mu_h)(1-4\mu_h)-\mu_Z^2,\\[0.15cm]
f_2=& \, (\mu_Z(y_3+\mu_Z-8\mu_h)-(1+\mu_Z)y_1y_2)(1+y_3+2\mu_Z)\\[0.15cm]
       & + y_1y_2(y_1y_2+1+\mu_Z^2+4\mu_h(1+\mu_Z))+4\mu_h\mu_Z(1+\mu_Z+4\mu_h)+\mu_Z^2,\\[0.15cm]
f_3=&\, y_1(y_1-1)(\mu_Z-y_1)-y_2(y_1+1)(y_1+\mu_Z)+2\mu_Z(1+\mu_Z-4\mu_h)\, .
\end{split}
\end{equation}
The kinematical boundaries of the phase space integration are defined by
\begin{equation}
\left| 2(1-x_1-x_2+2\mu_h-\mu_Z)+x_1x_2\right|\leq\sqrt{x_1^2-4\mu_h}\sqrt{x_2^2-4\mu_h}\, .
\end{equation}
Using 
\begin{equation}
x_1=\frac{s+m_h^2-m_{23}^2}{s},\qquad x_2=\frac{m_{12}^2+m_{23}^2-m_h^2-m_Z^2}{s}
\end{equation}
and
\begin{equation}
\frac{d\sigma}{d x_1 dx_2}=s^2 \frac{d\sigma}{d m_{12}^2 d m_{23}^2}\, , 
\end{equation}
one can obtain the differential cross section as a function of $m_{12}\equiv m_{hh}$ and $m_{23}\equiv m_{hZ}$.
Figure~\ref{dhsmhhinv} in particular is derived by integrating over $m_{23}$ and varying $m_{hh}$.

It is interesting to analyze the enhancement of the differential cross section  
$d\sigma/dm_{hh}$ near the kinematic boundary $m_{hh} \simeq \sqrt{s}$. 
As mentioned in footnote~\ref{ftn:softsingularity}, the enhancement is due to the singularity associated 
with the soft emission of a transversely-polarized $Z$ boson in the diagrams in the first row of Fig.~\ref{fig:dhs}. 
The leading singular behavior can be thus isolated by 
setting $d_3=0$,  to switch off the Higgs trilinear coupling, and by fixing the couplings $a$ and $b$ to their SM value ($\delta_b =0$).
It is useful to make a change of 
variables in eq.~(\ref{eq:DHSdiffxsec}) from $(x_1, x_2)$ to 
$(r\equiv m_{hh}^2/s, x_2)$, where  the energy of the $Z$ is related to the invariant mass of the two Higgses by 
\begin{equation}
r =1-x_3+\frac{m_Z^2}{s}\, .
\end{equation}
By integrating $x_2$ over the interval $r + \epsilon\leq x_2 \leq 1-\epsilon$ and expanding for $\epsilon = m_Z^2/s$ small, we obtain 
\begin{equation}
\frac{d\sigma}{d r}\simeq \frac{G_F^3 m_Z^6[1+(1-4 s_W^2)^2]}{192 \sqrt 2 \pi^3 s} \left[(1-r)+\frac{r}{1-r}\log \left(1-r\right)-\frac{r}{1-r}\log\epsilon\right]\, ,
\end{equation}
which shows the singularity at $r = 1$.  The logarithmic terms in the above formula follow from the collinear singularity 
also associated with the $Z$ emission.
Notice that events  with a final longitudinally-polarized $Z$ have no soft singularity. 
At very large c.o.m. energies the process $e^+ e^- \to hh Z_L$ dominates the total cross section and its leading contribution,  which is proportional to $\delta_b$, 
peaks at $m_{hh}/\sqrt{s} \sim 1/\sqrt{7}$, see eq.~(\ref{eq:diffsigmaL}).  The left plot of Fig.~\ref{dhsmhhinv} shows that for $\sqrt{s}=1\,$TeV the values of the differential 
cross section near the kinematic edge increases  when going from $\delta_b =0$ to $\delta_b =0.5$. This means that 
the contribution from transversely-polarized final $Z$ bosons is still large in this case, as also shown by the right plot of Fig.~\ref{fig:dhs2}. 
We have checked that for larger c.o.m. energies (or, similarly, much larger values of $\delta_b$), the differential
cross section $d\sigma/dm_{hh}$ eventually peaks at the intermediate values $m_{hh}/\sqrt{s} \sim 1/\sqrt{7}$.

\begin{figure}[t]
\begin{center}
\includegraphics[width=0.99\linewidth]{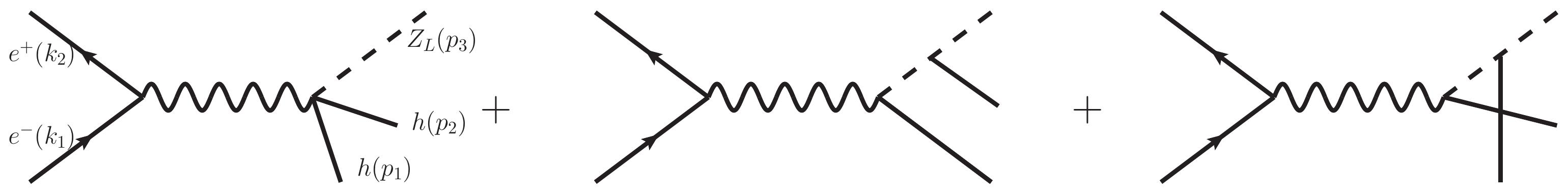}
\caption{\small Diagrams contributing to the leading high-energy behavior of the double Higgs-strahlung cross section.}
\label{diagdhseq}
\end{center}
\end{figure}

The high-energy limit of the double Higgs-strahlung total cross section can be easily calculated explicitly. 
In a gauge in which the equivalence theorem is manifest, the diagrams contributing to the leading high-energy behavior are those depicted in Fig.~\ref{diagdhseq}. 
%
%
%
%
The relevant vertices are found by expanding the Lagrangian of eq.~(\ref{eq:HiggsLag})
\be
\Delta\mathcal L= a\frac{h}{v}(\partial_\mu \chi^3)^2-m_Z \left(2 a \frac{h}{v}+b\frac{h^2}{v^2}\right)\partial_\mu \chi^3 Z^\mu.
\ee
Neglecting the masses of the initial state leptons, as well as  those of the Higgs and the $Z$ boson, the amplitude can be written as
\be
i\mathcal A(e^+e^-\to hhZ_L) \simeq (\sqrt 2 G_F)^{3/2} \frac{2m_Z^2}{s}(b-a^2)\bar v(k_2) \slashed p_3(g_V-g_A\gamma^5) u(k_2)\, ,
\ee
where $k_1$ and $k_2$ are the momenta of the incoming electron and positron, $p_3$ is the momentum of the outgoing $Z$ boson and $s$ is 
the center-of-mass energy. The vector and axial-vector couplings of the electron are given by
\be
g_V=-\frac{1}{2}+2 s_W^2,\quad g_A=-\frac{1}{2}.
\ee
Squaring and averaging the amplitude over the initial spins one gets
\be\label{sqamp}
\overline{|\mathcal A|^2} \simeq 8 m_Z^4(\sqrt 2 G_F)^3(g_V^2+g_A^2)(b-a^2)^2\frac{p_3\cdot k_1\, p_3\cdot k_2}{s^2}.
\ee
The total cross section is written as
\be
\sigma(e^+e^-\to hhZ_L) = \frac{1}{2}\times\frac{1}{2s}\int \overline{|\mathcal A|^2}\, d\Phi^{(3)}\, ,
\ee
where the extra $1/2$ factor accounts for the two Higgs particles in the final state.
The phase space integral can be done by using the recursive formula
\be\label{3bdph}
d\Phi^{(3)}(P;p_1,p_2,p_3)=\int \frac{d p_{12}^2}{2\pi} \, d\Phi^{(2)}(p_{12}; p_1,p_2) \, d\Phi^{(2)}(P; p_{12},p_3)\, .
\ee
Notice that $p_{12}^2$ is the invariant mass of the two Higgses. Since the amplitude in eq.~(\ref{sqamp}) does not depend on the momenta of the two Higgs bosons,
$p_1$ and $p_2$,  the first phase space integral in eq.~(\ref{3bdph}) is trivial and gives
\be
d\Phi^{(2)}(p_{12}; p_1,p_2)=\frac{1}{8\pi}\, .
\ee
Taking into account the energy and angular dependence of the amplitude one obtains
\begin{equation}
\label{eq:diffsigmaL}
\frac{d\sigma}{dm_{hh}^2} \simeq\frac{(\sqrt 2 G_F)^3(g_V^2+g_A^2)}{1536\pi^3} \, \frac{m_Z^4}{s} \, (b-a^2)^2\left(1-\frac{m_{hh}^2}{s}\right)^3\, , 
\end{equation}
and integrating over $0\leq m_{hh}^2\leq s$ it follows
\be
\sigma \simeq \frac{(\sqrt 2 G_F)^3(g_V^2+g_A^2)}{6144\pi^3} m_Z^4 (b-a^2)^2=0.15\,{\textrm{fb}} \, (b-a^2)^2\, .
\ee


\end{document}